\documentclass[prx,aps,superscriptaddress,nofootinbib,twocolumn,notitlepage,floatfix,10pt]{revtex4-2}
\usepackage{amsmath,mathtools,amsthm,amssymb,pifont}
\usepackage[utf8]{inputenc}
\usepackage[american]{babel}
\usepackage{graphicx,xcolor,bbold,titlesec}
\usepackage{braket}
\usepackage{MnSymbol}

\usepackage[colorlinks,
citecolor=red,
linkcolor=red,
urlcolor=red]{hyperref}
\usepackage{tikz,ifthen}
\usepackage{bbold}
\usepackage{tikz-network}
\usetikzlibrary{patterns,decorations.pathreplacing,calligraphy}
\usepackage{orcidlink}
\usepackage{color}
\usepackage{MnSymbol}

\begin{document}
\title{Fermionic Magic Resources of Quantum Many-Body Systems}

\author{Piotr Sierant~\orcidlink{0000-0001-9219-7274}}
\email{piotr.sierant@bsc.es}
\affiliation{Barcelona Supercomputing Center Plaça Eusebi G\"uell, 1-3 08034, Barcelona, Spain}

\author{Paolo Stornati~\orcidlink{0000-0003-4708-9340}}
\email{paolo.stornati@bsc.es}
\affiliation{Barcelona Supercomputing Center Plaça Eusebi G\"uell, 1-3 08034, Barcelona, Spain}

\author{Xhek Turkeshi~\orcidlink{0000-0003-1093-3771}}
\email{turkeshi@thp.uni-koeln.de}
\affiliation{Institut f\"ur Theoretische Physik, Universit\"at zu K\"oln, Z\"ulpicher Strasse 77, 50937 K\"oln, Germany}

\begin{abstract}
Understanding the computational complexity of quantum states is a central challenge in quantum many-body physics. 
In qubit systems, fermionic Gaussian states can be efficiently simulated on classical computers and hence can be employed as a natural baseline for evaluating quantum complexity. 
In this work, we develop a framework for quantifying fermionic magic resources, also referred to as fermionic non-Gaussianity, which constitutes an essential resource for universal quantum computation.
We leverage the algebraic structure of the fermionic commutant to define the fermionic antiflatness (FAF)--an efficiently computable and experimentally accessible measure of non-Gaussianity, with a clear physical interpretation in terms of Majorana fermion correlation functions.  
Studying systems in equilibrium, we show that FAF detects phase transitions, reveals universal features of critical points, and uncovers special solvable points in many-body systems. 
Extending the analysis to out-of-equilibrium settings, we demonstrate that fermionic magic resources become more abundant in highly excited eigenstates of many-body systems. We further investigate the growth and saturation of FAF under ergodic many-body dynamics, highlighting the roles of conservation laws and locality in constraining the increase of non-Gaussianity during unitary evolution. 
This work provides a framework for probing quantum many-body complexity from the perspective of fermionic Gaussian states and opens up new directions for investigating fermionic magic resources in many-body systems. 
Our results establish fermionic non-Gaussianity, alongside entanglement and non-stabilizerness, as a resource relevant not only to foundational studies but also to experimental platforms aiming to achieve quantum advantage.
\end{abstract}

\maketitle

\section{Introduction}
The exact classical representation of a pure quantum state of a system of $N$ qubits requires a number of parameters that scales exponentially with $N$. Consequently, the operation of a quantum device containing sufficiently many qubits may be intractable for classical computers, as indicated by complexity theory arguments~\cite{aaronson11the, bouland19on}. This insight forms the foundation for quantum simulators and quantum computers~\cite{40years}, and suggests the possibility of achieving computational quantum advantage~\cite{preskill12quantum, daley22practical}, where a quantum device outperforms classical computers in a specific computational task.

However, the description of quantum states is only \emph{at most} exponentially costly in $N$ for classical computers. Many physical problems exhibit structures that can be exploited to drastically reduce the cost of storing and processing the quantum states. For this reason, understanding the computational complexity of quantum many-body systems is a central goal in quantum information science, condensed matter, and high-energy physics~\cite{haferkamp2022linear, Ayral23, hangleiter2023computational, FausewehQMB2024, Baiguera2025complexity}. A key question in this endeavor is identifying which quantum states can be classically simulated efficiently, and which may exhibit genuine quantum computational power. Resource theory provides a unified and versatile framework to tackle these problems~\cite{chitambar2019quantumresourcetheories}.

A cornerstone example is the theory of entanglement. 
Weakly entangled states can be efficiently represented with tensor network methods at cost scaling polynomially in $N$~\cite{Vidal03a, Orus14,  Orus19, ran2020tensor, Banuls23}. 
This fact plays an essential role in our understanding of equilibrium many-body systems, since the ground states of local, gapped Hamiltonians obey area-laws for entanglement entropy~\cite{Eisert10area}, and, in 1D systems, can be efficiently represented using matrix product states (MPS)~\cite{Verstraete08, Schuch08, Schollwock2011}. 
In contrast, states with volume-law entanglement -- arising, for instance, in out-of-equilibrium scenarios~\cite{Dalessio16}, elude efficient descriptions in terms of tensor network states. 
As a result, extensive entanglement~\cite{amico2008entanglement, horodecki2009quantum} is considered a necessary condition for a quantum state to possess genuine computational power.

Other important classes of quantum states can be efficiently simulated with classical computational resources that scale only polynomially with the system size $N$. Two prominent examples are stabilizer states~\cite{Gottesman1998, aaronson2004improvedsimulationof} and fermionic Gaussian states~\cite{Valiant01, Terhal02, Bravyi05flo}. Remarkably, both classes admit efficient classical representations even when the states are strongly entangled -- highlighting that extensive entanglement alone is not sufficient to achieve quantum advantage.

The classical tractability of stabilizer states motivates the development of the resource theory of non-stabilizerness, often referred to as ``magic" resources~\cite{Veitch2014theresourcetheory, Bravyi16, Howard_2017, Wang_2019, Heimendahl21xtent, liu2022manybody}. 
Measures of non-stabilizerness remain non-increasing under stabilizer protocols~\cite{Heimendahl22}, which include the action of Clifford circuits and computational basis measurements. 
In contrast, non-stabilizerness can increase under non-stabilizer operations, such as the application of the $T$-gate~\cite{Bravyi_2005}, which, together with Clifford gates, form a universal set of quantum gates~\cite{Barenco95}. 
Measures of non-stabilizerness, such as the stabilizer rank~\cite{Bravyi16, Bravyi16improved}, stabilizer fidelity~\cite{Bravyi19simulatio}, or the magic robustness~\cite{Howard_2017, Heinrich19robustness, Sarkar20rom}, provide bounds on the increasing complexity of representing quantum states as they deviate further from the set of stabilizer states. 
However, computing these measures involves costly minimization procedures which become prohibitive already for systems containing fewer than $10$ qubits~\cite{Hamaguchi24handbook}. This highlights the importance of the recently introduced stabilizer R\'{e}nyi entropy (SRE)~\cite{leone2022stabilizerrenyientropy} which not only exhibits the desirable properties of a non-stabilizerness measure~\cite{leone2024stabilizer}, but also can be efficiently computed classically~\cite{haug2023quantifying, lami2023perfect, tarabunga23gauge, tarabunga2024mps}, and measured experimentally~\cite{Oliviero2022measuring, haug2024efficient, Haug23scalable}. 
The introduction of SRE has led to intensive activities aimed at understanding the non-stabilizerness and the ensuing complexity of many-body states in various quantum phases and at quantum phase transitions~\cite{White21, haug2023quantifying, tarabunga23gauge,Odavic23,  Passarelli24, Tarabunga24rk, Falcao25, Jasser2025, Bera2025SYK}, and in out-of-equilibrium scenarios~\cite{rattacaso2023stabilizer, Turkeshi25spectrum, turkeshi2024magic, Tirrito24anti, Odavic2025, haug2024saturation, kos2024exact, Santra25complexity, Passarelli25boundary, Hou2025highway}.

Fermionic Gaussian states are generated from the vacuum state by Gaussian unitaries, i.e., transformations generated by Hamiltonians quadratic in Majorana fermion operators~\cite{Majorana2006, Bravyi05flo}, or, equivalently, by matchgate circuits~\cite{Valiant01, Terhal02, Jozsa08}. 
The classical representation of fermionic Gaussian states requires resources scaling only polynomially with $N$, enabling insights into the equilibrium and non-equilibrium physics of systems that can be mapped to non-interacting fermionic models such as the transverse field Ising or XY models~\cite{Surace22, Mbeng24}. 
Non-Gaussian operations, associated, e.g., with interactions, drive the state away from the manifold of fermionic Gaussian states, rendering classical simulations inefficient~\cite{Oszmaniec22}. 

This motivates the development of the theory of fermionic magic resources, also known as the theory of fermionic non-Gaussianity.
Measures of fermionic non-Gaussianity, such as fermionic rank and fermionic Gaussian extent~\cite{Hebenstreit19, Cudby24gaussian, Reardon24extent, Bittel24optimal, Hakkaku22,Bittel24fermionic}, remain invariant under Gaussian unitaries, but may increase when non-Gaussian operations are applied, increasing the computational complexity of the state~\cite{ReardonSmith24improved, Dias24classical, Mele25learning}. 
However, computing the fermionic rank or the fermionic Gaussian extent, in analogy to their stabilizer counterparts, requires complex minimization procedures, restricting their usefulness in studies of many-body systems. 
Several works~\cite{Gottlieb05, Gottlieb07, Lumia24, Lyu24NGE,Coffman25magic} have proposed measures of fermionic magic resources, which can be classically computed without the need for costly minimization procedures.

Building on these foundations, this work introduces a systematic and physically motivated framework for constructing efficiently computable measures of fermionic magic resources grounded in the algebraic theory of the commutant of Gaussian unitaries. Inspired by the construction of SRE, we define fermionic antiflatness (FAF), a family of measures of fermionic non-Gaussianity. The FAF vanishes for Gaussian states, remains invariant under Gaussian operations, and increases with the strength of non-Gaussian correlations. Crucially, the FAF can be efficiently computed in large systems with tensor network techniques and can also be measured straightforwardly in experiments. Through this construction, we bring the power of resource-theoretic reasoning into the fermionic domain while retaining a clear physical interpretation of the proposed measure by its relation to correlations between Majorana operators.

We explore the behavior of the FAF in a variety of physically and computationally relevant settings, considering both equilibrium and out-of-equilibrium scenarios. After establishing the framework for constructing efficiently computable measures of fermionic non-Gaussianity and formally defining the FAF in Sec.~\ref{sec:methods}, we study its properties in representative model systems in Sec.~\ref{sec:modelSystems}. We begin with analytically tractable cases, such as product states and typical Haar-random states, which serve as reference points for our further investigations. We then examine the FAF in random matrix product states and in random quantum circuits. The former interpolate between the structureless Haar-random states and weakly entangled states with finite-range correlations, while the latter constitute minimal models of local many-body dynamics. These two settings provide the basis for studies of the FAF in ground states of local Hamiltonians and in out-of-equilibrium many-body systems, respectively. In Sec.~\ref{sec:equilibrium}, we investigate the FAF across distinct phases of matter and analyze how critical properties of quantum phase transitions are reflected in its behavior. Subsequently, in Sec.~\ref{sec:outofeq}, we examine the FAF in the context of out-of-equilibrium ergodic many-body systems. We investigate the properties of the FAF in highly excited eigenstates of ergodic systems and analyze the growth of the FAF under ergodic dynamics from initial fermionic Gaussian states. Section~\ref{sec:overview} contains a brief overview of our results, while in Sec.~\ref{sec:conclusion}, we provide an outlook and conclude our work.

\begin{figure}
    \centering
    \includegraphics[width=1\linewidth]{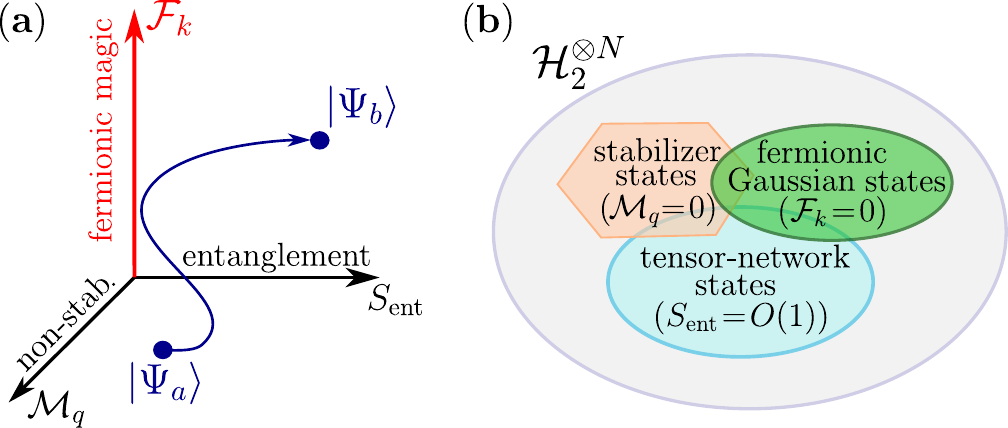}
    \caption{ 
    Resources characterizing the computational complexity of quantum many-body states $\ket{\Psi_{a,b}}$ of $N$ qubits. (a) Entanglement is a central tool for analyzing many-body phenomena and has recently been complemented by measures of non-stabilizerness (non-stab.). This work introduces a framework for studying \textit{fermionic magic resources}, proposing the \textit{fermionic antiflatness} (FAF), $\mathcal{F}_k$, as a diagnostic in many-body systems.
    (b) Hilbert space $\mathcal{H}_2^{\otimes N}$ for $N$ qubits, showing the classes of computationally tractable states: tensor-network states with limited entanglement, $S_{\mathrm{ent}}=O(1)$, stabilizer states with vanishing SRE, $\mathcal{M}_q=0$, and fermionic Gaussian states with vanishing FAF, $\mathcal{F}_k=0$.
    }
    \label{fig:sketch}
\end{figure}

In summary, this work serves two main purposes. First, it introduces FAF as a measure of fermionic magic resources, rooted in the algebraic properties of Gaussian unitaries, that is both efficient to compute and simple to interpret. Second, it examines the behavior of the FAF in various fundamental physical scenarios, including quantum phase transitions and ergodic dynamics, to establish benchmarks for future studies of fermionic non-Gaussianity in more complex settings. Altogether, this provides a coherent framework for quantifying fermionic magic resources and offers a new perspective on the analysis of the complexity of quantum many-body states, as shown in Fig.~\ref{fig:sketch}.

\section{Overview}
\label{sec:overview}
We consider a system of $N$ qubits, or equivalently, via the Jordan-Wigner transformation~\cite{Jordan1928}, a system of $N$ fermionic modes. The $2N$ Majorana operators $\gamma_k$, acting on the $d=2^N$ dimensional Hilbert space~$\mathcal{H}_2^{\otimes N}$ of the system, can be represented in terms of the Pauli operators $ X_i, Y_i, Z_i$ acting on the $i$-th qubit as
\begin{equation}
  \gamma_{2k-1} = \left(\prod_{m=1}^{k-1} Z_m\right)X_k\;,\,\,\,\,
   \gamma_{2k} = \left(\prod_{m=1}^{k-1} Z_m\right)Y_k\;.
\label{eq:JW}
\end{equation}
The Majorana operators are traceless, Hermitian, $\gamma^\dagger_k = \gamma_k $, and satisfy the anticommutation relations $\{ \gamma_k, \gamma_l \} = 2 \delta_{kl} \mathbb{1}$. The Jordan-Wigner transformation, Eq.~\eqref{eq:JW}, enables the identification of the states of $N$ qubits with those of $N$ fermions. In particular, the vacuum state is given by $\ket{\mathbf{0}} = \ket{0}^{\otimes N} \in \mathcal{H}_2^{\otimes N}$, and fermionic Gaussian unitaries, i.e., operators of the form $U_G=\exp\left[\frac{1}{4}\sum_{m,n=1}^{2N} H_{m,n} \gamma_m \gamma_n \right]$, where $H=(H_{m,n})$ is a $2N\times 2N$ antisymmetric matrix, generate Fermionic Gaussian states as $\ket{\Psi_G} = U_G\ket{\mathbf{0}}$. The covariance matrix $M=(M_{m,n})$ is defined for any state $\ket{\Psi}$ by its matrix elements $M_{mn}=-\frac{i}{2}\langle \Psi|[\gamma_m,\gamma_n]|\Psi\rangle$.  The covariance matrix of the fermionic Gaussian state $\ket{\Psi_G}$ encodes, through Wick's theorem~\cite{Negele18quantum}, the complete information about all correlation functions characterizing $\ket{\Psi_G}$. This fact, together with the observation that storing and processing the covariance matrix $M$ can be done classically with resources scaling polynomially with $N$, is the root cause for the simulability of fermionic Gaussian states.

A measure $\varphi$ of fermionic magic resources is a function mapping an arbitrary state $\ket{\Psi} \in \mathcal{H}_2^{\otimes N}$ to a non-negative real number, satisfying the following properties:  (i) faithfulness — $\varphi(\ket{\Psi}) = 0$ if and only if $\ket{\Psi}$ is a fermionic Gaussian state;  (ii) Gaussian invariance — $\varphi(U_G \ket{\Psi}) = \varphi(\ket{\Psi})$ for all Gaussian unitaries $U_G$; iii) (sub)additivity when $\ket{\Psi}$ is a product state. 
To find efficiently computable measures $\varphi$, we consider quantities of the form $\varphi(\ket{\Psi}) = \mathrm{tr}\left(W\,|\psi\rangle\langle \psi|^{\otimes k}\right)$, where $k$ is a fixed number and $W$ is an operator. In Sec.~\ref{sec:methods}, we argue that for $\varphi$ to be Gaussian invariant, the operator $W$ must belong to the commutant of the group of fermionic Gaussian unitaries. This condition yields a broad class of potential measures of fermionic non-Gaussianity, from which we select the fermionic antiflatness (FAF), defined as 
\begin{equation}
    \mathcal{F}_k(|\Psi\rangle) = N-\frac{1}{2}\mathrm{tr}\left[(M^T M)^k\right]\;.\label{eq:fafOVER}
\end{equation}
The FAF satisfies the desired properties of faithfulness, Gaussian invariance and additivity on tensor products of states with fixed fermionic parity $\mathcal{P}=(-i)^N\prod_{m=1}^{2N} \gamma_m$ (the FAF is subadditive on tensor products of states which are not eigenstates of $\mathcal P$). Moreover, the FAF is a simple function of the elements $M_{mn}$ of the covariance matrix, i.e., the 2-point correlation functions of Majorana operators, $\langle \Psi|\gamma_m \gamma_n|\Psi\rangle$. This property facilitates the interpretation of FAF in various physical contexts and enables its experimental measurements.

To put our investigations of fermionic non-Gaussianity in the broader context of resources characterizing the complexity of many-body states, we draw parallels and contrasts between the FAF and the behavior of entanglement entropy and SRE in various physical phenomena. The von Neumann entanglement entropy of a state $|\Psi\rangle$ for a bipartition of the system into a subsystem $A$ and its complement $B$ is given by
\begin{equation}
   S_{\mathrm{ent}}(|\Psi\rangle) = -\mathrm{tr}_A \left(  \rho_A \log_2( \rho_A)\right),\label{eq:ent}
\end{equation}
where $\rho_A = \mathrm{tr}_B (\ket{\Psi}\bra{\Psi})$ is the reduced density matrix in $A$. 
The SRE~\cite{leone2022stabilizerrenyientropy} is defined as
\begin{equation}
    \mathcal{M}_{q}(|\Psi\rangle) = \frac{1}{1-q}\log_2\left[\sum_{P\in P_N}
  \frac{\langle \Psi | P |\Psi\rangle^{2q}}{d} \right]\;,\label{eq:sre}
\end{equation}
where $P_N =\{ X_1^{r_1^x} Z_1^{r_1^z} \cdots X_N^{r_N^x} Z_N^{r_N^z} \;| \; r_k^\alpha \in \mathbb{Z}_2\}$ is the set of Pauli strings, $q \geq2$ is an integer, and $d=2^N$.

The additivity of FAF for product states with fixed fermionic parity $\mathcal{P}$ implies that $\mathcal{F}_k$ can scale extensively with the system size $N$, even when the entanglement is limited, $S_{\mathrm{ent}} = O(1)$. This suggests a generic linear scaling of FAF with system size,
\begin{equation}
\label{eq:NdepOverview}
\mathcal{F}_k(\ket{\Psi}) = D_k N + f_k,
\end{equation}
where $D_k$ and $f_k$ are, respectively, the leading and subleading terms, which we show to be relevant in various physical scenarios. For Haar-random states, the FAF is nearly maximal, with $D_k = 1$ and a subleading term that decays exponentially with system size, $f_k = O(d^{-k})$. This behavior is analogous to the near-maximal entanglement entropy~\cite{Page93} and SRE~\cite{Turkeshi25spectrum} observed in Haar-random states. Random matrix product states (RMPS)~\cite{Garnerone10, Garnerone10a, Haferkamp21, Lami25anti} with fixed bond dimension $\chi$ also exhibit the linear scaling of FAF as in Eq.~\eqref{eq:NdepOverview}, but with a leading term given by $D_k = 1 - a_k/\chi^{\beta_k}$, where $a_k$ and $\beta_k$ are constants (numerically, we find $\beta_1 \approx 2$, and $\beta_2$ is close to $4$). This shows that fermionic magic resources in RMPS with fixed $\chi$ are less abundant than in Haar-random states, providing a baseline for investigating FAF in many-body ground states.

The transverse field Ising model (TFIM), whose ground state is a fermionic Gaussian state, is the starting point for our study of FAF in equilibrium systems. The ground state departs from the manifold of fermionic Gaussian states when non-Gaussian impurity terms are introduced into the TFIM Hamiltonian. First, we consider a model with a single impurity term, and then extend the setup by adding an extensive number of impurity terms, resulting in the axial next-nearest-neighbor Ising (ANNNI) model~\cite{Selke1988}. The Ising phase transition occurs in both the impurity and ANNNI models, and the critical point is described by conformal field theory, as manifested, for example, by the logarithmic divergence in $N$ of the entanglement entropy $S_{\mathrm{ent}}$ at the transition~\cite{calabrese2009entanglement}.
The single impurity introduces only a finite amount of FAF to the ground state, the leading term is vanishing, $D_k = 0$, and $f_k$ is a constant dependent on the value of the field $h$. The phase transition is reflected in a logarithmic divergence of the derivative of $f_k$ with respect to $h$.
In contrast, the extensive number of non-Gaussian terms in the ANNNI model leads to the extensive scaling of FAF, as in Eq.~\eqref{eq:NdepOverview}, with the leading term $D_k$ varying smoothly with the field strength $h$. The behavior of the subleading term at the transition depends on the choice of boundary conditions (we consider periodic boundary conditions (PBC) or open boundary conditions (OBC)), and is given as
\begin{equation}
  f_k = 
  \begin{cases}
    a_k \ln(N) + c_k, & \text{for OBC},\\
    c_k, & \text{for PBC},
  \end{cases}
  \label{eq:f3Overview}
\end{equation}
where $a_k$ and $c_k$ are constants. The presence of a logarithmic divergence that depends on the boundary conditions is a characteristic feature of boundary-changing operators in the conformal field theory describing the critical point~\cite{Cardy08boundaryCFT}. Similar behavior, analogous to FAF in Eq.~\eqref{eq:f3Overview}, is also observed at critical points in the participation entropy~\cite{Luitz14}. The behavior of FAF, both away from the critical point and at the transition, can be understood from the generic scaling forms of the two-point Majorana correlation functions. As the final point in our study of many-body ground states, we note that FAF clearly identifies the Peschel–Emery point~\cite{Peschel81}, at which the ground state of the ANNNI model becomes exactly solvable, even though other quantities, such as correlation functions or entanglement entropy, vary smoothly in the vicinity of this point.

Numerical study of highly excited eigenstates of the impurity and ANNNI models indicates that FAF becomes extensive at any finite energy density, approaching the Haar-random state values in eigenstates near the middle of the spectrum as the system size $N$ increases. Under the dynamics of brick-wall Haar-random circuits, starting from an initial fermionic Gaussian state, FAF becomes extensive, $\mathcal{F}_k \propto N$, already at constant circuit depth $t = O(1)$. Similar to the behavior of SRE~\cite{turkeshi2024magic} and participation entropy~\cite{turkeshi2024hilbert}, FAF saturates, up to a fixed accuracy $\epsilon$, to the Haar-random value at times scaling logarithmically with system size, $t_{\mathrm{sat}} \propto \log N$, which is much shorter than the time $t \propto N$ required for entanglement entropy to saturate. These results for random circuit dynamics serve as a reference for analyzing time evolution in the ergodic impurity and ANNNI models. At short times, the ANNNI model exhibits behavior resembling that of random circuits, with a rapid increase in FAF to extensive values at $t = O(1)$. In contrast, the impurity model displays a slower, ballistic growth of fermionic magic resources, $\mathcal{F}_k \propto t$. The latter can be attributed to the ballistic propagation of information about the non-Gaussian impurity, while the former results from the presence of extensively many impurities. Despite these differences, at longer times both models exhibit saturation of $\mathcal{F}_k$ at times scaling linearly with system size, $t_{\mathrm{sat}} \propto N$, significantly longer than the saturation time of FAF observed in random circuits. We argue that this is a generic behavior originating from the energy conservation in dynamics of ergodic many-body systems.

\section{Methods}
\label{sec:methods}
In this section, we describe the framework for identifying efficiently computable measures of fermionic magic resources. We begin by fixing the notation and outlining the basic properties of fermionic Gaussian unitaries. It is shown that the $k$-th fermionic commutant provides a range of candidates for fermionic magic measures, from which we select the FAF. We then demonstrate that the FAF satisfies the desired properties of a fermionic non-Gaussianity measure and comment on its computational and experimental relevance.

\subsection{Preliminaries}
\label{sec:prelim}
\textit{Majorana Strings.---}
The Jordan-Wigner transformation, cf. Eq.~\eqref{eq:JW}, relates the Pauli operators $ X_i, Y_i, Z_i$ acting on $i$-th qubit to fermionic Majorana operators $\gamma_k$. 
Given an ordered subset $S=\{m_1,m_2,\dots,m_{|S|}\}\subset [2N]$~\footnote{We denote the set $[k]\equiv \{1,2,\dots,k\}$ for any positive integer $k$.}, we define the Majorana strings by
\begin{equation}
    \gamma_S\equiv \gamma_{m_1}\gamma_{m_2}\cdots\gamma_{m_{|S|}}
\end{equation}
with $\gamma_{\varnothing}=\mathbb{1}$ denoting the identity operator. 
Generally, throughout this work, we use uppercase letters for subsets and lowercase letters for individual indices.
Analogously to the Pauli strings, Majorana strings define an orthogonal basis in the space of operators acting on $\mathcal{H}_2^{\otimes N}$, with $\mathrm{tr}(\gamma_S \gamma_R^\dagger)= d\delta_{S,R}$, where $d=2^{N}$. 
Let us denote by $\binom{[2N]}{\kappa}$ the subsets $S\subset [2N]$ of cardinality $|S|=\kappa$. 
The Majorana strings with $\kappa$ Majorana modes define the independent subspaces
\begin{equation}
    \mathcal{B}_\kappa = \mathrm{span}\left\lbrace\gamma_S \;|\; S\in \binom{[2N]}{\kappa}\right\rbrace\;,
\end{equation}
for $\kappa \in \{0,1,\dots,2N\}$. 
Since Majorana strings define a basis in the operator space, any linear operator $A$ can be decomposed as
\begin{equation}
    A=\frac{1}{d}\sum_{\kappa=0}^{2N}  \sum_{S\in \binom{[2N]}{\kappa}} A_S \gamma_S,\quad A_S=\mathrm{tr}(\gamma_S^\dagger A)\;.
\end{equation}
We say an operator $A$ is (fermionically) even if all coefficients $A_S=0$ for $|S|$ odd. 
The fermionic parity operator
$\mathcal{P}=(-i)^N\prod_{m=1}^{2N} \gamma_m = \prod_{k=1}^{N} Z_k$
commutes with even operators.

\textit{Fermionic Gaussian Unitaries.---}
Fermionic Gaussian unitaries (FGU) are the subgroup of the unitary group $\mathrm{U}(d)$ comprising unitaries generated by operators quadratic in Majorana operators
\begin{equation}
    U_G=\exp\left[\frac{1}{4}\sum_{m,n=1}^{2N} H_{m,n} \gamma_m \gamma_n \right]\;,
    \label{eq:fgu}
\end{equation}
where $H=-H^T$ is a $2N\times 2N$ antisymmetric matrix.
Majorana fermions transform covariantly under FGU, namely
\begin{equation}
    U^\dagger_G \gamma_m U_G = \sum_{n=1}^{2N} G_{m,n}\gamma_n\;,
    \label{eq:cov}
\end{equation}
where $G\in \mathrm{SO}(2N)$ is a special orthogonal $2N\times2N$ matrix, determined by the matrix $H=(H_{m,n})$ as $G=\exp(H)$\footnote{Vice versa, for any rotation $G\in \mathrm{SO}(2N)$, there exists an anti-symmetric matrix $H$, such that $\exp(H)=G$, which generates the FGU $U_G$.}.
Using the anticommutation relations, it is easy to show that generic Majorana strings transform as 
\begin{equation}\label{eq:majonum}
    U_G^\dagger \gamma_S U_G = \sum_{S'\in \binom{[2N]}{|S|}} \det(G|_{S,S'}) \gamma_{S'},
\end{equation}
where $G|_{S,S'}$ denotes the restriction of $G$ on the rows indexed by $S$ and the columns indexed by $S'$. 
From Eq.~\eqref{eq:majonum}, it follows that the size $|S|$ of the Majorana string is invariant under FGU, implying that the subspaces $\mathcal{B}_\kappa$ are invariant under $U_G$ for any $G\in \mathrm{O}(2N)$. 
We denote the group of fermionic Gaussian transformations acting on $2N$ Majorana modes by $\mathfrak{G}_N$.

Matchgate circuits are the qubit representation of fermionic Gaussian unitaries under the Jordan-Wigner transformation.  
Matchgates are a particular class of two-qubit gates generated by two-qubit $X$ rotations $\exp[i\theta X\otimes X]$, and single-qubit $Z$ rotations $\exp[i\theta (I\otimes Z)]$ and $\exp[i\theta (Z\otimes I)]$.  
Eq.~\eqref{eq:JW} implies that
\begin{equation}
    [V_{mn}(\theta)]^\dagger \gamma_l V_{mn} = \begin{cases}
        \cos(\theta)\gamma_m + \sin(\theta)\gamma_n &m=l \\
        -\sin(\theta)\gamma_m + \cos(\theta)\gamma_n &m=n\\
        \gamma_l & \mathrm{otherwise},
    \end{cases}\label{eq:rotat}
\end{equation}
where $ V_{mn}(\theta)=\exp[ \theta \gamma_m \gamma_n/2]$ are rotations in the $(m,n)$-plane. 
Since $i X_m X_{m+1} = \gamma_{2m}\gamma_{2m+1}$ and $i Z_m=\gamma_{2m-1}\gamma_{2m}$, the $X\otimes X$ rotations and single-qubit $Z$ rotations implement fermionic Gaussian unitaries in the $(2m,2m+1)$ and $(2m-1,2m)$ planes, respectively. 
The set of these transformations for all allowed $m\in [2N]$ generates all rotations in $\mathrm{SO}(2N)$. 
Adding the reflection operator $X_N$ generates the whole $\mathrm{O}(2N)$, since this operator acts as $\gamma_m\mapsto \gamma_m$ for $m<2N$ and $\gamma_{2N}\mapsto -\gamma_{2N}$. For our purposes, the matchgate unitaries and fermionic Gaussian unitaries are used interchangeably, and $\mathfrak{G}_N$ also denotes the matchgate group.

\textit{Fermionic Gaussian States.---}
A pure fermionic Gaussian state, also referred to as a free fermionic state, is defined by $|\Psi_G\rangle = U_G|\mathbf{0}\rangle$, where $U_G\in \mathfrak{G}_N$ is the Gaussian unitary induced by $G\in \mathrm{O}(2N)$, and 
\begin{equation}
    |\mathbf{0}\rangle\langle \mathbf{0}| = \frac{1}{d}\prod_{m=1}^N ( \mathbb{1}+Z_m)=\frac{1}{d}\prod_{m=1}^N (\mathbb{1}-i \gamma_{2m-1}\gamma_{2m})\;,
\end{equation}
is the reference vacuum state $|\mathbf{0}\rangle=|0\rangle^{\otimes N}$. 
Since $\tilde\gamma_m=U_G^\dagger \gamma_m U_G =\sum_{n} G_{mn}\gamma_n$ with $G\in \mathrm{O}(2N)$, generic pure fermionic states $|\Psi_G\rangle = U_G|\mathbf{0}\rangle$ are given by
\begin{equation}
\begin{split}
        |\Psi_G\rangle\langle \Psi_G| &= \frac{1}{d}\prod_{m=1}^N \left(\mathbb{1}-i \sum_{n,l=1}^N G_{2m-1,n} G_{2m,l}\gamma_{n}\gamma_{l} \right) \\
        &= \exp\left[\frac{i}2\sum_{m,n} M_{mn}\gamma_m\gamma_n\right]\;,
\end{split}
\label{eq:fermionicG}
\end{equation}
where the second line follows from algebraic manipulations using the anticommutation relations, and we introduced a $2N\times 2N$ antisymmetric matrix $M$ known as the covariance matrix, with matrix elements 
\begin{equation}
M_{mn} = -\frac{i}{2}\langle \Psi_G|[\gamma_m,\gamma_n]|\Psi_G\rangle.
\label{eq:covMAT}
\end{equation}
The covariance matrix of the vacuum state $|\mathbf{0}\rangle$ is
\begin{equation}
\label{eq:covVAC}
    M_0= \bigoplus_{m=1}^N \begin{pmatrix}
        0 & 1\\ -1 & 0
    \end{pmatrix}\;,
\end{equation}
and the covariance matrix of any Gaussian state $\ket{\Psi} = U_G \ket{\mathbf{0}}$, due to Eq.~\eqref{eq:cov}, can be obtained as $M=G M_0G^T$.
The above discussion pinpoints how fermionic Gaussian states are fully characterized by their covariance matrix $M$. 
Indeed, a fundamental characterization of free fermionic states is the \textit{Wick theorem}: a state $|\Psi\rangle$ with covariance matrix $M$ is a fermionic Gaussian state if and only if 
\begin{equation}
    \langle \Psi | \gamma_S |\Psi\rangle = \sum_{P\in H_{2l}} \mathrm{sign}(P)\prod_{\{m,n\}\in P} M_{mn}
\end{equation}
for any $S$ of cardinality $|S|=2l$ even, where $H_{2l}$ denotes the set of perfect matchings of the indices $S = \{m_1, m_2, \dots, m_{2l}\}$, and $\mathrm{sign}(P)$ accounts for the sign of the permutation needed to reorder the operators into the matched pairs (due to the anticommutation relations).
All free fermionic states are generated by even operators, meaning that $\langle \Psi|\gamma_S |\Psi\rangle = 0$ for any $|S|$ odd. 

\textit{Choi representation.---}
In the following, we use the Choi representation~\cite{Jamiolkowski72,Choi1975}. Consider the Choi state $|\Phi_+\rrangle\equiv \sum_{x=0}^{d-1} |x\rangle\otimes |x\rangle/\sqrt{d}$, where $\{ \ket{x} \}$ are the computational basis states. For an operator $A$, we define its Choi representation by $|A\rrangle = (A\otimes I) |\Phi_+\rrangle$. 
Within this framework, $\mathrm{tr}(A^\dagger B) = \llangle A | B\rrangle$ and $U A U^\dagger \mapsto (U\otimes U^*)|A\rrangle\equiv \mathcal{U}|A\rrangle$. 
In particular, for any Majorana string $\gamma_S$, its Choi representation is denoted $|\gamma_S\rrangle$, and $\llangle \gamma_S|\gamma_{S'}\rrangle = \delta_{S,S'}/d$. 
For notational convenience, we denote $\mathcal{U}_G=U_G\otimes U_G^*$, the unitary adjoint action of FGU in the Choi representation. 

\subsection{The fermionic Gaussian commutant}
Our analysis of the measures of fermionic magic resources parallels the systematic construction of nonstabilizerness measures from the algebraic structure of the Clifford group, see \cite{turkeshi2024magic,Bittel25commutant}. 
We briefly discuss the algebraic structure of the fermionic Gaussian commutant theory~\cite{Wan23,Diaz2023plateau,braccia2024computing}, and defer a systematic mathematical discussion of the fermionic commutant for future work.

In analogy with the stabilizer formalism, our goal is to construct computable measures $\varphi$ of the non-Gaussianity that remain tractable both analytically and numerically. 
These measures are inherently nonlinear functions of the state and thus require access to multiple copies of the investigated state, $\ket{\Psi}\bra{\Psi}$. To linearize the problem, we consider quantities of the form  
\begin{equation}
\varphi(\ket{\Psi}) =\mathrm{tr}\left[W\, (|\Psi\rangle\langle \Psi|)^{\otimes k}\right]\;,\label{eq:obs}
\end{equation} 
where \(W\) is an observable defined on the replicated Hilbert space $(\mathcal{H}_2^{\otimes N})^{\otimes k}$. Such replica correlation functions can be calculated solely by evaluation of the trace in Eq.~\eqref{eq:obs}, i.e., they do not require any minimization procedures. 
When  $W$ admits an efficient classical representation such as a matrix product operator of limited bond dimension, tensor network methods may facilitate the evaluation of Eq.~\eqref{eq:obs}. 
Finally, since $\varphi(\ket{\Psi})$ is linear in \( (|\Psi\rangle\langle \Psi|)^{\otimes k}\), it can often be measured experimentally using multi-copy protocols, such as the randomized measurements~\cite{Elben2023}, classical shadows~\cite{Huang_2020}, or variants of SWAP tests~\cite{Buhrman01}.

To ensure that $\varphi(\ket{\Psi})$ captures genuinely non-Gaussian features, we further constrain \(W\) to be invariant under fermionic Gaussian unitaries $U_G$. This naturally leads to the study of the commutant of the Gaussian group, which provides a structured framework for constructing measures of fermionic non-Gaussianity. 
As discussed previously, fermionic Gaussian unitaries correspond (up to global phases) to orthogonal transformations \(G \in \mathrm{O}(2N)\), offering a direct bridge between symmetries of multi-copy Gaussian transformations and replica-invariant observables.
We define the $k$-th commutant of the fermionic Gaussian unitaries, in short $k$-th fermionic commutant, as the set of linear operators acting on $(\mathcal{H}_2^{\otimes N})^{\otimes k}$ such that
\begin{equation}
    \mathrm{Comm}_k(\mathfrak{G}_N) \equiv \{ W \;|\; [W,U_G^{\otimes k}]=0\;, \forall G\in\mathrm{O}(2N)\}\;.
\end{equation}
A natural family of these operators for $k >1$\footnote{For $k=1$ the only operator in $\mathrm{Comm}_1(\mathfrak{G}_N)$ is proportional to identity, see Appendix~\ref{app:COMMexampl}.} comes from 
\begin{equation}\label{eq:comop}
    |\Upsilon_{r_1,r_2,\dots,r_k}^{(k)}\rrangle \equiv \sum_{\substack{A_1,\dots,A_k\subset [2N] \\ A_m\cap A_n=\varnothing\; \forall m\neq n \\ |A_m|=r_m \; \forall m\in [k]}} \bigotimes_{m=1}^{k}|\gamma_{A_m}\gamma_{A_{m+1}}\rrangle 
\end{equation}
where $A_{k+1}\equiv A_1$~\cite{Wan23}. 
We now show that each of these operators belongs to the $k$-th fermionic commutant $\mathrm{Comm}_k(\mathfrak{G}_N)$. 
The condition $[W,U_G^{\otimes k}]=0$ rephrases to $\mathcal{U}_G^{\otimes k}|W\rrangle = |W\rrangle$ in the Choi formalism. Thus, we must check
\begin{equation}\label{eq:condition}
    \mathcal{U}_G^{\otimes k}|\Upsilon_{r_1,r_2,\dots,r_k}^{(k)}\rrangle = |\Upsilon_{r_1,r_2,\dots,r_k}^{(k)} \rrangle\;,
\end{equation}
for any orthogonal transformation $G\in \mathrm{O}(2N)$. 
A proof of this statement, together with additional remarks on  $|\Upsilon_{r_1,r_2,\dots,r_k}^{(k)}\rrangle$ is detailed in the Appendix~\ref{app:commutant}.

For even $k$, we introduce the fermionic overlap
\begin{equation}
\label{eq:fermOV1}
    \zeta_{r_1,r_2,\dots,r_k} \equiv \llangle \Upsilon_{r_1,r_2,\dots,r_k}^{(k)}|\rho^{\otimes k}\rrangle\;,
\end{equation}
for any $r_1,r_2,\dots,r_k$ such that $0<\sum_{j} r_j\le 2N$, where $\rho^{\otimes k}=(|\Psi\rangle\langle \Psi|)^{\otimes k}$. 
It is important to stress that for any Gaussian state $|\Psi\rangle = U_G|\mathbf{0}\rangle$, the corresponding fermionic overlap is just a combinatorial number
\begin{equation}
    \zeta_{r_1,r_2,\dots,r_k}= \llangle \Upsilon_{r_1,r_2,\dots,r_k}^{(k)}| \left(|\mathbf{0}\rrangle^{\otimes 2k}\right) \equiv \mathcal{N}_{r_1,r_2,\dots,r_k}\;,
\end{equation}
where we used $|\Upsilon_{r_1,r_2,\dots,r_k}^{(k)}\rrangle \in \mathrm{Comm}_k(\mathfrak{G}_N)$. 
On the other hand, for a generic non-Gaussian state $|\Psi\rangle$, we have
\begin{equation}
     0 < \zeta_{r_1,r_2,\dots,r_k}^{(k)} \le \mathcal{N}_{r_1,r_2,\dots,r_k} \;.\label{eq:conditionpos}
\end{equation}

We define the non-Gaussianity measure
\begin{equation}
   \varphi_{r_1,r_2,\dots,r_k}^{(k)}(|\Psi\rangle) \equiv \mathcal{N}_{r_1,r_2,\dots,r_k} - \zeta_{r_1,r_2,\dots,r_k}(|\Psi\rangle) \;,\label{eq:gfa}
\end{equation}
which, by construction, is \textit{invariant} under fermionic Gaussian transformations. 
This construction holds for arbitrary choices of $r_1, r_2, \dots, r_k$, providing a variety of possible measures of fermionic non-Gaussianity. While the Gaussian invariance of each $\varphi_{r_1,r_2,\dots,r_k}^{(k)}$ is ensured by construction, other desirable properties for measures of fermionic magic resources, such as faithfulness and (sub)additivity, must be proven separately, see Appendix~\ref{app:commutant}. The multitude of candidate measures $\varphi_{r_1,r_2,\dots,r_k}^{(k)}$ raises questions about whether some of them satisfy additional desirable properties, such as monotonicity under measurements of the fermionic number operator, which is proportional to $\gamma_{2m-1}\gamma_{2m}$. We leave these questions for future investigation.

In the following, we focus on the specific case $r_1 = r_2 = \dots = r_k = 1$. 
As we discuss in detail in the next section, this choice offers several advantages, as it admits a simple representation in terms of the Majorana correlation matrices. This leads us to define $\mathcal{F}_k \equiv \varphi_{1,1,\dots,1}^{(2k)}$ as a meaningful measure of fermionic magic resources.

\subsection{Fermionic antiflatness}
We define the fermionic antiflatness (FAF) as 
\begin{equation}
    \mathcal{F}_k(|\Psi\rangle)\equiv \varphi^{(2k)}_{1,1,\dots,1}(|\Psi\rangle) = N-\frac{1}{2}\mathrm{tr}\left[(M^T M)^k\right]\;,\label{eq:faf}
\end{equation}
where the right-hand side is obtained by direct inspection using the definition of the covariance matrix $M$, Eq.~\eqref{eq:covMAT}, while $k\geq1$ is an integer.
The FAF  fulfills the key properties of the fermionic non-Gaussianity measure, namely: (i) it is faithful, meaning $\mathcal{F}_k( \ket{\Psi})\ge 0$ with equality holding if and only if the state $|\Psi\rangle$ is fermionic Gaussian; (ii) it is invariant under FGU; (iii) it is subadditive under tensor products for generic states $\mathcal{F}_k(|\Psi\rangle\otimes |\Phi\rangle)\le \mathcal{F}_k(|\Psi\rangle) + \mathcal{F}_k(|\Phi\rangle)$ and additive, 
$\mathcal{F}_k(|\Psi\rangle\otimes |\Phi\rangle)= \mathcal{F}_k(|\Psi\rangle) + \mathcal{F}_k(|\Phi\rangle)$, if at least one of the states $\ket{\Psi}, \ket{\Phi}$ has a well-defined fermionic parity $\mathcal{P}$. 
Of these properties, the faithfulness (i) follows from the well-known result that a pure state $\ket{\Psi}$ is a fermionic Gaussian state if and only if $M^T M= \mathbb{1}$~\cite{Bravyi05flo}.

The Gaussian invariance, (ii), is immediately implied by the fact that $\Upsilon_{r_1,r_2,\dots,r_k}^{(k)}$ belongs to the $k$-th fermionic commutant.
Alternatively, when $\ket{\Psi} \mapsto U_G \ket{\Psi}$ under a fermionic Gaussian unitary $U_G$, the covariance matrix transforms as $M\mapsto G M G^T$, and the value of $\mathcal{F}_k$ remains unchanged due to the cyclic invariance of trace in Eq.~\eqref{eq:faf}.
 
Before proving (iii), we first comment on the relation between $\mathcal{F}_k$ and the spectrum of the covariance matrix.
The covariance matrix $M$ is a $2N\times 2N$ antisymmetric matrix, and can be brought to the canonical form
\begin{equation}
\label{eq:canon}
   M = Q^T \bigoplus_{m=1}^N \begin{pmatrix}
        0 & \lambda_m\\ -\lambda_m & 0
    \end{pmatrix}Q
\end{equation}
where $\lambda_i>0$ are the so-called Williamson eigenvalues of $M$~\cite{Williamson36, Zumino62}, and $Q$ is an orthogonal matrix. 
In terms of Williamson eigenvalues, the FAF can be written as 
\begin{equation}
\label{eq:fafWilliam}
    \mathcal{F}_k(|\Psi\rangle) = N-\sum_{i=1}^N \lambda_i^{2k}\;,
\end{equation}
and can be interpreted as a measure of the flatness of the distribution of $\{\lambda_i\}$. When $\ket{\Psi}$ is a fermionic Gaussian state, the spectrum is completely flat, with $\lambda_i=1$. In contrast, when $\ket{\Psi}$ deviates from the manifold of fermionic Gaussian states, the Williamson eigenvalues decrease, with $\lambda_i\in[0,1]$, and are no longer equal, which is indirectly measured by $\mathcal{F}_k$, motivating the name of FAF.

\subsubsection{Additivity and subadditivity}
We now prove the (sub)additivity of FAF, (iii), for a product state of the form  $
|\Psi\rangle = |\psi_A\rangle \otimes |\psi_B\rangle$ with $|\psi_{A/B}\rangle \in \mathcal{H}_{A/B}$, the Hilbert space of $N_A$ and $N_B=N-N_A$ qubits, respectively. When at least of of the states $|\psi_{A/B}\rangle$  has a definite fermionic parity $\mathcal{P}$, the covariance matrix of $|\Psi\rangle$ is block diagonal, and FAF is additive. 
In the generic case, using the explicit covariance matrix structure implied by the product state, we show that FAF is subadditive.

\paragraph{Fixed fermionic parity}
Recall that the Jordan-Wigner transformation, cf. Eq.~\eqref{eq:JW}, is defined in the combined system of $N$ qubits, starting from the first site in the subsystem $A$. 
Hence, via the natural embedding, Majorana strings restricted to the subsystem $A$ are defined by the same Jordan-Wigner transformation on the first $N_A$ qubits.
As a result, the correlation matrix elements $M_{ij}=\langle \Psi|\gamma_i\gamma_j|\Psi\rangle$ when both $i\neq j$ lie in the subsystem $A$ is given by $M^A_{ij}=\langle \psi_A|\gamma^A_i\gamma^A_j|\psi_A\rangle$, where $\gamma^A_i$ are Majorana operators defined on $\mathcal{H}_A$. 
On the other hand, a Majorana operator with the index $i\in \{2N_A+1,\dots,2N\}$ is implicitly given by $\mathcal{P}_A\gamma^B_i=\gamma_i$, where $\mathcal{P}_A=\prod_{i=1}^{N_A} Z_i$, the parity operator on $A$, $\gamma^B_i$ the Majorana string as if the Jordan-Wigner is defined from the site $N_A+1$, and $\gamma_i$ the Majorana string in the full system. 
Thus, when $i\neq j$ both are in $B$, the parity operation simplifies and one recovers $M_{ij}=M^B_{ij}\equiv \langle \psi_B|\gamma^B_i\gamma^B_j|\psi_B\rangle$. 
On the other hand, when $i\in A$ and $j\in B$ we have $M_{ij}=  \langle \psi_A|\bar{\gamma}^A_i|\psi_A\rangle \langle \psi_B|\gamma_j|\psi_B\rangle$ where $\bar{\gamma}^A_i=-i\mathcal{P}_A \gamma^A_i$. Conversely, when $i\in B$ and $j\in A$, we have $M_{ij}=  \langle \psi_A|\bar{\gamma}^A_j|\psi_A\rangle \langle \psi_B|\gamma_i|\psi_B\rangle$.
Collecting these results, we observe that the covariance matrix admits a block form
\begin{equation}
\begin{split}
    M(|\Psi\rangle) &= M_\mathrm{d} + \Delta\;, \\
    M_\mathrm{d} &= \begin{pmatrix}
        M^A & 0 \\
        0   & M^B
    \end{pmatrix}, \quad 
    \Delta = \begin{pmatrix}
        0 &  \vec{v}\vec{w}^T\\
        \vec{w}\vec{v}^T & 0
    \end{pmatrix}, \label{eq:correlation-block}
\end{split}
\end{equation}
where $M^A$ and $M^B$ denote the correlation matrices for $|\psi_A\rangle$ and $|\psi_B\rangle$, while the off-diagonal blocks are given by the real vectors $v_i=\langle \psi_A|\bar{\gamma}^A_i|\psi_A\rangle $ for $i=1,\dots,2N_A$ and $w_j=\langle \psi_B|\gamma^B_j|\psi_B\rangle$ for $j=2N_{A}+1,\dots,2N_A+2N_B$. 
Note that if either $|\psi_A\rangle$ or $|\psi_B\rangle$ is fermion-parity even, then $\Delta = 0$. In that case, 
\begin{equation}
    \mathrm{tr}[(M^TM)^k]= \mathrm{tr}[((M^A)^TM^A)^k] + \mathrm{tr}[((M^B)^TM^B)^k]\;,
\end{equation}
and the additivity of FAF follows from noting that $\mathcal{F}_k(|\psi_{A/B}\rangle)=N_{A/B}-\mathrm{tr}[((M^{A/B})^TM^{A/B})^k] $. 

\paragraph{Generic case}
On the other hand, generic states do not have a well-defined fermionic parity. As a result, $\Delta$ is non-vanishing. 
Consider the decomposition $M^T M = D + O$, where we define the block-diagonal matrix $D = M_\mathrm{d}^T M_\mathrm{d} + \Delta^T\Delta$ and the block anti-diagonal matrix $O=\Delta^T M_\mathrm{d} + M^T_\mathrm{d}\Delta$. 
We note that 
\begin{equation}
    \Delta^T \Delta= \begin{pmatrix}
        \|\vec{w}\|^2 \vec{v} \vec{v}^\dagger & 0 \\ 
        0 &  \|\vec{v}\|^2 \vec{w} \vec{w}^\dagger
    \end{pmatrix}
\end{equation}
Expanding the $k$-th power of $M^T M$ gives:
\begin{equation}
    (M^T M)^k = \sum_{i_1, \dots, i_k \in \{0,1\}} D^{i_1} O^{1 - i_1} \cdots D^{i_k} O^{1 - i_k}.\label{eq:summm}
\end{equation}
Due to the block structure of $O$, any term in the expansion with an odd number of $O$ matrices has vanishing trace.
Moreover, due to the anti-symmetry of $M_A$ and $M_B$, we have $\vec{w}^\dagger M_B \vec{w}=0=\vec{v}^\dagger M_A \vec{v}$. 
This implies that any string with individual $A$ multiplied by the left and right with $O$ is identically zero. 
On the other hand, we have
\begin{equation}
    O^2=\begin{pmatrix}
        (\vec{w}^\dagger M_B^\dagger M_B\vec{w}) \vec{v}\cdot\vec{v}^\dagger &0 \\ 
       0 &  (\vec{v}^\dagger M_A^\dagger M_A\vec{v}) \vec{w}\cdot\vec{w}^\dagger
    \end{pmatrix} \;.
\end{equation}
From these building blocks, one finds by direct inspection that each operator in Eq.~\eqref{eq:summm} has a non-negative trace, implying the inequality
\begin{equation}
    \mathrm{tr}[(M^T M)^k]\ge \mathrm{tr}[(M_A^\dagger M_A)^k]+ \mathrm{tr}[(M_B^\dagger M_B)^k]\;,
\end{equation}
from which we conclude that 
\begin{equation}
    \mathcal{F}_k(|\Psi\rangle) \leq \mathcal{F}_k(|\psi_A\rangle) + \mathcal{F}_k(|\psi_B\rangle).
\end{equation}

The FAF defined in Eq.~\eqref{eq:faf} will serve as the primary measure of fermionic non-Gaussianity throughout the remainder of this work. A complementary perspective, based on the non-Gaussian entropy considered in \cite{Lumia24, Lyu24NGE, Coffman25magic}, is discussed in Appendix~\ref{app:relation}.

\subsubsection{Experimental relevance}
We conclude this section by commenting on the efficient computability and the experimental relevance of FAF. Since $\mathcal{F}_k$ depends only on the covariance matrix $M$, its evaluation requires calculation of $N(2N-1)$ independent expectation values $\braket{\Psi|\gamma_m \gamma_n|\Psi}$ and straightforward algebraic manipulations to evaluate the trace in~Eq.~\eqref{eq:faf}. The calculation of the expectation values can be efficiently performed when $\ket{\Psi}$ is represented as a state vector, or when $\ket{\Psi}$ admits an efficient tensor network representation, for instance, when the state is given as an MPS. 
The two-point Majorana correlation functions can be efficiently measured in experiments through shadow tomography protocols~\cite{Zhao21, Wan23} or related methods~\cite{Denzler24, Majsak2025}, enabling measurement of two-point Majorana correlators to additive precision $\epsilon$ using $O( N^2 \ln(N) / \epsilon^2)$ measurement rounds. Once the $N(2N-1)$ independent expectation values are measured, constructing the covariance matrix and computing $\mathcal{F}_k$ is straightforward. 

\section{Fermionic magic resources in model systems}
\label{sec:modelSystems}
In this section, we compute the FAF for several classes of quantum states to build intuition about how fermionic non-Gaussianity manifests in multi-qubit systems and to establish reference points for its behavior in equilibrium and non-equilibrium settings. We begin with analytically tractable examples, such as product states and Haar-random states. We then proceed to numerical studies of FAF: first in random matrix product states (RMPS), which model many-body states with finite-range correlations~\cite{Lio24corrRMPS}; and then in random quantum circuits, which serve as minimal models of unitary, local dynamics.

\subsection{Simple examples}
\label{subsec:simple}

\subsubsection{Fermionic magic product state}
\label{subsec:prod1}
The subspace of $\mathcal{H}_2^{\otimes N}$ containing states with even fermionic parity ($\mathcal{P}=1$) is $d_e=2^{N-1}$ dimensional. 
For $N=1$, this subspace is one-dimensional, spanned by $\ket{0}$, which is a fermionic Gaussian state. Interestingly, it was shown~\cite{Bravyi05capacity} that \textit{all} $N=2$ and $N=3$ qubit states in the positive fermionic parity subspace are fermionic Gaussian states. 
For all these states, the FAF is vanishing: $\mathcal{F}_k=0$. The first example of a state with fixed $\mathcal{P}=1$ which is not a fermionic Gaussian state can be constructed for $N=4$. One such example is
\begin{equation}
\label{eq:psi_theta}
\ket{\Psi_{\theta}}=\frac{1}{2}\left( \ket{0000}+\ket{0011}+\ket{1100}+e^{i \theta }\ket{1111} \right),
\end{equation}
where $\theta$ is an arbitrary angle. 
Let $\braket{i\gamma_m \gamma_n} \equiv \braket{ \Psi_\theta | i\gamma_m \gamma_n|\Psi_\theta }$, by a direct computation, we find that the only non-vanishing 2-point Majorana correlation functions are $\braket{i\gamma_1 \gamma_3} =\braket{i\gamma_2 \gamma_4}= \sin(\theta)/2$, $\braket{i\gamma_1 \gamma_4}=\braket{i\gamma_2 \gamma_3} = - \cos^2(\theta/2)$, and the correlators after a translation by two sites, $\braket{i \gamma_{m+4}\gamma_{n+4}} = \braket{i \gamma_{m}\gamma_{n}}$ (here $m,n\in[1,4]$), which together with $\braket{i\gamma_m \gamma_n} =-\braket{i\gamma_n \gamma_m}$ fixes the entire covariance matrix $M$. Through Eq.~\eqref{eq:faf}, we evaluate FAF, which reads $\mathcal{F}_k( \ket{\Psi_{\theta}} ) = 4[1- \cos^{2k}\left(\theta/2\right)]$. For generic $\theta$, the state $\ket{\Psi_{\theta}}$ is indeed not Gaussian. 
The value of $\mathcal{F}_k$ increases smoothly with $\theta$ from the free-fermionic limit at $\theta=0$ with $\mathcal{F}_k=0$, and admits the maximal possible value at $\theta=\pi$.

We now consider a system of $N$ qubits, assuming that $N$ is a multiple of 4, and take a product state of the form
\begin{equation}
\ket{\Psi_{\theta, N}}=\left[\frac{1}{2}\left( \ket{0000}+\ket{0011}+\ket{1100}+e^{i \theta }\ket{1111} \right)\right]^{\otimes \frac{N}{4}}.
\label{eq:psitheta_N}
\end{equation}
The state $\ket{\Psi_{\theta, N}}$ is a product of states with even fermionic parity. 
Therefore, by the additivity of the FAF, we immediately find that 
\begin{equation}
\mathcal{F}_k( \ket{\Psi_{\theta,N}}) = N \left(1- \cos^{2k}\left(\frac{\theta}{2}\right)\right).
\label{eq:FkProdN}
\end{equation}
The example illustrates how the generic extensive scaling of FAF, Eq.~\eqref{eq:NdepOverview}, arises from the additivity of FAF on products of states with fixed fermionic parity. In that case, the leading term, $D_k=1- \cos^{2k}\left(\theta/2\right)$, is associated with the value of FAF in the states that enter the product, while the subleading term $f_k$ vanishes. 

\subsubsection{Action of local non-Gaussian gates}
\label{subsec:localGate}
The state $\ket{\Psi_{\theta, N}}$ can be obtained by acting with $N/4$ non-Gaussian gates on the vacuum state $\ket{\mathbf{0}}=\ket{0}^{\otimes N}$, with each gate producing the state $\ket{\Psi_{\theta}}$ on a block of $4$ subsequent qubits. If, instead of acting with $N/4$ gates, one acted only with $n$ gates, the resulting state would be $\ket{\Psi_{\theta}}^{\otimes n} \otimes \ket{0}^{\otimes (N - 4n)}$. The FAF of this state, $\mathcal{F}_k = 4n[1- \cos^{2k}\left(\theta/2\right)]$, is a constant proportional to the number of sites on which the non-Gaussian gates acted. In the following, we generalize this result.

Let us consider a fermionic Gaussian state $\ket{\Psi_G}= U_G \ket{\mathbf{0}}$, and choose a local non-Gaussian gate $O_a$ acting on at most $a$ qubits and preserving the fermionic parity. Due to the fermionic Gaussianity of $\ket{\Psi_G}$, the FAF is vanishing, $\mathcal{F}_k(\ket{\Psi_G})=0$. 
How much does the FAF increase upon the action of $O_a$, and what are the possible values of $\mathcal{F}_k( O_a \ket{\Psi_G})$?

To calculate $\mathcal{F}_k( O_a \ket{\Psi_G})$, we examine the transformation induced by the action of the fermionic Gaussian unitary $U_G$ followed by the non-Gaussian gate $O_a$ on the Majorana operators. We find that 
\begin{equation}
\label{eq:nonG}
U_G^{\dagger} O_a^{\dagger} \gamma_m O_a U_G = U_G^{\dagger}  \gamma_m  U_G + C_m,
\end{equation}
where $C_m = U_G^{\dagger}  [O^{\dagger}_a, \gamma_m ] O_a U_G$ is non-vanishing only for $m \in A$ where $A$ is the subset of $[2N]$ of size $|A|=2a$ associated with the indices of Majorana operators corresponding to sites at which $O_a$ acts non-trivially\footnote{We note that $C_m$ is vanishing for $m\in A$ only when  $O_a$ preserves fermionic parity -- otherwise, the commutator in $C_m$ is non-vanishing for any $m$, giving rise to an extensive scaling of $\mathcal{F}_k( O_a \ket{\Psi_G})$.}. Eq.~\eqref{eq:nonG} implies that the covariance matrix of state $O_a \ket{\Psi_G}$ is given as
\begin{equation}
M = G M_0G^T + \kappa,
\label{eq:mloc1}
\end{equation}
where $M_0$ is the covariance matrix of the vacuum state, cf. Eq.~\eqref{eq:covVAC}, $G$ is the orthogonal transformation associated with the fermionic Gaussian unitary $U_G$, and the matrix $\kappa$ with entries $\kappa_{mn}$ defined by
\begin{equation}
\kappa_{mn}= i \braket{\mathbf{0}|  \left( [\tilde\gamma_m, C_n] + [C_m, \tilde\gamma_n] +[C_m,C_n] \right)|\mathbf{0}}
\end{equation}
with $\tilde \gamma_m = U_G^{\dagger}  \gamma_m  U_G  $. Since the matrix elements of $\kappa$ are vanishing for any $m,n$ not belonging to the set $A$, the matrix $\kappa$ has rank no greater than $|A|=2a$, $\mathrm{rank}(\kappa) \leq 2a$. Eq.~\eqref{eq:mloc1}, shows that
\begin{equation}
M^TM = \mathbb{1} + \kappa^{T} G M^{(0)}G^T + (G M^{(0)}G^T + \kappa^T)\kappa,
\label{eq:mloc2}
\end{equation}
is a sum of the identity matrix, and two matrices of rank $2a$ at most. This, due to the sub-additivity of rank, is a perturbation of rank at most $4a$ on the identity matrix. General arguments for restricted rank modifications of the symmetric eigenvalue problem~\cite{Arbenz88} indicate that only at most $4a$ eigenvalues $\tilde{\lambda}_i^2$ of the matrix $M^TM$ differ from unity. This, due to $M$ being a covariance matrix of a pure normalized state, translates to $0\leq\tilde{\lambda}_i^2 <1$, with the other $2N-4a$ eigenvalues remaining $\lambda_i^2=1$. Therefore, Eq.~\eqref{eq:fafWilliam}, gives our final result
\begin{equation}
\mathcal{F}_k( O_a \ket{\Psi_G})  \leq 2a,
\label{eq:locNON}
\end{equation}
showing that the action of a local, fermionic parity-preserving, non-Gaussian gate $O_a$ on a fermionic Gaussian state results in a state with a finite, that is, not extensive, amount of fermionic magic resources. We emphasize that this result holds for an arbitrary initial fermionic Gaussian state $\ket{\Psi_G}= U_G \ket{\mathbf{0}}$, even when the initial state is strongly entangled.

\subsubsection{Product of single-qubit states}
\label{subsec:producsingle}
The preceding examples considered states with fixed fermionic parity $\mathcal{P}$. When the value of $\mathcal P$ is not defined, even a single-qubit state becomes non-Gaussian, and the FAF becomes subadditive. In the following, we consider a system of $N$ qubits and a product of single-qubit states, given by
\begin{equation}
    \ket{T}^{\otimes N} = \left( \cos\left(\frac{\theta}{2}\right)\ket{0} + \sin\left(\frac{\theta}{2}\right) e^{i\phi}\ket{1} \right)^{\otimes N}
\label{eq:prodEX}
\end{equation}
where $\theta$ and $\phi$ are two angles parameterizing the single-qubit state. The state $\ket{T}^{\otimes N}$ is a product state, and the entanglement entropy $S_{\mathrm{ent}}$ vanishes for any bipartition of the system. At the same time, $\ket{T}^{\otimes N}$ is not an eigenstate of $\mathcal{P}$, and hence, it is not a fermionic Gaussian state.

To calculate $\mathcal{F}_k(\ket{T}^{\otimes N})$, we observe that $\ket{T}^{\otimes N}$ is obtained from the vacuum $\ket{\mathbf{0}}$ by action of the $x$-axis rotation $R_x(\theta) = \prod_k e^{i \theta X_k}$, and the $z$ rotation, $R_z(\theta) = \prod_k e^{i \phi Z_k}$. The $z$ rotation is a fermionic Gaussian unitary since $R_z(\theta) =e^{\phi \sum_k \gamma_{2k-1}\gamma_{2k} } $ and it does not affect the value of $\mathcal{F}_k$. Therefore, for simplicity, we fix $\phi=0$ in Eq.~\eqref{eq:prodEX}.  Using the fact that  Eq.~\eqref{eq:prodEX} is a product state, we can directly read off the covariance matrix $M$ of the state $\ket{T}^{\otimes N})$. The non-zero elements of the covariance matrix are given by
\begin{equation}
\begin{split}
    \braket{i\gamma_{2k-1}\gamma_{2k}} &= \cos(\theta) \;, \\
\braket{i\gamma_{2k}\gamma_{2 l-1}}& = \sin^2(\theta) \cos(\theta)^{k-l-1}, \label{eq:prodMAJ}   
\end{split}
\end{equation}
and $M_{lk}=-M_{kl}$ for $k<l$.
We perform a fermionic Gaussian transformation that permutes Majorana operators as $\gamma_{2k-1} \mapsto \gamma_{k}$ and $\gamma_{2k} \mapsto \gamma_{N+k}$ (which does not change the value of FAF), and use Eq.~\eqref{eq:prodMAJ}, to obtain a transformed covariance matrix that has the following block structure
\begin{eqnarray}
M =  \begin{pmatrix}
0 & C \\
-C^T & 0
\end{pmatrix},
\end{eqnarray}
where $C$ is a $N\times N$ matrix with entries $C_{ij} = \cos(\theta)\delta_{ij}+(1-\delta_{ij})\sin^2(\theta) \cos(\theta)^{i+j-1}$ for $j \geq i$ and $C_{ij}=0$ otherwise. To calculate $\mathcal{F}_k$, we evaluate the matrix $M^TM$ which is block-diagonal $M^TM = \mathrm{diag}(-CC^{T},-C^{T} C)$. The matrix $CC^{T}$ has a simple structure:
one can verify analytically that $CC^{T}-1$ is a rank-1 matrix, with $N-1$ eigenvalues vanishing, and one non-trivial eigenvalue equal to $\cos^{2N}(\theta)$. This gives the final result
\begin{equation}
\mathcal{F}_k( \ket{T}^{\otimes N} ) = 1-\cos^{2kN}( \theta).
\end{equation}
While FAF increases monotonically with the system size $N$, its value does not exceed unity. This result is consistent with the subadditivity of FAF on states without fixed fermionic parity. However, it shows that the interpretation of FAF for such states is less straightforward than the one appearing from the typical extensive scaling of states with well-defined $\mathcal{P}$.

\subsection{Typical many-body states}
\label{subsec:typical}
We now analyze the behavior of FAF in typical quantum states.
We consider two relevant cases: (i) the situation without any symmetry, (ii) the typical state in the even fermionic parity subspace. 
To model typical quantum states, we take an $N$-qubit state $|\Psi\rangle=U|\mathbf{0}\rangle$ where $U\in \mathrm{U}(2^N)$ is a unitary drawn from the Haar measure on $\mathrm{U}(2^N)$.
For fixed $k$, the computation of the average value of $\mathcal{F}_k$ can be performed exactly using the Weingarten calculus. 
Writing $\rho_0=|\mathbf{0}\rangle\langle \mathbf{0}|$, the FAF can be expressed in the replica space as the expectation value
\begin{equation}
\label{eq:fafreplica}
    \mathcal{F}_k=N-\llangle \Upsilon_{1,1,\dots,1}^{(2k)}|(U\otimes U^*)^{\otimes 2k}|\rho_0^{\otimes 2k}\rrangle\;.
\end{equation}
In the following, we compute the average ${\mathcal{F}}^{\mathrm{typ}}_k\equiv \mathbb{E}_{U\in \mathrm{U}(2^N)}[\mathcal{F}_k]$.
We employ the Weingarten formula~\cite{Roberts2017, Kostenberger21, Collins22}
\begin{equation}
    \mathbb{E}_U(U\otimes U^*)^{\otimes 2k}|\rho_0^{\otimes 2k}\rrangle=\sum_{\sigma,\pi}\mathrm{Wg}_{\sigma,\pi}|T_\sigma\rrangle \llangle T_\pi|\rho_0^{\otimes 2k}\rrangle\;,
\end{equation}
where the sum runs over the permutations $\sigma \in S_{2k}$ of $2k$ replicas, with action $T_{\sigma}|x_1,\dots,x_{2k}\rangle=|x_{\sigma^{-1}(x_1)},\dots,x_{\sigma^{-1}(x_{2k})}\rangle$  on the computational basis, and $\mathrm{Wg}$ is the inverse of the matrix with elements $G_{\pi,\sigma}=\mathrm{tr}(T_\pi^\dagger T_\sigma)$. 
This expression simplifies for pure states, in which case $\llangle T_\pi|\rho_0^{\otimes 2k}\rrangle=1$. Furthermore, a marginal sum over one index of the Weingarten matrix gives $\sum_\pi \mathrm{Wg}_{\pi,\sigma}=[d(d+1)\dots(d+k-1)]^{-1}$, where $d$ is the dimension of the Hilbert space. 
Collecting these results, we have the general expression, holding for pure states and for arbitrary $k$
\begin{equation}
    \mathcal{F}^{\mathrm{typ}}_k= N-\frac{(d-1)!}{(d+2k-1)!}\sum_{\sigma\in S_{2k}}\llangle \Upsilon_{1,1,\dots,1}^{(2k)}| T_\sigma\rrangle\;.\label{eq:typfaf}
\end{equation}
This expression can be explicitly computed for a few $k$, and corresponds to a bookkeeping exercise for the Majorana strings in $|\Upsilon^{(2k)}_{1,\dots,1}\rrangle$. 
We can already, however, anticipate that, in the scaling limit $N\to\infty$, the average FAF of Haar random state is $\mathcal{F}^{\mathrm{typ}}_k\simeq N$, since $\llangle \Upsilon^{(2k)}_{1,\dots,1}|T_\sigma\rrangle\le N^{2k} 2^{kN}$ for any permutation $\sigma$, which is suppressed by a denominator $O(2^{2kN})$ in Eq.~\eqref{eq:typfaf}. 
This result confirms the heuristic expectation that generic Haar-random pure states exhibit maximal fermionic non-Gaussianity.

In the following sections, we will compare how fast in bond dimension $\chi$ or circuit depth $t$, a random tensor network state or a quantum circuit reaches the Haar value Eq.~\eqref{eq:typfaf}, which motivates us to calculate $\mathcal{F}^{\mathrm{typ}}_k$ at any finite system size $N$. 
To that end, we note that the only nonzero contributions in the sum Eq.~\eqref{eq:typfaf} arise from permutations $\sigma$ whose cycle structure $\lambda(\sigma)$ consists entirely of cycles of even length, i.e., $\lambda_i = 2m_i$ with $m_i \in \mathbb{N}$. 

For $k=1$, the sum over Majorana strings extends over terms $i_1,i_2\in [2N]$ such that $i_1\neq i_2$. The only nontrivial contribution to the sum in Eq.~\eqref{eq:typfaf} comes from the transposition $\sigma = (12)$, since $\mathrm{tr}[(\gamma_{i_1}\gamma_{i_2})^2] \neq 0$, while all other permutations yield vanishing trace. 
Direct inspection gives the fundamental building block 
\begin{equation}
    \llangle T_\sigma| (|\gamma_{i_1}\gamma_{i_2} \rrangle \otimes|\gamma_{i_2}\gamma_{i_1}\rrangle) = \mathrm{tr}(\gamma_{i_1}^2\gamma_{i_2}^2)=2^N\;,
\end{equation}
for any allowed choice of the indices $i_1,i_2$. There are $2N(2N-1)/2$ such index pairs, yielding for $N\ge 2$,
\begin{equation}
\label{eq:faf1typ}
    \mathcal{F}^{\mathrm{typ}}_1=N-\frac{1}{d+1}N(2N-1)\;,
\end{equation}
whereas for $N=1$, $\mathcal{F}^{\mathrm{typ}}_1=0$. 
A similar calculation can be performed for Haar-random states in the even-parity sector, where the Hilbert space dimension changes from $d=2^N$ to $d_e=2^{N-1}$.  
Replacing $d\mapsto d_e$ in Eq.~\eqref{eq:faf1typ},  we find that $\mathcal{F}^{\mathrm{typ,e}}_1=0$ for $N=1,2,3$ and becomes non-trivial only when $N\ge 4$ -- consistently with the fact that all even states of $N\leq3$ qubits are fermionic Gaussian states~\cite{Bravyi05capacity}. 
A similar computation can be performed for $k=2$, leading for $N\ge 3$ to 
\begin{equation}
\label{eq:faf2typ}
    \mathcal{F}^{\mathrm{typ}}_{2}=N-
    \frac{N(2N-1)}{(d+1)(d+2)(d+3)}
    (-8N^2+4N(d+7)-d-14),
\end{equation}
where $d=2^N$ for Haar-random states and $d\mapsto d_e$ for Haar-random states in the $\mathcal{P}=1$ subspace. (For $N\le 3$, again, one finds that $\mathcal{F}^{\mathrm{typ}}_2=0$ for even Haar-random states.)

For $k > 2$, the computation becomes increasingly intricate, but the leading contribution can be isolated.
At leading order, the expression $\sum_\sigma\llangle \Upsilon^{(2k)}_{1,\dots,1}|T_\sigma\rrangle$, by dimensional analysis, is dominated by the permutation with cycle structure $\lambda(\sigma)=(2,2,\dots,2)$ of length $k$, i.e., pairwise transpositions forming a perfect matching over the $2k$ replicas.
Among these, only the non-crossing matchings contribute. 
These are matchings for which no two pairs $(i_1, j_1)$ and $(i_2, j_2)$ satisfy the crossing condition $i_1 < i_2 < j_1 < j_2$ — such crossing pairings do not contribute at leading order~\cite{fava2024designsfreeprobability}. 
This results in 
\begin{equation}
    \mathcal{F}^{\mathrm{typ}}_k = N-C_k \frac{2^kN^{k+1}}{d^k}+O\left(\frac{N^k}{d^k}\right)\label{eq:leadingtyp}
\end{equation}
with $d=2^N$ for Haar random states and $d=2^{N-1}$ for the Haar random states in the even parity sector, where $C_k = \frac{1}{k+1} \binom{2k}{k}$ is the $k$-th Catalan number, counting the number of non-crossing pairwise matchings among $2k$ elements. 

An alternative way of deriving Eq.~\eqref{eq:leadingtyp} relies on viewing the covariance matrix $M$ of a typical state $\ket{\Psi}$ as a \textit{random matrix}. 
Denoting by $x=\braket{\Psi | P | \Psi}$ the expectation value of a fixed non-identity Pauli string $P$ in a Haar random state $\ket{\Psi}$, typicality arguments indicate that the distribution of $x$ has a Gaussian form~\cite{Turkeshi25spectrum}
\begin{equation}
    \Pi_\mathrm{typ}(x) = \frac{e^{-x^2/(2 \sigma^2)}}{\sqrt{2 \pi \sigma^2}},
    \label{eq:pheno}
\end{equation}
with variance $\sigma^2=1/d$ and $\sigma^2=2/d$ respectively for Haar-random states without and with $\mathbb{Z}_2$ symmetry.
The value $\sigma^2$ is fixed by the normalization condition $\sum_P \braket{\Psi | P | \Psi}^2 =d$ stemming from the normalization of $\ket{\Psi}$ and the fact that Pauli strings define an orthogonal basis in the operator space. The Gaussian distribution $\Pi_\mathrm{typ}(x)$ is an excellent approximation of the exact distribution of $x=\braket{\Psi | P | \Psi}$ in Haar-random states derived in~\cite{Turkeshi25spectrum}.
The entries of the covariance matrix are 2-point Majorana fermion correlation functions, which are particular examples of Pauli strings. For sufficiently large $N$, we assume that the entries of the covariance matrix $M_{ij}$ for $i<j$, in a fixed Haar random state $\ket{\Psi}$ are uncorrelated random variables drawn from the probability distribution $\Pi_\mathrm{typ}(x)$, while $M_{ij}$ for $i\geq j$ are fixed by the antisymmetry of the covariance matrix: $M_{ii}=0$ and $M_{ji}=-M_{ij}$. As shown in the following, these assumptions enable a compact derivation of the leading term of FAF for Haar-random states. The random covariance matrix $M$ created in this manner belongs to the Gaussian Ensemble of anti-symmetric Hermitian matrices, with 
the joint probability distribution for its eigenvalues $\{\theta_j\}$ given by~\cite{Mehtabook}
\begin{equation}
    P(\{\theta_j\})=C\prod_{1\le j<k\le N} (\theta_j^2-\theta_k^2)^2\exp\left[-\sum_{j=1}^N \frac{\theta_j^2}{\sigma^2}\right]\;, 
    \label{eq:RMTjpdf}
\end{equation}
where $C$ is a normalization constant. 
This distribution can be recast using the wavefunction 
\begin{equation}
    \varphi_j(\theta)=\sqrt{\frac{(2j)!}{(2j-1)!!\;\sqrt{\pi}\; \sigma}} e^{\theta^2/(2\sigma^2)}(-\sigma\partial_\theta)^j e^{-\theta^2/\sigma^2}\;
\end{equation}
in terms of the kernel $K_N(x,y)=2\sum_{j=0}^{N-1}\varphi_{2j}(x)\varphi_{2j}(y)$, as
\begin{equation}
    P(\{\theta_j\})=\frac{1}{N!}\det[K_N(\theta_j,\theta_k)]_{j,k=1,\dots,N}.
\end{equation}
From this, marginalizing over $\theta_2,\dots,\theta_N$, we obtain the spectral density for $\theta\equiv \theta_1$,
\begin{equation}
\label{eq:density}
    \varrho(\theta) = 2 \sum_{j=0}^{N-1} \varphi_{2j}^2(\theta)\;\mapsto_{N\gg 1} \varrho_\mathrm{sc}(\theta)=\frac{\sqrt{8\sigma^2 N-x^2}}{4\pi \sigma^2 N}\;.
\end{equation}
This function allows for the computation of the scaling limit 
\begin{equation}
\begin{split}
    \frac{1}{2}\mathrm{tr}[(M^TM)^k]&=N\int_{-\sqrt{8N\sigma^2}}^{\sqrt{8N\sigma^2}} \varrho_\mathrm{sc}(\theta) \theta^{2k} d\theta=
    C_k\frac{2^kN^{k+1}}{d^k}\;,
\label{eq:RMTc}
\end{split}
\end{equation}
where we used a change of variable and the moments of the semicircle distribution $\int_{-2}^2dy\sqrt{4-y^2} y^{2k}=C_k$~\cite{Mehtabook}. Recalling the definition of FAF, Eq.~\eqref{eq:faf}, we reproduce the expression in Eq.~\eqref{eq:leadingtyp}, where again $d=2^N$ in the generic case, while $d=2^{N-1}$ for states in the even fermionic parity sector. Due to the concentration of the Haar measure, the average values of FAF, $\mathcal{F}^{\mathrm{typ}}_k$, accurately approximate the value of FAF in the individual realizations of typical states. Our numerical estimates indicate that the fluctuations of $\mathcal{F}_k$ around the average  $\mathcal{F}^{\mathrm{typ}}_k$ decay polynomially with the system size $N$.

Finally, we comment on the role of the index \(k\) in \(\mathcal{F}_k\).
The results for the states considered in Sec.~\ref{subsec:simple}, as well as for the typical states studied in this section, indicate similar behavior of \(\mathcal{F}_k\) for all integers \(k\ge 1\).
At the same time, higher-index FAF can serve as more sensitive discriminators of many-body states in the following sense.
Consider an ensemble \(E_{2t}\) of states forming a \(2t\)-design, i.e., averages of observables involving up to \(2t\) copies agree with the Haar-typical values~\cite{mele2024introductiontohaar}.
Because \(\mathcal{F}_k\) is defined by an operator acting on \(2k\) replicas, its ensemble average over \(E_{2t}\) matches the typical-state result in Eq.~\eqref{eq:leadingtyp} for all \(k\le t\); deviations can appear only for \(k>t\).
Nevertheless, as we show below for classes of states relevant to quantum many-body physics, the qualitative behavior of \(\mathcal{F}_k\) is similar across all \(k\ge 1\).

\subsection{Random matrix product states}
\label{subsec:RMPS}

Quantum states prepared by low-depth quantum circuits and Haar-random states can be viewed as representing two extremes in the landscape of quantum states: the former are efficiently preparable but highly structured, the latter are maximally random but experimentally inaccessible due to 
scaling of the circuit depth with system size $N$.
The random matrix product states (RMPS) offer a middle ground by providing an ensemble of states that, while requiring linear-depth circuits to prepare, remain highly structured and physically motivated~\cite{Garnerone10, Garnerone10a, Haferkamp21}. The RMPS with polynomial bond dimension $\chi$ has been shown to reproduce several statistical features of Haar-random states, such as spectral properties and entanglement scaling, while retaining efficient classical descriptions. In the following, we investigate the FAF of RMPS.

\begin{figure}
\centering
\includegraphics[width=1\linewidth]{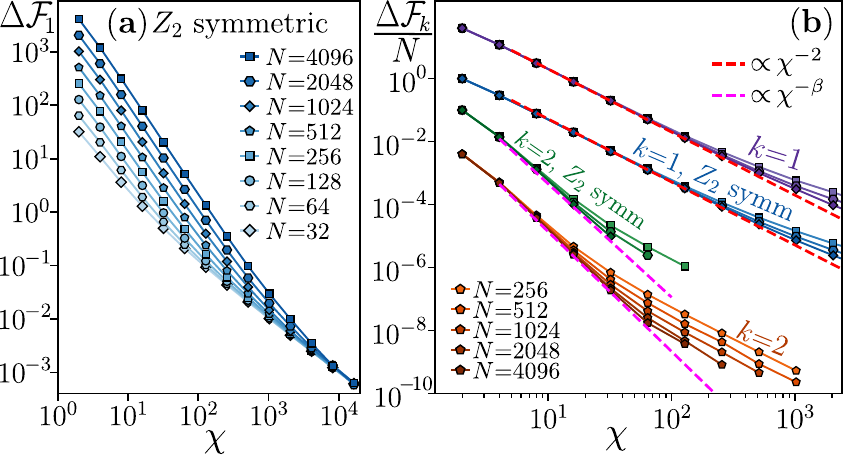}
\caption{Fermionic antiflatness $\mathcal{F}_k$ in random MPS. (a) The difference $\Delta \mathcal{F}_1$, Eq.~\eqref{eq:dF}, between FAF of RMPS and FAF of typical state $\mathcal{F}^{\mathrm{typ}}_k$ as a function of bond dimension $\chi$ in $\mathbb{Z}_2$-symmetric RMPS (i.e., with fixed $\mathcal P =1$) in a system of $N$ qubits. (b) At large $N$,  $\Delta \mathcal{F}_k/N$ decays according to a power-law $\Delta \mathcal{F}_k/N \propto \chi^{-\beta}$, for both RMPS with and without the $\mathbb{Z}_2$ symmetry for $k=1,2$. The exponent $\beta=2.00(5)$ for $k=1$ and $\beta=3.7(3)$ for $k=2$. For clarity of presentation, data for $k=1$ are rescaled by a factor $80$, while data for $k=2$ are rescaled by factors $1/10$ and $1/80$, respectively, for RMPS with and without the $\mathbb{Z}_2$ symmetry.
}
    \label{fig:rmps}
\end{figure}

The MPS form a widely studied class of many-body states on $N$ qudits whose wavefunction is expressed as 
\begin{equation}
    \langle x_1,\dots,x_N|\psi\rangle = \sum_{\substack{x_1, \dots, x_N \\ \alpha, \beta, \dots, \gamma}} A^{(1)}_{\alpha}(x_1) A^{(2)}_{\alpha \beta}(x_2) \dots A^{(N)}_{\gamma}(x_N),
\end{equation}
where $x_i \in \{0, 1\}$ indexes the local basis of the $i$-th qubit, and $\alpha, \beta, \dots, \gamma \in \{1, \dots, \chi\}$ are auxiliary indices associated with a virtual space of dimension $\chi$, known as the bond dimension~\cite{Schollwock2011}. 
The tensors $A^{(i)}_{\alpha \beta}(x_i)$ can be viewed as $\chi \times \chi$ matrices that depend on the local basis element $x_i$. This structure is commonly visualized as
\begin{equation}\label{eq:mps_bulk}
\begin{tikzpicture}[baseline=(current  bounding  box.center),scale=0.8]
\definecolor{mycolor}{rgb}{0.86, 0.08, 0.24}
    \foreach \x in {1,...,5}{
        \draw[thick, black] (1.1*\x,0.8) -- (1.1*\x,1.6);
    }
    \draw[line width=0.8mm, gray!60!black, dotted] (-0.3,0.8) -- (0.2,0.8);
    \draw[line width=0.8mm, gray!60!black] (0.2,0.8) -- (6.4,0.8);
    \draw[line width=0.8mm, gray!60!black, dotted] (6.4,0.8) -- (6.9,0.8);
    \foreach \x in {1,...,5}{
        \draw[thick, fill=mycolor, rounded corners=2pt] (1.1*\x-0.4,0.4) rectangle (1.1*\x+0.4,1.2);
        \pgfmathsetmacro{\y}{int(\x - 3)}
        \node[scale=0.4] at (1.1*\x,0.8) {}; 
    }
\end{tikzpicture}
\end{equation}
with vertical legs corresponding to physical degrees of freedom and horizontal links representing contractions over the virtual space of dimension $\chi$.

The RMPS arise when the tensors $A^{(i)}$ are induced by the action of a Haar random unitary~\cite{Garnerone10}. 
Specifically, consider a system with bond dimension $\chi=2^r$ and we focus on Haar random unitaries $U\in \mathrm{U}(2^{r+1})$\footnote{An analogous argument applies for unitaries that preserve the parity of the system, replacing $\mathrm{U}(2^{r+1})$ with $\{U\in \mathrm{U}(2^{r+1})| [U,Z^{\otimes (r+1)}]=0\} $.}. 
We reshape the unitary matrix as a rank-4 tensor $U_{p,\alpha}^{s,\beta}$ with $p,s\in {0,1}$ and $\alpha,\beta\in \{0,\dots,2^{r}-1\}$, and identify $A^{(i)}_{\alpha,\beta}(s)=U_{0,\alpha}^{s,\beta}$.  
Graphically, the RMPS structure can be  represented as
\begin{equation}\label{eq:rmps_bulk}
\begin{tikzpicture}[baseline=(current  bounding  box.center),scale=0.8]
\definecolor{mycolor}{rgb}{0.86, 0.08, 0.24}
    \foreach \x in {1,...,5}{
        \draw[thick, black] (1.1*\x,0) -- (1.1*\x,1.6);
        \draw[thick, fill=white] (1.1*\x,0) circle (0.2);
        \node[scale=0.9] at (1.1*\x,-0.45) {$|0\rangle$};
    }
    \draw[line width=0.8mm, gray!60!black, dotted] (-0.3,0.8) -- (0.2,0.8);
    \draw[line width=0.8mm, gray!60!black] (0.2,0.8) -- (6.4,0.8);
    \draw[line width=0.8mm, gray!60!black, dotted] (6.4,0.8) -- (6.9,0.8);
    \foreach \x in {1,...,5}{
        \draw[thick, fill=mycolor, rounded corners=2pt] (1.1*\x-0.4,0.4) rectangle (1.1*\x+0.4,1.2);
        \pgfmathsetmacro{\y}{int(\x - 3)}
        \node[scale=0.4] at (1.1*\x,0.8) {};
    }
\end{tikzpicture}
\end{equation} 
where the red squares with four indices are commonly referred to as Matrix Product Operators. 
This architecture can be naturally recast into a sequential circuit form, of a "staircase" structure, where each gate acts on $r+1$ consecutive sites~\cite{Lami23ann, Sauliere25universality}, with $r = \log_d \chi$,
\begin{equation}\label{eq:rmpsOBC}
   \begin{tikzpicture}[baseline=(current  bounding  box.center),scale=0.6]
\definecolor{mycolor}{rgb}{0.86, 0.08, 0.24}
    \foreach \x in {1,...,6}{
        \draw[thick, black] (\x,0) -- (\x,6);
        \draw[thick, fill=white] (\x,0) circle (0.2);
        \draw[line width=0.8mm, gray!60!black] (\x,\x+0.2-1) -- (\x,\x);
        \node[scale=0.75] at (\x,-0.6) {\large $|0\rangle$};
    }
    \foreach \x in {1,...,5}{
        \draw[thick, fill=mycolor, rounded corners=2pt] (\x-0.2,\x-0.4) rectangle (\x+1+0.2,\x+0.4);
        \node[scale=0.4] at (\x+0.5,\x) {};
    }
\end{tikzpicture}
\end{equation}
The Haar-random gate applied to \( r+1 \) qubits is constructed using a deep even-odd brickwall circuit of two-qubit Haar-random gates, explicitly shown in~Eq.~\eqref{eq:rmpsBrickwall}. To ensure convergence to a Haar-random unitary on \( r+1 \) qubits, we use at least \( 2r \) layers in the brickwall circuit. 
To quantify the deviation from the Haar value, we introduce our primary observable 
\begin{equation}
\label{eq:dF}
\Delta \mathcal{F}_k = \mathcal{F}^{\mathrm{typ}}_k - \mathcal{F}_k,
\end{equation}
which represents the difference between \( \mathcal{F}_k \) of the considered state and the average FAF of a Haar-random state, \( \mathcal{F}^{\mathrm{typ}}_k \), given in Eqs.~\eqref{eq:faf1typ},~\eqref{eq:faf2typ}, and ~\eqref{eq:leadingtyp}. 

Our numerical analysis is based on two complementary techniques. The first method is to explicitly construct the RMPS by sampling unitaries in $\mathrm{U}(2^{r+1})$, and compute the covariance matrix $M$ by contracting matrix product operators representing pairs of Majorana operators with the RMPS tensor network. From this, we get the FAF via Eq.~\eqref{eq:faf}. Averaging over more than $200$ realizations, these simulations allow us to explore up to $N\le 1024$ qubits with bond dimension $\chi=2^r\le 1024$ and up to $N=4096$ for bond dimension $\chi=2^r\le 256$. 

We can outperform the direct RMPS construction method for $k=1$. 
In this case, as shown by Eq.~\eqref{eq:fafreplica}, the FAF computations involve $2$ copies of the system. Therefore, any 2-design, i.e., an ensemble of random gates that matches the statistical properties of the Haar ensemble $\mathrm{U}(2^{r+1})$ for up to two replicas, will give the averaged results~\cite{mele2024introductiontohaar}.
Concretely, we sample the 2-qubit Clifford gates with uniform probability over the Clifford group and apply them according to the staircase circuit in Eq.~\eqref{eq:rmpsOBC}. 
The resulting stabilizer state, and the covariance matrix $M$, can be efficiently computed with tableaux formalism~\cite{aaronson2004improvedsimulationof, Gidney2021stim} with computational resources scaling polynomially with system size $N$. 
Averaging over at least $10^4$ realizations of the circuits, this method enables us to compute the average of $\mathcal{F}_1$ over the ensemble of the RMPS with system sizes up to $N=4096$ and with bond dimensions $\chi \leq16384$. 

In Fig.~\ref{fig:rmps}, we present a numerical analysis of the fermionic antiflatness for random matrix product states. 
The left panel, Fig.~\ref{fig:rmps}~(a), shows the decay of the difference $\Delta \mathcal{F}_1$, given by Eq.~\eqref{eq:dF}, with the bond dimension $\chi$ of RMPS restricted to the even fermionic parity sector $\mathcal{P}=1$ (which we also refer to as the $\mathbb{Z}_2$-symmetric case). 
As $\chi$ increases, $\Delta \mathcal{F}_1 \to0$, confirming that RMPS converge to Haar-random behavior in the large $\chi$ limit.
To quantify this convergence, Fig.~\ref{fig:rmps}~(b) analyzes the decay of the rescaled FAF $\Delta \mathcal{F}_k/N$ for $k = 1$ and $k = 2$. We observe that, in both the generic and $\mathbb{Z}_2$-symmetric RMPS ensembles, the rescaled FAF exhibits data collapse onto universal curves in the large-N limit, following the power-law scaling
\begin{equation}
    \label{eq:FAF1_scaling}
    \Delta \mathcal{F}_k/N \propto \chi^{-\beta},
\end{equation}
where $\beta$ is an exponent depending on the value of $k$. 
For $k=1$, we find $\beta = 2$ with good accuracy. For $k=2$, the results suggest $\beta = 3.7(3)$, although this estimate is based on data that are not yet fully converged in $N$ and $\chi$.
The observed upward trend of $\beta$ with increasing system size and bond dimension suggests that the true asymptotic value may be $\beta=4$, making our estimate $\beta = 3.7(3)$ a conservative lower bound. 

Analyzing the two-point Majorana correlation functions provides further insight into the scaling behavior of RMPS described in Eq.~\eqref{eq:FAF1_scaling2}.
For a fixed bond dimension $\chi = 2^r$, in the regime of $r \ll N$, the squared expectation value of the two-point Majorana correlator decays exponentially with distance $x$
\begin{equation}
    \label{eq:Majorana_decay}
    \langle i \gamma_n \gamma_{n+x} \rangle^2 = C(r)\, e^{-x/\xi(r)},
\end{equation}
where $\xi(r)$ is the correlation length of the RMPS, and $C(r)$ is an $r$-dependent constant~\cite{Lio24corrRMPS}.
For $k=1$, the definition of FAF, Eq.~\eqref{eq:faf} translates to 
\begin{equation}
    \mathcal{F}_1=N-\sum_{m<n} \braket{i\gamma_m \gamma_n}^2.
\end{equation}
To obtain the leading behavior, it suffices to approximate $\mathcal{F}_1$ as $\mathcal{F}_1 \approx N(1-\sum_{x} \braket{i\gamma_n \gamma_{n+x}}^2)=D_k N$. The latter equation, through Eq.~\eqref{eq:Majorana_decay}, together with our numerical observation that $\xi(r)=O(1)$ is an $r$-independent constant, implies that $D_k = 1-c(\xi)C(r)$, where $c(\xi)$ is a constant. This is consistent with Eq.~\eqref{eq:FAF1_scaling2}, since $C(r)$ features and exponential decay with $r$, $C(r)\propto 2^{-2r}$.

More precisely, the result~Eq.~\eqref{eq:FAF1_scaling}, together with the leading term behavior $\mathcal{F}^{\mathrm{typ}}_k \approx N$, implies that the average FAF of RMPS is given by
\begin{equation}
    \label{eq:FAF1_scaling2}
    \mathcal{F}_k = \left( 1- \frac{a_k}{\chi^\beta} \right) N + b_k,
\end{equation}
where $a_k$ and $b_k$ are constants. This is another example of the extensive scaling of FAF with system size, cf.~Eq.~\eqref{eq:NdepOverview}, with a prefactor $D_k= 1- a_k \chi^{-\beta}$ approaching smoothly unity with an increase of the bond dimension $\chi$.
This behavior mirrors that of the stabilizer Rényi entropy (SRE), Eq.~\eqref{eq:sre}, in RMPS ensembles, which also approaches the Haar-random value with increasing $\chi$ in a power-law fashion~\cite{Lami25rmps}. 
This similarity is particularly striking because FAF and SRE characterize quantum complexity from two fundamentally distinct perspectives: fermionic Gaussian states and stabilizer states, respectively. 
Taken together, these results suggest that RMPS with increasing bond dimension approximate typical, featureless states not only in terms of entanglement but also with respect to other structural features. They highlight that entanglement, non-stabilizerness, and non-Gaussianity capture complementary aspects of the emergent randomness in shallow-circuit ensembles.

We conclude by highlighting a key connection between RMPS with polynomial bond dimension $\chi = \mathrm{poly}(N)$ and random circuits of logarithmic depth~\cite{magni2025anticoncentrationcliffordcircuitsbeyond}.
RMPS can be efficiently prepared via sequential two-site unitaries using tensor network algorithms such as time-evolving block decimation (TEBD), which apply unitary gates layer by layer to build up the entanglement structure. 
The circuit depth required to prepare an RMPS is proportional to the number of layers needed to generate the desired bond dimension $\chi$. 
Since each layer can at most double the effective bond dimension, achieving $\chi = \mathrm{poly}(N)$ requires only $\mathcal{O}(\log N)$ layers of two-site unitaries. 
Our result Eq.~\eqref{eq:FAF1_scaling2} shows that RMPS with $\chi = \mathrm{poly}(N)$ exhibit, for $N\gg1$, fermionic magic properties analogous to those of Haar-random states. This suggests that local random circuits of depth $\log(N)$ may suffice to approximate the FAF of Haar-random states. 
In the next section, we test this conjecture by analyzing the dynamics of local brick-wall circuits with tunable depth $t$.

\begin{figure*}
    \centering
    \includegraphics[width=1.\linewidth]{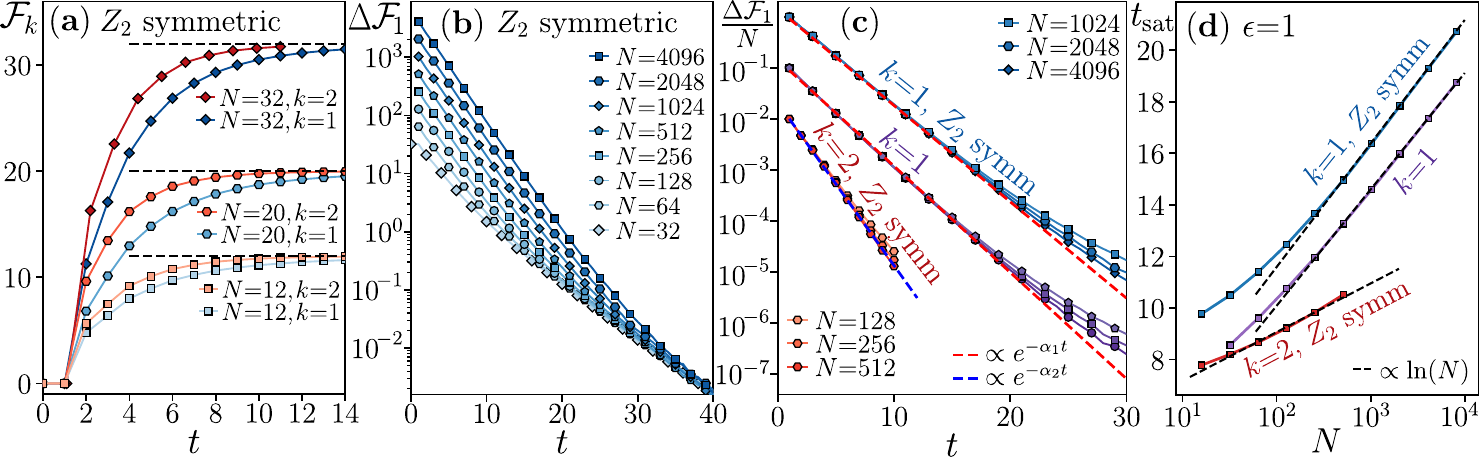}
    \caption{Fermionic antiflatness $\mathcal{F}_k$ growth in random quantum circuit dynamics acting on $N$ qubits.
    (a) $\mathcal{F}_k$ (shown for $k = 1, 2$) increases rapidly under the dynamics of a $\mathbb{Z}_2$-symmetric circuit of depth $t$.
    (b) The difference $\Delta \mathcal{F}_1$ decays exponentially in time $t$, signaling the saturation of FAF toward its Haar-random value $\mathcal{F}^{\mathrm{typ}}_k$.
    (c) The rescaled gap $\Delta \mathcal{F}_k/N$ collapses across system sizes onto a universal curve decaying as $\propto e^{-\alpha_k t}$, with rates $\alpha_1 = 0.45(2)$ and $\alpha_2 = 0.73(3)$. For clarity, data for $k = 1$ (without $\mathbb{Z}_2$ symmetry) are rescaled by a factor $1/10$, and for $k = 2$ (with $\mathbb{Z}_2$ symmetry) by $1/100$.
    (d) The saturation time $t_{\mathrm{sat}}$, defined by the condition $\Delta \mathcal{F}_k = \epsilon$ (with $\epsilon = 1$), scales logarithmically with system size: $t_{\mathrm{sat}} \propto \log(N)$.}
    \label{fig:circuits}
\end{figure*}

\subsection{Dynamics of random quantum circuits}
\label{subsec:circ}
Here, we study the dynamics of fermionic non-Gaussianity in random circuits acting on $N$ qubits. 
Specifically, we investigate the growth of FAF starting from the vacuum state $\ket{\Psi_0} = \ket{\mathbf{0}}$, evolved under a depth-$t$ brick-wall quantum circuit $U_t = \prod_{r=1}^t U^{(r)}$, schematically illustrated as
\begin{equation}\label{eq:rmpsBrickwall}
\begin{tikzpicture}[baseline=(current bounding box.center),scale=0.6]
\definecolor{mycolor}{rgb}{0.86, 0.08, 0.24}

\def\nqubits{6}
\def\ndepth{5}
\foreach \x in {1,...,\nqubits}{
    \draw[thick] (\x,0) -- (\x,\ndepth+0.5);
    \draw[thick, fill=white] (\x,0) circle (0.2);
    \node[scale=0.75] at (\x,-0.6) {\large $|0\rangle$};
}
\foreach \d in {0,...,\numexpr\ndepth-1\relax} {
    \pgfmathsetmacro{\y}{\d + 1} 
    \ifodd\d
        \foreach \x in {2,4} {
            \draw[thick, fill=mycolor, rounded corners=2pt]
                (\x-0.2,\y-0.3) rectangle (\x+1.2,\y+0.3);
        }
    \else
        \foreach \x in {1,3,5} {
            \draw[thick, fill=mycolor, rounded corners=2pt]
                (\x-0.2,\y-0.3) rectangle (\x+1.2,\y+0.3);
        }
    \fi
}
\end{tikzpicture}\;.
\end{equation}
Specifically, each layer of the circuit is constructed as a product of independent, identically distributed two-qubit gates, arranged in an alternating brick-wall pattern
\begin{equation}
\label{eq:brickwallCirc}
    U^{(2m)}=\prod_{i=1}^{N/2-1} U_{2i,2i+1}\;,\quad U^{(2m+1)}=\prod_{i=1}^{N/2} U_{2i-1,2i}\;.
\end{equation}
In the above equation, each two-qubit gate $U_{i,j}$ is sampled uniformly from $\mathrm{U}(4)$,or from the matrix ensemble $\mathrm{U}_{\mathbb{Z}_2}(4)\equiv\{U\in \mathrm{U}(4)\:| [Z^{\otimes 2},U]=0\}$. 
In the latter case, the fermionic parity $\mathcal{P}$ of the state is conserved, and we refer to the resulting circuits as $\mathbb{Z}_2$-symmetric.

How do fermionic magic resources increase under the circuit dynamics?
Initially, the system is in the fermionic Gaussian state $|\Psi_0\rangle = \ket{\mathbf{0}}$, and the non-Gaussianity of the state $\ket{\Psi_t} = U_t |\Psi_0\rangle$ increases with the circuit depth $t$, which we also refer to as time.
Analogously to the RMPS analysis, we consider FAF averaged over the circuit realizations, $\mathbb{E}[\mathcal{F}_k(\ket{\Psi_t})]$, which, for simplicity, we denote as $\mathcal{F}_k(t)$.
In parallel to Sec.~\ref{subsec:RMPS}, we consider two numerical techniques. For $k=1$, we employ stabilizer simulations by replacing the gates $U_{i,j}$ in Eq.~\eqref{eq:brickwallCirc} with Clifford gates selected uniformly from the group of all $11520$ two-qubit Clifford gates, or, in the $\mathbb{Z}_2$-symmetric case, from the subgroup of two-qubit gates that commute with the $Z^{\otimes2}$ operator.
This method allows us to obtain the average $\mathcal{F}_1(t)$ for system sizes up to $N=8192$ and times $t\le 40$.
For $k=2$, we instead employ standard tensor network evolution techniques, representing the time-evolved state $\ket{\Psi_t}$ as an MPS, which allows us to reach system sizes $N\le 512$ at moderate depths up to $t=10$.

The growth of $\mathcal{F}_k(t)$ for $k=1,2$ as a function of the circuit depth $t$ for different system sizes is shown in Fig.~\ref{fig:circuits}~(a).
We observe that $\mathcal{F}_k(t)$ increases rapidly under the action of the $\mathbb{Z}_2$-symmetric brick-wall circuit, 
scaling extensively, $\mathcal{F}_k(t)\propto N$ already at $2\le t=O(1)$.  At $t=1$, the state $\ket{\psi_1}=U_1\ket{\psi_0}$ is a product of two qubit states, which is a fermionic Gaussian state in the $\mathbb{Z}_2$-symmetric case, $\mathcal{F}_k(\ket{\Psi_1})=0$, cf. Sec.~\ref{subsec:prod1}.
To quantify the convergence of $\mathcal{F}_k$ to the FAF of Haar-random states, we analyze the deviation $\Delta \mathcal{F}_k = \mathcal{F}_k^\mathrm{typ} - \mathcal{F}_k(t)$ as a function of time $t$.
We observe an exponential decay of $\Delta \mathcal{F}_k$, signaling fast saturation of FAF towards the typical value, as shown for $k=1$ in Fig.~\ref{fig:circuits}~(b). 
Universal trends in the data are visible in Fig.~\ref{fig:circuits}~(c), which shows that $\Delta \mathcal{F}_k/N$ for $k=1$ and $k=2$ for different system sizes collapse onto master curves dependent on $k$ and the symmetry of the circuit. We observe an exponential decay, 
\begin{equation}
\label{eq:FdecayCirc}
    \Delta \mathcal{F}_k/N \propto e^{-\alpha_k t},
\end{equation}  
with decay rates $\alpha_1 = 0.45(2)$ and $\alpha_2 = 0.73(3)$. The FAF with higher $k$ saturates faster to the long-time value, and the saturation rate is similar regardless of whether the circuit possesses $\mathbb{Z}_2$ symmetry or not. 
We define the saturation time $t_\mathrm{sat}$ as the depth $t$ at which $\Delta \mathcal{F}_k$ drops below a fixed threshold $\epsilon$. Fig.~\ref{fig:circuits}~(d) shows that $t_\mathrm{sat}$ scales logarithmically with system size, $t_\mathrm{sat} \propto \log(N)$, for both $k=1,2$ and independently of whether the circuit has $\mathbb{Z}_2$ symmetry.

The presented results provide direct numerical support for the conjecture put forward at the end of Sec.~\ref {subsec:RMPS}, namely, that logarithmic-depth random circuits are sufficient to generate states with FAF values comparable to those of Haar-random states.  
The observed exponential relaxation of $\mathcal{F}_k$, Eq.~\eqref{eq:FdecayCirc}, confirms that the scaling is controlled by circuit depth rather than system size, aligning with similar depth-controlled behaviors observed for other typicality indicators, such as anticoncentration~\cite{turkeshi2024hilbert} and SRE~\cite{turkeshi2024magic}.
Importantly, the operator $|\Upsilon_{r_1,r_2,\dots,r_k}^{(k)}\rrangle$, Eq.~\eqref{eq:comop}, whose expectation value gives $\mathcal{F}_k$ is \textit{not} a positive semidefinite operator. 
Existing rigorous results show that random circuits of logarithmic depth form $k$-designs~\cite{Laracuente24approximate, Schuster2024random}, which apply to quantities such as the frame potential or linear cross-entropy benchmarking. However, these results do not extend to FAF, due to the lack of positive semidefiniteness of  $|\Upsilon_{r_1,r_2,\dots,r_k}^{(k)}\rrangle$. 
This makes our findings a genuinely non-trivial manifestation of typicality, and calls for further theoretical developments to better understand the conditions under which quantities like FAF, corresponding to non-positive replica operators, saturate at time scales at which the considered ensembles start forming approximate $k$-designs.

\section{Fermionic magic resources in equilibrium}
\label{sec:equilibrium}
The basis for our investigations of fermionic magic resources in quantum many-body systems is provided by the transverse-field Ising model (TFIM)
\begin{equation}
    \label{eq:Ising0}
H_{0}= -h_z\sum_{m=1}^{N} Z_m -  J \sum_{m=1}^{N-1}X_m X_{m+1} -g X_{N} X_{1}  \;,
\end{equation} 
where $h_z$, $J$ and $g$ are parameters of the model, with the latter, $g$, specifying boundary conditions: $g=0$ for open boundary conditions (OBC) and $g=J$ for periodic boundary conditions (PBC). 
In the following, we set the energy scale as $J=1$. While $H_0$ is a Hamiltonian describing a system of $N$ spins, that is, an operator acting on the $d=2^N$ dimensional Hilbert space~$\mathcal{H}_2^{\otimes N}$, all physically relevant properties of the TFIM can be determined by solving computational problems that scale only \textit{polynomially} with system size $N$. 
The techniques enabling this drastic simplification are standard in the literature~\cite{Lieb61,Suzuki13,Mbeng24,Surace22} and are rooted in the fact that Eq.~\eqref{eq:Ising0} describes a system of non-interacting fermions. To provide context for our investigations of FAF when the free-fermionic description breaks down due to introduction of interactions~\cite{Turner17free}, we briefly discuss the role played by fermionic Gaussian states in solving the TFIM.

Employing the Jordan-Wigner transformation~Eq.~\eqref{eq:JW}, the Hamiltonian~Eq.~\eqref{eq:Ising0} can be rewritten as
\begin{equation}
    \label{eq:IsingMajo}
H_{0}=i h_z  \sum_{m=1}^{N} \gamma_{2m-1} \gamma_{2m} +i \sum_{m=1}^{N-1}  \gamma_{2m}\gamma_{2m+1} + i g \mathcal{P}    \gamma_{2N}\gamma_{1}\;.
\end{equation} 
The TFIM conserves fermionic parity, $[H_0,\mathcal{P}]=0$. In the following study, we will focus on the even parity subspace of the Hilbert space, $\mathcal{P}=1$, in which the Hamiltonian reduces to 
\begin{equation}
    \label{eq:IsingPositve}
H_{+}=i h_z  \sum_{m=1}^{N} \gamma_{2m-1} \gamma_{2m} +i  \sum_{m=1}^{N-1}  \gamma_{2m}\gamma_{2m+1} - i g   \gamma_{1} \gamma_{2N}\;,
\end{equation} 
becoming a quadratic expression in terms of Majorana operators 
\begin{equation}
    \label{eq:quad}
H_{+}= \frac{i}{4}\sum_{m,n} H_{m,n} \gamma_m \gamma_n,
\end{equation} 
where $(H_{m,n})=H=-H^T$ is an antisymmetric $2N\times 2N$ matrix.
The matrix $H$ can be brought to its block-diagonal canonical form
\begin{equation}
\label{eq:canon2}
   H = G^T \bigoplus_{m=1}^N \begin{pmatrix}
        0 & \epsilon_m\\ -\epsilon_m & 0
    \end{pmatrix}G
\end{equation}
where $\epsilon_i>0$ are the so-called Williamson eigenvalues of $H$ (compare with Eq.~\eqref{eq:canon}), and $G$ is an orthogonal $2N\times 2N$ matrix. The unitary $U_G$ induced by the transformation $G$, defines a new set of Majorana fermions $\tilde{\gamma}_{k} = U^{\dagger}_G \gamma_k U_G$, allowing us to rewrite 
Eq.~\eqref{eq:quad} as  
\begin{equation}
    \label{eq:quadDiag}
H_{+}= \frac{i}{2}\sum_{m} \epsilon_m \tilde{\gamma}_{2m-1} \tilde{\gamma}_{2m}.
\end{equation} 
The operators $i \tilde{\gamma}_{2k-1} \tilde{\gamma}_{2k}$ commute for different choices of $k$, and each of them has eigenvalues $\pm 1$. Consequently, the ground state of the TFIM in the positive parity sector has energy $E_0=-\sum_m \epsilon_m/2$ and can be obtained from the vacuum state as $\ket{\Psi_0} = U_G\ket{\mathbf{0}}$. Similarly, higher excited eigenstates $\ket{\Psi_n}$ of $H_+$ are obtained by acting with $U_G$ on other computational basis states $\ket{\sigma}$ is the $\mathcal{P}=1$ subspace. The ground state  $\ket{\Psi_0}$, as well as the excited states $\ket{\Psi_n}$, result from the action of a fermionic Gaussian unitary $U_G$ on computational basis states, and hence, these states are fermionic Gaussian states, cf.~Eq.~\eqref{eq:fermionicG}. 

Despite being a free-fermionic state, the TFIM ground state, $\ket{\Psi_0}$, plays an essential role in our understanding of quantum many-body physics, and in particular, of quantum phase transitions~\cite{Sachdev99, Suzuki13}. 
The TFIM ground state undergoes a phase transition at the critical field strength $h_z=h_c=1$, in the vicinity of which the correlation length $\xi$ diverges according to a power-law, $\xi \propto |h_z-h_c|^{-\nu}$, with a critical exponent $\nu=1$. The system is in ferromagnetic phase for $h_z<h_c$, in which the spin-spin coupling terms $X_m X_{m+1}$ enforce polarization along the $\pm x$ direction, while for $h_z >h_c$, the system system is in paramagnetic phase with spins polarized into $-z$ direction. The development of the long-range quantum correlations in the vicinity of the transition leads to a particular sensitivity of the ground state to changes in the parameters of the model~\cite{Rams11, Damski13}, and is reflected in the entanglement~\cite{amico2008entanglement, horodecki2009quantum} of the ground state. 
In particular, at the transition, $h=h_c$, the entanglement entropy diverges logarithmically~\cite{Vidal03, Peschel04, Its05}, 
\begin{equation}
    \label{eq:entlog}
S \propto \frac{c}{3}\ln(l),
\end{equation} 
where $l$ is the subsystem size, and the prefactor depends on the central charge $c$ of the conformal field theory describing the transition~\cite{DiFrancesco97, Holzhey94, Calabrese04, calabrese2009entanglement}. (For the Ising universality class, the value is $c = 1/2$.)

Obtaining these insights into physics of quantum phase transition in TFIM, and finding a ground state of any Hamiltonian quadratic in Majorana operators, Eq.~\eqref{eq:quad}, requires finding the orthogonal matrix $G$ which brings the Hamiltonian to the diagonal form, cf.~Eq.~\eqref{eq:quadDiag}, which is analytically tractable in some cases and, in general, can be solved with computational cost scaling as $N^3$. 
The resulting ground state $\ket{\Psi_0} = U_G\ket{\mathbf{0}}$ is a fermionic Gaussian state, and contains no fermionic magic resources: the FAF vanishes $\mathcal{F}_k(\ket{\Psi_0}) =0$. The latter reflects the computational simplicity of $\ket{\Psi_0}$. In the following, we analyze the behavior of fermionic magic resources in the ground states of many-body systems obtained by introducing terms to the TFIM Hamiltonian~Eq.~\eqref{eq:Ising0} which go beyond quadratic terms in Majorana operators and introduce interactions between fermionic modes.


We conclude this section with a remark on basis dependence. The choice of Majorana operators is not unique: local rotations
\((X_i,Y_i,Z_i)\mapsto (U_i X_i U_i^\dagger,\, U_i Y_i U_i^\dagger,\, U_i Z_i U_i^\dagger)\),
where \(U_i \in \mathrm{U}(2)\) are on-site unitaries, define inequivalent sets of Majorana operators. In the absence of system-specific knowledge, these choices are a priori equivalent, and FAF may be optimized over \(\{U_i\}\). For parity-conserving Hamiltonians \([H,\mathcal{P}]=0\) (e.g., the TFIM considered below), a strategy leading to the simplest physical picture is to avoid local rotations that break parity; this restricts \(\{U_i\}\) to rotations about the \(Z\) axis. Such \(U_i\) are fermionic Gaussian unitaries and therefore leave FAF invariant.

\subsection{Fermionic antiflatness induced by a single impurity}
We begin our analysis by perturbing the TFIM with a single impurity term, which breaks down the free-fermionic description of the system. Our primary focus is the behavior of FAF in ground states, particularly in the vicinity of the ferromagnetic–paramagnetic phase transition. Hence, we choose the impurity term to commute with the fermionic parity operator $\mathcal{P}$, ensuring that the Ising transition still persists in the system. Specifically, the impurity is taken as $H_{\mathrm{imp}} = \lambda X_{\ell_0} X_{\ell_0+2} $, with the site $\ell_0 = N/2$ located at the center of the system. Upon Jordan-Wigner transformation~Eq.~\eqref{eq:JW}, the resulting Hamiltonian, referred to hereafter as the \emph{impurity model}, reads 
\begin{equation}
    \label{eq:imp1}
H_{\mathrm{imp}} = H_{+} - \lambda \gamma_{2 \ell_0} \gamma_{2\ell_0+1} \gamma_{2\ell_0+2} \gamma_{2\ell_0+3}
\end{equation}  
within the even fermionic parity subspace, $\mathcal{P} = 1$.
The impurity term is quartic in Majorana operators and $H_{\mathrm{imp}}$ at $\lambda \neq 0$ can no longer be diagonalized via a quadratic transformation~Eq.~\eqref{eq:quadDiag}. We find the ground state $\ket{\Psi_0}$ of Eq.~\eqref{eq:imp1} by the standard density matrix renormalization group algorithm (DMRG)~\cite{White1992, White1993, Schollwock2011} finding $\ket{\Psi_0}$ approximated as an MPS with a fixed bond dimension~$\chi$. We verify convergence of all results with respect to bond dimension $\chi$ by confirming that they remain unchanged when $\chi$ is doubled.

\begin{figure}
    \centering
    \includegraphics[width=1\linewidth]{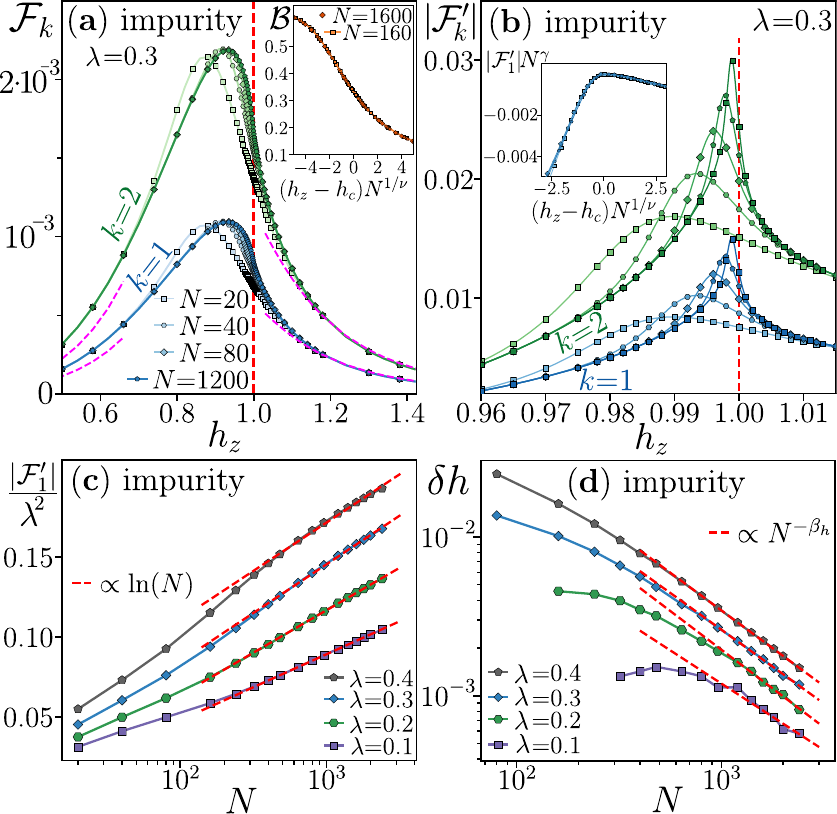}
    \caption{
    Fermionic antiflatness $\mathcal{F}_k$ in the impurity model~Eq.~\eqref{eq:imp1}. 
    (a) The FAF (shown for $k=1,2$) is enhanced in the vicinity of the ferromagnetic–paramagnetic transition, and the slope of the $\mathcal{F}_k(h_z)$ curves increases near the critical point $h_z = h_c = 1$, indicated by the red dashed line. The transition belongs to the Ising universality class with critical exponent $\nu = 1$, as shown by the collapse of the Binder cumulant $\mathcal{B}$ in the inset. 
    (b) The absolute value of the derivative $|\mathcal{F}'_k|$ with respect to $h_z$ exhibits a maximum at field $h = h_m$. The inset shows the scaling collapse of $|\mathcal{F}'_1| - F_1$ near the transition.
    (c) The value of $|\mathcal{F}'_{k=1}|$ at the maximum, denoted $F_1$, increases logarithmically with $N$. 
    (d) The location $h_m$ of the maximum converges to $h_c$, with $\delta h = |h_m - h_c| \propto N^{-\beta_h}$ and $\beta_h = 0.98(4)$.
    }
    \label{fig:impGS}
\end{figure}
To confirm that the ferromagnetic–paramagnetic phase transition of the TFIM persists in the presence of the impurity term, we evaluate the Binder cumulant~\cite{Binder81, Binder81a},
\begin{equation}
    \mathcal{B} = 1-\frac{\braket{\Psi_0|(\sum_m X_m)^4|\Psi_0}} {3(\braket{\Psi_0|(\sum_m X_m)^2|\Psi_0})^2}.
\end{equation}
The Binder cumulant approaches a step-like profile near the critical field $h_z = h_c = 1$ as the system size $N$ increases, as expected for the ferromagnetic–paramagnetic phase transition. Furthermore, the data collapse under the one-parameter scaling hypothesis, $\mathcal{B} = f\big((h_z - h_c) N^{1/\nu}\big)$, with critical exponent $\nu = 1$, as illustrated in the inset of Fig.~\ref{fig:impGS}(a). This confirms that the impurity term is irrelevant in the renormalization group sense and does not alter the universality class of the transition.

Consistent with the expectation that the ground state $\ket{\Psi_0}$ of the impurity model is not a fermionic Gaussian state, the FAF, $\mathcal{F}_k$, is nonzero.
As presented in Fig.~\ref{fig:impGS}(a), $\mathcal{F}_k$ is enhanced in the vicinity of the transition, $h_z=h_c=1$, and decreases smoothly away from the critical point.
Away from the critical point, $\mathcal{F}_k$ quickly saturates with system size $N$, i.e., $\mathcal{F}_k = \mathrm{const}$, and the ground state of $H_{\mathrm{imp}}$, Eq.~\eqref{eq:imp1}, can be approximated by the perturbative expansion
\begin{equation}
\label{eq:pert}
     \ket{\Psi_0} \approx \ket{\phi_0} + \lambda \sum_{n\neq 0} \delta^{(1)}_n \ket{\phi_n} + \lambda^2\sum_{n} \delta^{(2)}_n \ket{\phi_n},
\end{equation}
where $H_+ \ket{\phi_n} = E_n \ket{\phi_n}$, and  $\delta^{(1,2)}_n$ are the first and the second order perturbation theory correction coefficients, whose explicit form is immaterial for the present discussion and can be found in standard quantum mechanics textbooks~\cite{sakurai20modern}.
To understand the implications of Eq.~\eqref{eq:pert} for FAF, we compute the covariance matrix of $\ket{\Psi_0}$ as 
\begin{equation}
\label{eq:pert2}
   M_{ab} =  M^{(0)}_{ab}   +  \lambda \sum_n  \delta^{(1)}_n \Xi^{[n]}_{ab} + O(\lambda^2),
\end{equation}
where $M^{(0)}_{ab} =  -\frac{i}{2}\bra{\phi_0} [\gamma_a , \gamma_b ]\ket{ \phi_0}$ and we introduce $\Xi^{[n]}_{ab} =-\frac{1}{2} \Re( \bra{\phi_0} i[\gamma_a,\gamma_b]\ket{\phi_n})$. This implies that FAF, cf. Eq.~\eqref{eq:faf}, is given by
\begin{equation}
\label{eq:pert3}
   \mathcal{F}_1 =    \lambda \sum_{n\neq0}  \delta_n  \mathrm{tr}[\Xi^{[n]} M^{(0)} ]+ O(\lambda^2),
\end{equation}
where $\Xi^{[n]}$ denotes the $2N\times 2N$ matrix with elements $\Xi^{[n]}_{ab}$, and used the fact that $N-\frac{1}{2}\mathrm{tr}[ (M^{(0)})^T  M^{(0)}]=0$ since $\ket{\phi_0}=U_G \ket{\mathbf{0}}$ is a fermionic Gaussian state. Using $\ket{\phi_n} = U_G\ket{\mathbf{n}}$ with Eq.~\eqref{eq:cov} and the cyclic property of the trace, we find that
\begin{equation}
  \mathrm{tr}[\Xi^{[n]} M^{(0)} ] =\frac{1}{4}\sum_{a\neq b}  \bra{ \mathbf{0}} [\gamma_a , \gamma_b ]\ket{ \mathbf{0} } \Re( \bra{ \mathbf{0} } i[\gamma_a,\gamma_b]\ket{ \mathbf{n}  } ),
\end{equation}
which vanishes for all excited states $\mathbf{n}\neq \mathbf{0}$. 
This implies, through Eq.~\eqref{eq:pert3}, that the lowest order contribution in $\lambda$ to FAF\footnote{The absence of the first order correction in $\lambda$ and the dependence $\mathcal{F}_k \propto \lambda^2$ is a direct consequence of the unperturbed states $\{ \ket{\phi_n} \}$ being fermionic Gaussian states. The impurity model, as well as the other models studied here share this property.} is $\mathcal{F}_1 \propto \lambda^2$, consistently with our numerical results for various $\lambda \in(0,0.3)$. Numerically computing the approximate ground state according to Eq.~\eqref{eq:pert}, and keeping the terms up to second order in $\lambda$, enables us to accurately reproduce the value of $\mathcal{F}_k$ even for $\lambda=0.3$, away from the transition, as shown by the magenta dashed lines in Fig.~\ref{fig:impGS}(a).

The system size dependence of FAF is more pronounced in the vicinity of the transition. The maximum of $\mathcal{F}_k$ as a function of $h_z$ shifts towards $h_z=h_c$ with increasing $N$. Nevertheless, even in the large $N$ limit, the maximum occurs at field strength $h_z < h_c$. 
The transition instead manifests itself by the increase of the slope of $\mathcal{F}_k(h_z)$ at $h_z=h_c$ with system size $N$ growth. To quantify this behavior, we focus on the derivative of FAF with respect to the magnetic field, $\mathcal{F}'_k = \frac{d\mathcal{F}_k}{d h_z}$. The behavior of the absolute value of the derivative, $|\mathcal{F}'_k|$, is shown in Fig.~\ref{fig:impGS}(b). The maximum of $|\mathcal{F}'_k|$ sharpens up with the increase of $N$ and occurs at a field strength $h_m$. The absolute value of the derivative at $h_z=h_m$ increases logarithmically with system size,
\begin{equation}
\label{eq:Fimpu}
\max_{h_z}|\mathcal{F}_k'| =F_1 \propto \ln(N),
\end{equation}
as presented in Fig.~\ref{fig:impGS}(c). At the maximum $|F'_1|$ no longer scales proportionally to $\lambda^2$ indicating the breakdown of the perturbative argument Eq.~\eqref{eq:pert3} in the vicinity of the transition. The position of the maximum $h_m$ converges to the critical point, as evidenced by the power-law decrease $\delta h = |h_m - h_c| \propto N^{-\beta_h}$, with $\beta_h \approx 1$ as shown in Fig.~\ref{fig:impGS}~(d). 
This behavior of  $|\mathcal{F}'_k|$ in the vicinity of the critical point is analogous to the behavior of entanglement entropy~\cite{Calabrese04} and stabilizer R\'{e}nyi entropy~\cite{haug2023quantifying}. To investigate the emergence of universal behavior, we plot $(|\mathcal{F}'_k|-F_1)N^{\gamma_{L,R}}$ as function of the rescaled variable $(h_z-h_c)N^{1/\nu}$. An excellent data collapse is observed, see the inset in Fig.~\ref{fig:impGS}(b), provided that an \emph{asymmetric} scaling is employed, with $\gamma_L=0.08(2)$ and $\gamma_R=-0.23(4)$ respectively below, $h_z<h_c$, and above, $h_z > h_c$, the transition.

Summing up, the numerical results for the impurity model show that a single non-Gaussian impurity term introduces a constant value of FAF to the ground state, similarly to the example of a local non-Gaussian gate acting on a fermionic Gaussian state, considered in Sec.~\ref{subsec:localGate}. While the FAF is enhanced in the vicinity of the transition, it is not the FAF, but rather its derivative over the tuning parameter, $|\mathcal{F}_k'|$, which admits a maximum diverging logarithmically with system size $N$,~Eq.~\eqref{eq:Fimpu}.

\subsection{Fermionic magic resources in the ANNNI model}
We now consider a situation in which non-Gaussian impurity terms appear throughout the entire 1D system, leading to
\begin{equation}
    \label{eq:ani}
H_{\mathrm{ANNNI}} = H_{0} - \lambda  \sum_{m=1}^{N-2 }X_m X_{m+2} - \lambda g X_{N-1}X_{1}- \lambda g X_{N}X_{2}.
\end{equation}  
This defines the so-called Axial Next-Nearest-Neighbour Ising (ANNNI) model~\cite{Elliott61, Selke1988}, with $g=0$ for OBC and $g=1$ for PBC. 
The ground state of ANNNI model exhibits a rich phase diagram~\cite{Fisher80, Arizmendi1991, Sen1992,Rieger96,  Allen01, Dutta03, Chandra07, Beccaria06, Beccaria2007, Fumani21, Ferreiramartins23, Cea24, Damerow25}, featuring a ferromagnetic-paramagnetic phase transition for $0< \lambda <0.5$, and a floating phase emerging for $\lambda>0.5$ and sufficiently small $h_z$. 
In the following, we restrict our analysis to the even-parity subspace and focus on the former transition in the $0 < \lambda < 0.5$ regime, aiming to characterize the behavior of FAF across the ferromagnetic and paramagnetic phases and to identify universal features emerging at the transition.

\begin{figure}
    \centering
    \includegraphics[width=1\linewidth]{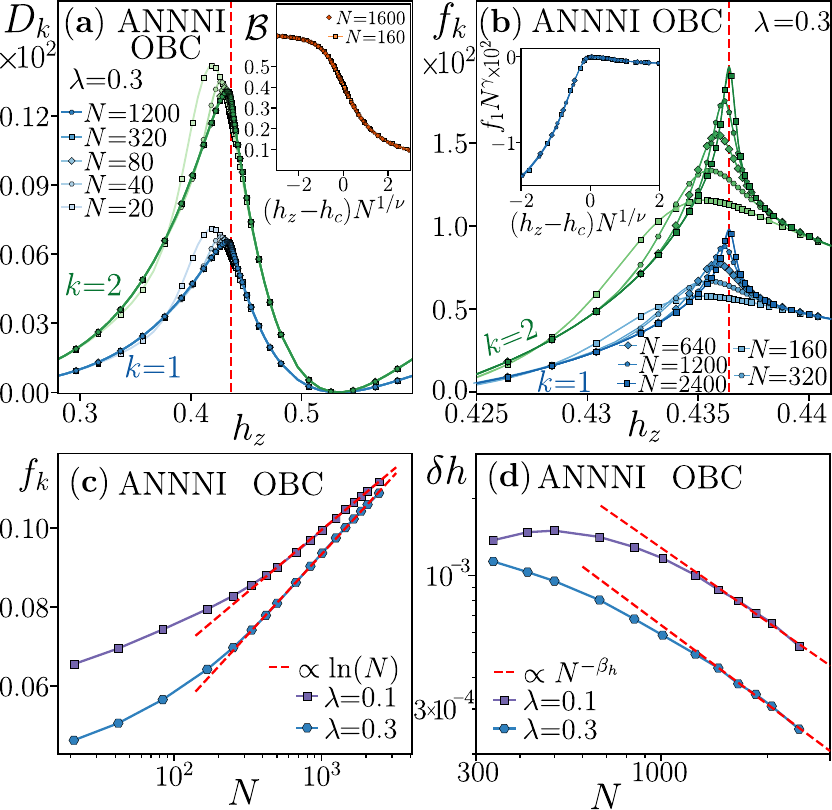}
    \caption{Fermionic antiflatness $\mathcal{F}_k$ in ANNNI model~Eq.~\eqref{eq:ani} with OBC. (a) The leading term $D_k$ in the FAF system size dependence (shown here for $k=1,2$) is enhanced in the vicinity of the ferromagnetic-paramagnetic transition. The latter belongs to the Ising universality class with the exponent $\nu=1$, as shown in the inset by the collapse of the Binder around the critical point $h_z=h_c=0.4367(1)$ for $\lambda=0.3$. (b) The subleading term $f_k$ admits a maximum at field $h=h_m$, and $f_k(h_m)$ is increasing logarithmically with system size $N$, as presented in (c), while its position is converging to $h_c$, as shown by $\delta h = |h_m-h_c|\propto N^{-\beta_h}$ (with $\beta_h=0.99(5)$), shown in panel (d). The inset in (b) shows the collapse of $f_k-f_k(h_m)$ at the transition.
    }
    \label{fig:aniGS}
\end{figure}

There are extensively many non-Gaussian, quartic in Majorana operators, terms $X_m X_{m+2} = -\gamma_{2 m} \gamma_{2m+1} \gamma_{2m+2} \gamma_{2m+3}$ in the ANNNI Hamiltonian~Eq.~\eqref{eq:ani}. This implies that  $\delta^{(1)}_n \propto N$ in Eq.~\eqref{eq:pert}, suggesting the extensive scaling of FAF with the system size $N$, given by Eq.~\eqref{eq:NdepOverview}. 
We note that extensive system size dependence analogous to Eq.~\eqref{eq:NdepOverview} is found for participation entropies~\cite{luitz2014universal, mace2019multifractal, Pausch21, sierant2022universal} and for stabilizer R\'{e}nyi entropy~\cite{haug2023quantifying} in various quantum many-body ground states. Computing the ground state of ANNNI model with DMRG algorithm, we confirm the extensive scaling of FAF, $\mathcal{F}_k \propto N$ for any non-trivial choice of the parameters of the system. 
For a general, i.e., not necessarily linear dependence of $\mathcal{F}_k$ on $N$, Eq.~\eqref{eq:NdepOverview} can be thought of as a local approximation of $\mathcal{F}_k(N)$. In that case, both $D_k$ and $f_k$ may depend non-trivially on $N$. To obtain $D_k$ and $f_k$, we compute $\mathcal{F}_k(N)$ for two close system sizes $N_1$ and $N_2= N_1+\Delta N$ (where $\Delta N < N_1/8$) and use Eq.~\eqref{eq:NdepOverview} to calculate $D_k=[F_k(N_2)-F_k(N-1)]/\Delta N$ and $f_k=[N_1F_k(N_2)-N_2F_k(N-1)]/\Delta N$.

We start our analysis by considering the ANNNI model with OBC, setting $g=0$ in Eq.~\eqref{eq:ani}. The leading term $D_k$ as a function of $h_z$ is shown in Fig.~\ref{fig:aniGS}(a). 
We observe that the shape of the $D_k(h_z)$ curve near the paramagnetic–ferromagnetic transition qualitatively resembles the behavior of $\mathcal{F}_k$ in the impurity model shown in Fig.~\ref{fig:impGS}(a).
Intuitively, the extensive scaling of FAF, $\mathcal{F}_k \propto D_k N$, can be interpreted as arising from the sum of contributions due to the extensively many impurity terms, each contributing a constant amount—analogous to our findings for the impurity model in Eq.~\eqref{eq:imp1}.
Moreover, in the limit $\lambda = 0$, corresponding to the free-fermionic model, and consistent with the reasoning of Eqs.~\eqref{eq:pert}–\eqref{eq:pert3}, the first-order term in $\lambda$ vanishes. Consequently, we find $D_k \propto \lambda^2$ for sufficiently small $\lambda$ away from the critical point $h_z = h_c$.

The position of the critical point, at which the ANNNI model undergoes a ferromagnetic–paramagnetic phase transition, decreases continuously with $\lambda$. It starts from the TFIM case, $h_c=1$ at $\lambda=0$ and approaching $h_c \to 0$ as $\lambda \to 1/2$. In particular, for $\lambda=0.3$, the critical point is located at $h_c=0.4367(1)$~\cite{Beccaria06}, as confirmed by the collapse of the Binder cumulant $\mathcal{B}$ in the inset in Fig.~\ref{fig:aniGS}(a). The transition belongs to the Ising universality class with the critical exponent $\nu=1$. 
While the leading term $D_k$ is enhanced in the vicinity of the transition, its maximum \emph{does not} coincide exactly with the critical point, mirroring the behavior of $\mathcal{F}_k$ for the impurity model.

Instead, the presence of the transition is reflected in the behavior of the subleading term $f_k$, shown in Fig.~\ref{fig:aniGS}(b). The subleading term admits a maximum $f^{(\mathrm{m})}_k$ at field strength $h_z=h_m$. In Fig.~\ref{fig:aniGS}(c), we observe a clear logarithmic increase
\begin{equation}
  \max_{h_z} f_k =   f^{(\mathrm{m})}_k \propto \ln(N) 
\end{equation} 
with system size $N$, analogous to the divergence of entanglement entropy at the critical point~Eq.~\eqref{eq:entlog}. 
The distance between the maximum of $f_k$ and the critical point, $\delta h = |h_m-h_c|$, decreases with the system size approximately as $N^{-1}$. To reveal the emergence of universal behavior in the subleading term, we show a data collapse of $(f^{\mathrm{m}}_k - f_k)N^{\gamma_{L,R}}$ as a function of $(h_z-h_c)N^{1/\nu}$ in the inset in Fig.~\ref{fig:aniGS}(b), using asymmetric scaling exponents with $\gamma_{L}=0.2$ and $\gamma_R=-0.2$ for $h_z<h_c$ and $h_z > h_c$, respectively. 
Outside of the critical regime of $h_z$, whose size is shrinking with the increase of system size as $N^{-1/\nu}$, both the leading and subleading terms saturate to system size independent values, $D_k=\mathrm{const}$, and $f_k=\mathrm{const}=c_k$, consistent with the phenomenology observed for RMPS with a fixed bond dimension $\chi$, cf. Eq.~\eqref{eq:FAF1_scaling2}.

\begin{figure}
    \centering
    \includegraphics[width=1\linewidth]{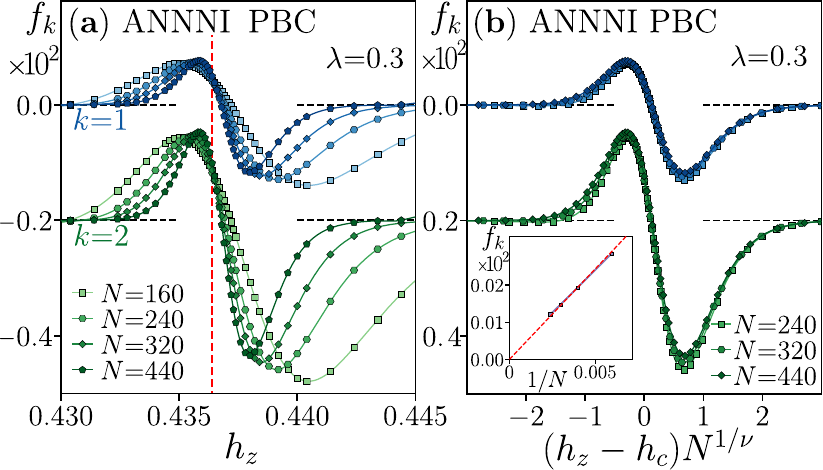}
    \caption{Fermionic antiflatness (FAF) $\mathcal{F}_k$ in ANNNI model~Eq.~\eqref{eq:ani} with PBC. (a) The subleading term $f_k$ in the FAF system size dependence (shown here for $k=1,2$, data for $k=2$ shifted by $-0.2$ for clarity) vanishes with increasing system size $N$ away from the transition, approaching a step-like behavior at the critical point $h_z=h_c$ denoted by the dashed line (here, for $\lambda=0.3$, $h_c=0.4367(1)$). (b) Collapse of $f_k$ plotted as a function of the scaling variable $(h_z-h_c)N^{1/\nu}$ onto universal, $k$-dependent, curves. The inset shows the subleading term at the crossing points of $f_k(h_z)$ curves for neighboring values of $N$.
    }
    \label{fig:aniPBC}
\end{figure}

The behavior of the leading term $D_k$ is largely independent of the choice of boundary conditions in the ANNNI model. In contrast, the subleading term $f_k$ exhibits quantitatively different behavior at the transition when the PBC ($g=1$ in Eq.~\eqref{eq:ani}) are imposed. 
The behavior of $f_k$ in the vicinity of the ferromagnetic-paramagnetic phase transition in the ANNNI model is shown in Fig.~\ref{fig:aniPBC}(a).
Away from the transition, the subleading term $f_k$ approaches zero, and the FAF becomes simply proportional to the system size, $\mathcal{F}_k = D_k N$. In the vicinity of the critical point, $f_k$ crosses over between positive and negative values within a shrinking critical region as $N$ increases. 
This crossover is associated with the emergence of universal behavior in $f_k$, which follows a single-parameter scaling form and exhibits data collapse as $f_k = g[(h_z-h_c)N^{1/\nu}]$, as demonstrated in Fig.~\ref{fig:aniPBC}(b).
The parameters $h_c=0.4367(1)$ and $\nu=1$ are obtained from the finite-size scaling analysis of the Binder cumulant $\mathcal{B}$.

Concluding this section, we summarize the behavior of FAF in the ground states of the impurity and ANNNI models. In the limit of large system size ($N \gg 1$), the FAF in the impurity model behaves as
\begin{equation}
  \mathcal{F}_k= c_k(h_z),
  \label{eq:f1}
\end{equation} 
where $c_k(h_z)$ is a constant independent of system size and dependent on the magnetic field $h_z$.  Its slope diverges logarithmically at the critical point, $|\mathcal{F}_k'|_{h_z=h_c} \propto \ln(N)$. 

For the ANNNI model, away from the critical regime, we find
\begin{equation}
  \mathcal{F}_k= D_k N + f_k,
  \label{eq:f2}
\end{equation} 
where the leading and subleading terms, $D_k$ and $f_k$, converge to finite constants that vary smoothly with $h_z$ In the case of periodic boundary conditions (PBC), the subleading term vanishes, $f_k = 0$.
At the critical point $h_z = h_c$ and for $N \gg 1$, the subleading term exhibits distinct scaling behavior depending on the boundary conditions
\begin{equation}
  f_k = 
  \begin{cases}
    a_k \ln(N) + c_k, & \text{for OBC },\\
    c_k, & \text{for PBC},
  \end{cases}
  \label{eq:f3}
\end{equation}
where $a_k$ and $c_k$ are constants.
This behavior of FAF  closely resembles the phenomenology of the participation entropy in ground states of many-body systems~\cite{Luitz14}. In particular, logarithmically scaling subleading terms appear in the system-size dependence of participation entropy when PBC are replaced by OBC~\cite{Stephan09, Stephan11,Zaletel11, Luitz14}. 
These logarithmic corrections in the participation entropy can be traced back to the influence of boundary condition changing operators in the conformal field theory describing the critical point~\cite{Cardy89, Affleck97, Cardy08boundaryCFT}. 
This connection motivates the following section, where we attempt to understand the universal aspects of the phenomenology of FAF in ground states, Eq.~\eqref{eq:f1}-Eq.~\eqref{eq:f3}.

\subsection{Fermionic antiflatness and critical correlations}
\label{subsec:critical}
A broader understanding of FAF in many-body ground states can be achieved by noting that $\mathcal{F}_k$ is a combination of correlation functions. Specifically, Eq.~\eqref{eq:faf} for $k = 1$\footnote{For simplicity, we set $k = 1$ in the following discussion. Our numerical results indicate that analogous reasoning applies for any $k > 1$.} can be rewritten as
\begin{equation}
\mathcal{F}_1(\ket{\Psi}) = N - \frac{1}{2}\sum_i \sum_{j \neq i} |\braket{\Psi  | \gamma_i \gamma_j |\Psi}|^2,
    \label{eq:fafCOR}
\end{equation}
implying that the FAF in ground state $\ket{\Psi_0}$ can be traced back to the behavior of the two-point Majorana operator correlation functions $\braket{\gamma_i \gamma_{j}} \equiv \braket{\Psi_0  | \gamma_i \gamma_{j} |\Psi_0}$. This connection provides deeper insight into the phenomenology of fermionic magic resources in many-body ground states.

Away from the critical point, the ground state is characterized by a finite correlation length $\xi$, which governs the exponential decay of the correlation function with the distance $r$
\begin{equation}
    \braket{\gamma_i \gamma_{i+r} } \propto e^{-r/\xi}\;.
    \label{eq:exp}
\end{equation}
As the critical point is approached, the correlation length diverges as $\xi\propto| h_z-h_c|^{-\nu}$, and the correlation functions decay according to a power-law. In particular, in the 1+1-dimensional Ising conformal field theory, the Majorana field has scaling dimension $\Delta=1/2$, which leads to  $\braket{\gamma_i \gamma_{i+r} } =c_0/r$ at criticality~\cite{Francesco87}, where $c_0$ is a constant. 
When TFIM is perturbed by the non-Gaussian terms $\lambda \sum_m X_m X_{m+2}$, which conserve the fermionic parity and, for $\lambda \leq 1/2$, constitute an irrelevant perturbation to the critical point, the correlation function is modified to
\begin{equation}
    \braket{\gamma_i \gamma_{i+r} }= \frac{c_0}{r}\left( 1+\frac{A}{r^{\Delta_{\mathrm{irr} } -1}}\right),
    \label{eq:pow}
\end{equation}
where $A$ is a constant, and $\Delta_{\mathrm{irr}}=2$ for the considered perturbation~\cite{Mussardo10}. 

\subsubsection{Periodic boundary conditions}
We begin our analysis with the ANNNI model under PBC. In this case, the translational invariance of $H_{\mathrm{ANNNI}}$, together with the fixed fermionic parity condition $\mathcal{P} \ket{\Psi_0}= \ket{\Psi_0}$, imply that $\braket{\gamma_i \gamma_{i+r}}$ is independent of $i$\footnote{This property of $\braket{\gamma_i \gamma_{j}}$ follows from the translational invariance of $\ket{\Psi_0}$ under PBC and holds only if $\ket{\Psi_0}$ has a well-defined value of $\mathcal{P}$. In that case, the strings of $Z_i$ operators from the Jordan-Wigner transformation,~Eq.~\eqref{eq:JW}, simplify.}. Consequently, Eq.~\eqref{eq:fafCOR} reduces to
 \begin{equation}
    \mathcal{F}_1(\ket{\Psi_0}) = N \left(1 - \frac{1}{2}\sum_{r=2}^{2N} |\braket{ \gamma_1 \gamma_{r}}|^2 - \frac{1}{2}\sum_{r=1, r\neq2}^{2N} |\braket{ \gamma_2 \gamma_{r}}|^2\right).
    \label{eq:corrTRANS}
\end{equation}
Away from the critical point, when $\xi \ll N$, the exponential decay of the correlation function, Eq.~\eqref{eq:exp}, implies that the expression in parentheses in Eq.~\eqref{eq:corrTRANS} becomes independent of system size, up to corrections decaying exponentially with $N$. This leads to $\mathcal{F}_1=D_1N$, with $D_1=\mathrm{const}$ and a vanishing subleading term $f_1=0$, consistent with our numerical results summarized in~Eq.~\eqref{eq:f2}. 
Moreover, the perturbative argument presented in Eqs.~\eqref{eq:pert}–\eqref{eq:pert3} shows that the first-order contribution in $\lambda$ to FAF vanishes when $\lambda = 0$ corresponds to a free-fermionic model, which is the case for $H_{\mathrm{ANNNI}}$.

In the vicinity of the critical point, where the correlation length becomes comparable to the system size, $\xi \approx N$, the expression in parentheses in Eq.~\eqref{eq:corrTRANS} acquires a non-trivial dependence on $N$. This results in a non-vanishing subleading term $f_k$ near the transition. 
This behavior is confined to the critical region, i.e., when $\xi \approx N$, which occurs within a magnetic field window $|h_z - h_c| \propto N^{-1/\nu}$ that shrinks as the system size increases.

At the critical point $h_z = h_c$, the power-law behavior in Eq.~\eqref{eq:pow} implies that the leading contributions to the sum in Eq.~\eqref{eq:fafCOR} for $\mathcal{F}_1$ involve a term proportional to $\sum_{r=1}^{2N}r^{-2}$ and a term $\propto \sum_{r=1}^{2N}r^{-{\Delta_{\mathrm{irr}}-1}}$. The former corresponds to the free-fermionic limit $\lambda=0$, where $\mathcal{F}_1=0$. This implies that the sum $\sum_{r=1}^{2N}r^{-2}$, 
is equal to 1 (see Appendix~\ref{app:IsingCrit} for further discussion).  
Therefore, the only non-trivial contribution to FAF at the critical point arises from the latter term $\mathcal{F}_{1} \propto N \sum_{r=1}^{2N}r^{-{\Delta_{\mathrm{irr}}-1}}$. For the case of $\Delta_{\mathrm{irr}}=2$, this gives $\mathcal{F}_{1} \propto N(\mathrm{const}-1/N^2)$, implying that $\mathcal{F}_1=D_1 N+O(1/N)$ at the paramagnetic-ferromagnetic phase transition in the ANNNI model with PBC. This result is consistent with data shown in Fig.\ref{fig:aniPBC}. The subleading term at the critical point $h_z=h_c$ exhibits a clean residual drift, $f_k \propto 1/N$, as shown in the inset of Fig.\ref{fig:aniPBC}(b).

The above reasoning shows how the finite size scaling ansatz, $f_k = g[(h_z-h_c)N^{1/\nu}]$, observed numerically in Fig.~\ref{fig:aniPBC}(b), emerges from the universal features of the two-point Majorana fermion correlation function decay in the ground state of ANNNI model with PBC. 
The appearance of the logarithmically diverging subleading term $f_k$ for OBC, as seen in Eq.\eqref{eq:f3}, can be understood within the conformal field theory description of the critical point as the effect of a boundary changing operator~\cite{Cardy89, Affleck97, Cardy08boundaryCFT}.

\subsubsection{Open boundary conditions}
For OBC, the system is no longer translationally invariant, and the correlation function $\braket{\gamma_i \gamma_{i+r}}$ depends both on $r$ and $i$.
To understand the behavior of FAF in ground states without translational symmetry, we divide the sites $i, j$ in Eq.~\eqref{eq:fafCOR} into bulk and boundary regions. When both $i,j$ belong to the bulk, the two-point correlation functions are approximately translationally invariant, and $\braket{\gamma_i\gamma_j}$ depends only on the distance $r=|i-j|$, similarly to the PBC case; see Eq.~\eqref{eq:exp} and Eq.~\eqref{eq:pow}. 
In contrast, if one of the sites belongs to the boundary, $i \in \partial V$, the correlation function has contributions stemming from the enhanced and oscillatory correlations close to the boundary. 
Thus, we can express the sum in Eq.~\eqref{eq:fafCOR} as $\sum_{j \neq i} |\braket{ \gamma_i \gamma_j}|^2 = N s_{\mathrm{bulk}} + \delta s_i$, where $s_{\mathrm{bulk}}$ is a constant and $\delta s_i$ decays to zero as site $i$ moves away from the boundary.
This enables us to write the FAF for the ANNNI model with OBC as
\begin{equation}
    \mathcal{F}_1(\ket{\Psi_0}) = N\left(1 - \frac{1}{2}s_{\mathrm{bulk}}\right) + \sum_{i=1}^{2N} \delta s_i.
    \label{eq:corrOBC}
\end{equation}
In passing, we note that both $s_{\mathrm{bulk}}$ and $\delta s_i$ are non-zero already for the free-fermionic case, $\lambda=0$, in which their net contribution yields $\mathcal{F}_1(\ket{\Psi_0})=0$. For non-zero $\lambda$, both $s_{\mathrm{bulk}}$ and $\delta s_i$ are modified, leading to a non-vanishing value of $\mathcal{F}_1(\ket{\Psi_0})$.
In particular, away from the critical point, the boundary contributions $\delta s_i$ decay exponentially with the distance from the system's edge. This yields a constant subleading term $f_k$ in Eq.~\eqref{eq:f2}. At the critical point, the correlation functions are enhanced near the edges of the system, resulting in $\delta s_i \propto \max\{1/i, 1/(N-i)\}$. Upon the summation over $i$, this behavior leads to the logarithmic divergence of $f_k$ at the paramagnetic-ferromagnetic phase transition in the ANNNI model with OBC, as shown in Eq.~\eqref{eq:f3}. 

\subsubsection{Single impurity term}
The presence of a single impurity term in the TFIM, as in the impurity model $H_{\mathrm{imp}}$, cf. Eq.\eqref{eq:imp1}, modifies the correlation function $\braket{\gamma_i \gamma_{i+r}}$ in a manner analogous to the presence of a boundary, both in the gapped phases and at the critical point~\cite{Affleck09, Roy22}.
Away from the critical point, the effect of the impurity term $\lambda X_{\ell} X_{\ell+2}$ on $\braket{\gamma_i \gamma_{j}}$ decays exponentially with the distance of sites $i,j$ from the impurity. Therefore, the introduction of a non-zero $\lambda$ leads, via Eq.~\eqref{eq:fafCOR}, to a constant contribution $\mathcal{F}_1=c_k$, consistent with Eq.~\eqref{eq:f1}. 
This effect is enhanced near the critical point. However, the perturbation described by Eq.~\eqref{eq:pow}, with the amplitude $A$ now decreasing with the distance from the impurity, still results only in a system-size-independent contribution to $\mathcal{F}_1$.
The derivative of $\mathcal{F}_1$ with respect to $h_z$ depends, through Eq.~\eqref{eq:fafCOR}, not only on the correlation function $\braket{\gamma_i \gamma_{j}}$, but also on its derivative $\frac{d}{d h_z} \braket{\gamma_i \gamma_{j}}$.
The critical behavior of this derivative can be quantitatively understood within conformal perturbation theory~\cite{Ginsparg88, Amoretti17}, and it involves three-point correlation functions of the Ising conformal field theory. These may decay more slowly with distance from the impurity, consistent with our numerical observation of a logarithmic divergence in $|\mathcal{F}_k'|$ at the critical point of the impurity model.

\subsubsection{Conclusion}
We
demonstrated how the universal features of two-point Majorana fermion correlation functions provide insights into the FAF and fermionic magic resources of ground states in many-body systems. 
We have observed that FAF exhibits distinct behavior across different phases of matter. 
Moreover, the enhancement of correlations in the vicinity of critical points leads to an increase in FAF, which can be interpreted as an increase in the complexity of the ground state from the perspective of fermionic Gaussian states. However, this is not the only noteworthy aspect of the FAF in ground states. 
In the following, we explore regions of the phase diagram where the ground state of the interacting many-body system becomes particularly simple from the point of view of fermionic magic resources.

\begin{figure}
    \centering
    \includegraphics[width=1\linewidth]{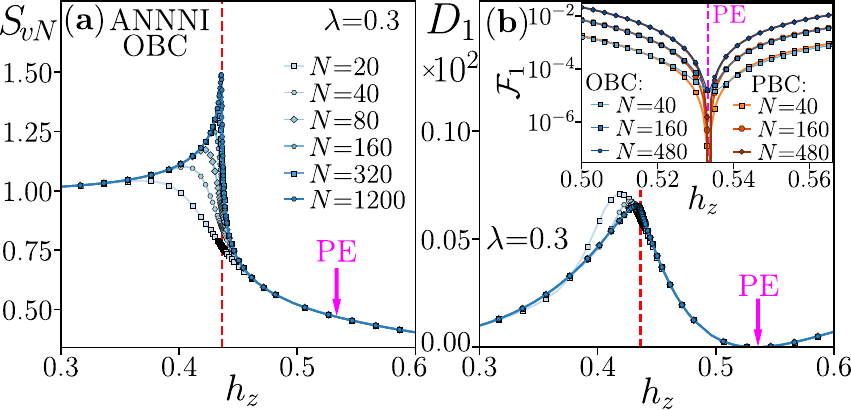}
    \caption{Peschel-Emery point in the ANNNI model. (a) Entanglement entropy $S_{vN}$ diverges logarithmically with $N$ at the critical point $h_z = h_c$ (denoted by the red dashed line), and is otherwise a smooth function of $h_z$, vanishing only in the limit $h_z \to \infty$. (b) The leading term $D_1$ in the FAF scaling is enhanced near the critical point and vanishes at the Peschel-Emery point~\cite{Peschel81}, corresponding to $h_z = h_{\mathrm{PE}}$.
    The inset shows that $\mathcal{F}_1$ vanishes at $h_z = h_{\mathrm{PE}}$ for the ANNNI model with PBC, while $\mathcal{F}_1(h_{\mathrm{PE}})$ is a system size-independent constant for OBC.
    }
    \label{fig:aniPE}
\end{figure}

\subsection{The Peschel-Emery line in ANNNI model}
The entanglement entropy in the ANNNI model diverges logarithmically at the critical point $h_z = h_c$ as a function of magnetic field strength, see Eq.\eqref{eq:entlog} and Fig.\ref{fig:aniPE}.
Away from the critical point, the entanglement entropy monotonically approaches the limiting values: $S_{vN} \to 0$ for $h_z\to \infty$, corresponding to the $z$-polarized ground state $\ket{\Psi_0}=\ket{11\dots1}$, and $S_{vN} \to 1$ for $h_z \to 0$, corresponding to the state $\ket{\Psi_0} = (\ket{++\ldots+}+\ket{–\ldots-})/\sqrt{2}$.
In both limits, $h_z\to 0,\infty$, the ground state is a fermionic Gaussian state and the FAF vanishes.
In contrast to the entanglement entropy, the FAF does not interpolate monotonically between its maximum near the critical point and \( \mathcal{F}_k \to 0 \) for \( h_z \to \infty \). 
Instead, the FAF exhibits a minimum at a specific value of the magnetic field, $h_{\mathrm{PE}}$.
As shown in Fig.~\ref{fig:aniPE}, for OBC, the minimum of $\mathcal{F}_k$ corresponds to a system-size independent constant. For PBC, the FAF vanishes, $\mathcal{F}_k(h_{\mathrm{PE}})=0$. The latter implies that the ground state $\ket{\Psi_0}$ for $h_z = h_{\mathrm{PE}} $ is a fermionic Gaussian state even though the Hamiltonian $H_{\mathrm{ANNNI}}$ is an interacting quantum many-body system. The small, constant value of $ \mathcal{F}_k(h_{\mathrm{PE}})$ for OBC indicates that the ground state is close to a fermionic Gaussian state.

The described behavior is a specific property of the ANNNI model. A line in parameter space, defined by $h_z=\frac{1}{4\lambda}-\lambda$, along which the ground state of the ANNNI model is exactly solvable, was discovered in~\cite{Peschel81}.
This line is known as the Peschel-Emery (PE) line. 
To better understand the fermionic magic resources of the ANNNI model, we briefly outline the derivation from~\cite{Peschel81}.
The first step is to perform a duality transformation. In the language of Majorana fermions, this corresponds to a fermionic Gaussian unitary $U_G$ satisfying $U^{\dagger}G \gamma_m U_G = \gamma_{m-1}$ for $m\in[2,2N]$, and $U^{\dagger}G \gamma_1 U_G = \gamma_{2N}$. 
Under this transformation, the ANNNI Hamiltonian with PBC becomes $U^{\dagger}_G H_{\mathrm{ANNNI}} U_G  = H'$ with $H' = \sum_m h_m$, where
\begin{eqnarray}
h_m = -i h_z \gamma_{2m }\gamma_{2m+1} - &\frac{i}{2}  \gamma_{2m-1} \gamma_{2m}- \frac{i}{2}  \gamma_{2m+1} \gamma_{2m+2} 
\nonumber \\ 
-&\lambda \gamma_{2m-1}\gamma_{2m} \gamma_{2m+1} \gamma_{2m+2}.
 \end{eqnarray}
Each of the operators $h_m$ acts only on neighboring qubits $m$ and $m+1$. However, these operators do not commute, $[h_m,h_n]\neq0$, for $m=n\pm1$. 
Analytic progress is possible when $h_z=\frac{1}{4\lambda}-\lambda$.
In that case, the ground states of the local two-qubit operator $h_m$ in the positive and the negative parity sectors are degenerate~\cite{Katsura15}.
This degeneracy is equivalent to the "no frustration" criterion of~\cite{Peschel81, Kurmann82, Muller85}, and it enables the ground state of $H'$\footnote{A straightforward calculation shows that $\ket{\Psi_0}$ is an eigenstate of $H'$ at the PE line, $h_z=\frac{1}{4\lambda}-\lambda$. A proof that $\ket{\Psi_0}$ is the ground state of the model can be found in~\cite{Peschel81, Katsura15}, while the impact of boundary conditions is discussed in~\cite{Kawabata17}.} in the positive parity sector to be written as
\begin{equation}
    \ket{\Psi_0}=\sqrt{a_N /2 }(\ket{\theta}^{\otimes N}+\ket{-\theta}^{\otimes N} ),
    \label{eq:PEgs}
\end{equation}
where $a_N = (1+\cos^N(\theta))^{-1}$ is a normalization constant, 
 $\ket{\theta}=(\cos(\theta/2)\ket{0} - \sin (\theta/2)\ket{1})$, and $\cos(\theta)=-[2(h_z+\lambda)]^{-1}$.
The two-point Majorana correlation functions $\braket{\Psi_0|\gamma_i \gamma_j|\Psi_0} = \braket{\gamma_i \gamma_j} $ decay exponentially with distance and can be calculated analytically from Eq.~\eqref{eq:PEgs} as
\begin{eqnarray}
\langle i \gamma_1 \gamma_2 \rangle &=& a_N \left( \cos(\theta) + \cos^{N-1}(\theta) \right) \nonumber \\
\langle i \gamma_1 \gamma_{2k+2} \rangle &=& a_N \sin^2(\theta)\, \cos^{N-k-1}(\theta)\nonumber  \\
\langle i \gamma_2 \gamma_{2k+1} \rangle &=& a_N \sin^2(\theta)\, \cos^{k-1}(\theta).
\end{eqnarray}
While the correlation functions are nontrivial, their analytic form enables calculation of the FAF, which can be readily shown to vanish, $\mathcal{F}_k=0$. 
This indicates that $\ket{\Psi_0}$ is a fermionic Gaussian state. 
Since $H'$ and $H_{\mathrm{ANNNI}}$ are related by a fermionic Gaussian unitary, the ground state of the ANNNI model with PBC at the PE line is a fermionic Gaussian state. Interestingly, the frustration-free criterion only allows the determination of the ground state of $H’$.
Even at the PE line, the Hamiltonian $H'$ remains an interacting system, and its excited eigenstates are generically \textit{not} fermionic-Gaussian states, exhibiting nonvanishing FAF\footnote{We note that some of the excited states of $H_{\mathrm{ANNNI}}$ at the PE line can be constructed analytically~\cite{Mahyaeh18}.}.

For OBC, the transformed ANNNI Hamiltonian $U^{\dagger}_G H_{\mathrm{ANNNI}} U_G$ differs from $H'=\sum_m h_m$ by $O(1)$ terms associated with the system boundaries. 
Although the ground state at the PE line for OBC cannot be written in the form Eq.~\eqref{eq:PEgs}, the presence of a finite correlation length ensures that boundary effects contribute only a small, nonzero value to $\mathcal{F}_k$. This contribution remains independent of the system size.

The above discussion explains why fermionic magic resources are particularly limited in the ground state of the ANNNI model at the PE line, as observed in Fig.~\ref{fig:aniPE}. While the PE line is a well-known result in equilibrium many-body physics, our findings highlight the potential of $\mathcal{F}_k$ to reveal hidden structure in quantum states—namely, the simplicity of their description as fermionic Gaussian states.

\section{Fermionic antiflatness in out-of-equilibrium systems}
\label{sec:outofeq}
While fermionic magic resources encode information about quantum phases of matter and exhibit singular behavior at phase transitions, the FAF in many-body ground states remains limited -- much smaller than in the typical states described in Sec.~\ref{subsec:typical}.
In this section, we consider fermionic magic resources in interacting ergodic many-body systems in out-of-equilibrium settings, where the system's behavior depends crucially on the properties of all its eigenstates.

In Sec.~\ref{sec:equilibrium}, we showed that all the eigenstates of the TFIM, Eq.~\eqref{eq:Ising0}, defined by $H_+ \ket{\Psi_n} = E_n\ket{\Psi_n}$, are fermionic Gaussian states of the form $\ket{E_n} = U_G\ket{\mathbf{n}}$. Here, $U_G$ is an appropriate fermionic Gaussian unitary, which can be found with computational cost scaling polynomially in the system size $N$.
When a system governed by a quadratic Hamiltonian Eq.~\eqref{eq:quad} is initialized in a fermionic Gaussian state $\ket{\Psi_{\mathrm{in}}}$, the time-evolved state $\ket{\Psi(t)}=e^{-iH_+t} \ket{\Psi_{\mathrm{in}}}$ remains fermionic Gaussian. This follows immediately from the fact that the time-evolution operator $e^{-iH_+t}$ for a quadratic Hamiltonian $H_+$ is itself a fermionic Gaussian unitary~Eq.~\eqref{eq:fgu}. Consequently, $\mathcal{F}_k(\ket{\Psi(t)})=0$ at all times $t$, and the time-evolved state is fully determined by its covariance matrix, which, through Eq.~\eqref{eq:cov}, transforms during time evolution as $M\mapsto G M G^T$.

This computational simplicity, reflected in the vanishing fermionic magic resources, enables classical simulation of free-fermionic systems at large scales and has led to insights into a variety of non-equilibrium many-body phenomena, including quench dynamics~\cite{Rossini09, Rossini10, calabrese2006timedependenceof,Ziraldo13, Calabrese11, Sierant22chal}, the Kibble-Zurek mechanism~\cite{Zurek05, Dziarmaga05, Caneva07, Bialonczyk20}, dynamical quantum phase transitions \cite{Heyl13, Heyl18}, the dynamics of periodically driven systems \cite{Russomanno12, Lazarides14}, or entanglement transitions~\cite{Cao19, Turkeshi21clicks,Alberton21, Piccitto22, Tirrito23full}. Nevertheless, the time evolution of free-fermionic systems is not generic~\cite{Lydzba23, Lydzba21}. In particular, features such as noisy entanglement growth~\cite{nahum2017quantum} and the inability to support volume-law entanglement~\cite{Cao19} are specific to evolution under fermionic Gaussian unitaries and do not represent generic many-body dynamics.

Generic interacting quantum many-body systems, when initialized in an out-of-equilibrium state, are expected to thermalize. During the thermalization process, local information encoded in the initial state is dispersed into entanglement and quantum correlations that spread throughout the entire system~\cite{Rigol_2008, Dalessio16}. 
Systems that thermalize in accordance with the Eigenstate Thermalization Hypothesis (ETH)~\cite{Deutsch91,Srednicki94,Feingold86,Srednicki99,foini2019eigenstatethermalizationand,foini2019eigenstatethermalizationhypothesis,dymarsky2019newcharacteristicof,pappalardi2022eigenstatethermalization,dymarsky2022boundoneigenstate,wang2023emergence,wang2022eigenstatethermalizationhypothesis,Pappalardi25full} are hereafter termed \emph{ergodic}.
The ETH provides the following ansatz for matrix elements $A_{mn}= \braket{\Psi_n|A|\Psi_m}$ of a few-body observable $A$ in the eigenbasis $\{ \ket{\Psi_n}\}$ of an ergodic system: 
\begin{equation}
 A_{mn} = \mathcal{A}(\bar E) \delta_{mn} + 
 \rho(\bar{E})^{-\frac{1}{2}}
 f_{\mathcal A} (\bar E, \omega_{mn}) R_{mn},
 \label{eq:ETH1}
\end{equation}
where  $\bar E = (E_m+E_n)/2$ is the mean energy, $\omega_{mn} = E_m-E_n$ is the energy difference, $\rho(\bar{E})$ is the density of states~\cite{Burke23} and $R_{mn}$ is a random variable with zero mean and unit variance, while $\mathcal{A}(E)$ and $f_{\mathcal A} (E, \omega)$ are smooth functions of their arguments. 

In the following, we compute highly excited eigenstates $\ket{\Psi_n}$ and time-evolved states $\ket{\Psi(t)}$ of interacting many-body systems as superpositions of $2^{N-1}$ computational basis states\footnote{We consider systems that conserve fermionic parity $\mathcal{P}$ and focus on the $\mathcal{P} = 1$ sector.}. This approach incurs a computational cost that scales exponentially with system size $N$. We perform full exact diagonalization of Hamiltonian matrices up to $N=17$, and extract mid-spectrum eigenstates for $18 \leq N \leq 22$ using the POLFED algorithm~\cite{Sierant20polfed}. Our goal is to understand the fermionic magic resources of highly excited eigenstates in ergodic systems and to examine the implications of the ETH ansatz Eq.~\eqref{eq:ETH1} for FAF. We then proceed to study the growth of FAF under ergodic many-body dynamics, using the results for typical quantum states from Sec.~\ref{subsec:typical} and for random quantum circuits from Sec.~\ref{subsec:circ} as reference points.

\subsection{Fermionic magic resources across the many-body spectrum}

\begin{figure}
    \centering
    \includegraphics[width=1\linewidth]{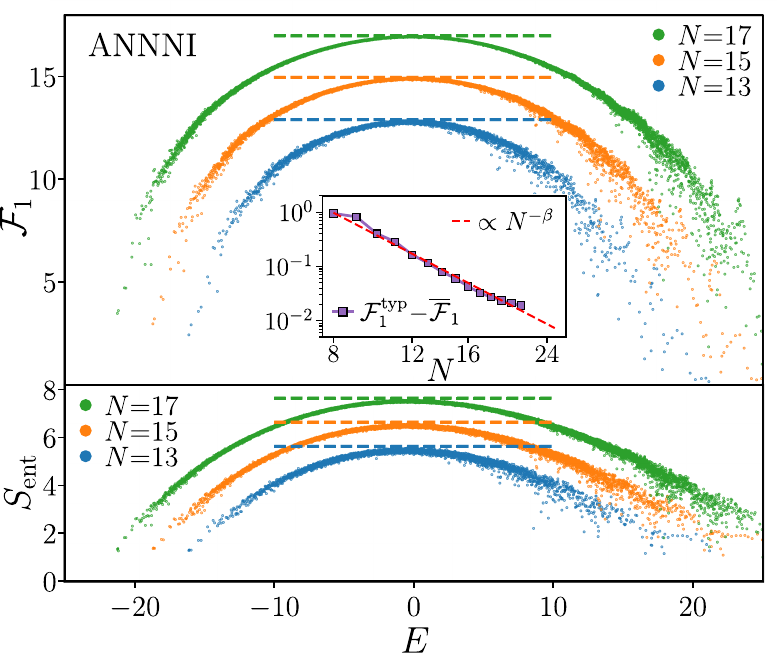}
    \caption{Fermionic antiflatness $\mathcal{F}_1$ (top panel) as function of energy $E$ of eigenstates $\ket{\Psi_n}$ of the ANNNI model, Eq.~\eqref{eq:ani}. The system size is $N$ and $\lambda=1$. The FAF is compared with eigenstate entanglement entropy $S_{\mathrm{ent}}$, shown in the bottom panel. 
    In the middle of the spectrum, the FAF, similarly to $S_{\mathrm{ent}}$, approaches the typical state result $\mathcal{F}^{\mathrm{typ}}_1$, cf. Eq.~\eqref{eq:typfaf}, indicated by dashed lines. 
    The inset shows the system size decay of the difference between the FAF of Haar-random states and the average FAF in the middle of the spectrum, $\mathcal{F}^{\mathrm{typ}}_1 - \overline{\mathcal{F}}_1 \propto N^{-\beta}$, where $\beta=4.3(2)$.
    }
    \label{fig:aniED}
\end{figure}

\begin{figure}
    \centering
    \includegraphics[width=1\linewidth]{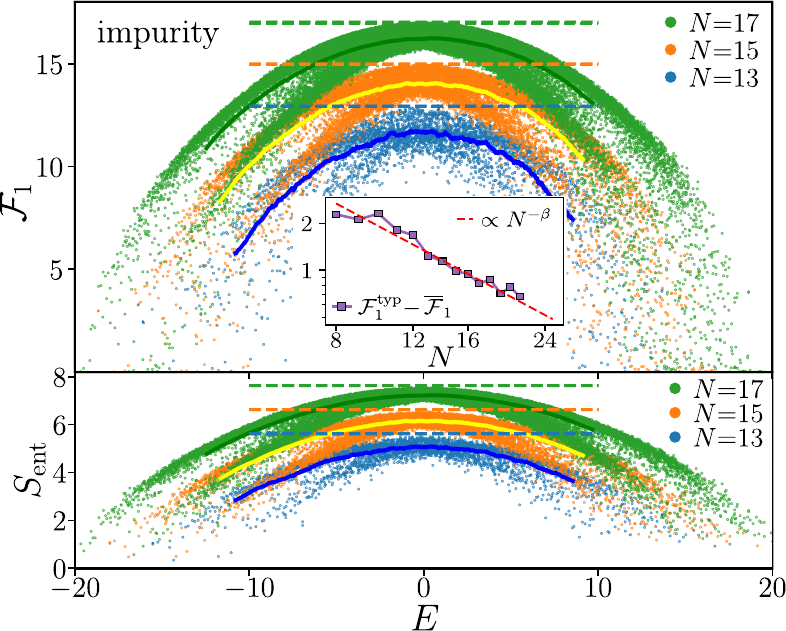}
    \caption{Fermionic antiflatness $\mathcal{F}_1$ (top panel) across the spectrum of 
    the impurity model, cf. Eq.~\eqref{eq:imp1}, with system size $N$ and $\lambda=1$, compared with eigenstate entanglement entropy $S_{\mathrm{ent}}$ (bottom panel). The FAF in the middle of the spectrum, similarly to $S_{\mathrm{ent}}$, approaches the typical state results given by $\mathcal{F}^{\mathrm{typ}}_1$, Eq.~\eqref{eq:typfaf}, denoted by the dashed lines, and its spread around the running average (darker lines) decreases with $N$. The inset shows the decay of the difference $\mathcal{F}^{\mathrm{typ}}_1 - \overline{\mathcal{F}}_1 \propto N^{-\beta}$ with $\beta=1.5(1)$.
    }
    \label{fig:impuED}
\end{figure}

The ETH ansatz, Eq.~\eqref{eq:ETH1}, implies that the highly excited eigenstates of ergodic systems near a fixed energy $E$ share common, narrowly distributed, typical properties~\cite{reimann2007typicality,gemmer2009dynamical}, reflected, for instance, in the expectation values $\mathcal{A}(E)$ of few body observables at the energy $E$. 
As a result, highly excited eigenstates of ergodic many-body systems are, to a large extent, featureless~\cite{backer2019multifractal, Haque22}, and resemble the eigenstates of random matrices. Indeed, the distribution of expectation values $x=\braket{\Psi_n|P|\Psi_n}$ of Pauli strings in eigenstates $\ket{\Psi_n}$ in mid-spectrum eigenstates of ergodic systems is Gaussian, with the same variance as for the Haar-random states~\eqref{eq:pheno}, up to tails whose weight is exponentially suppressed with system size $N$~\cite{Turkeshi25spectrum}. This suggests that $\mathcal{F}_k$ for middle-spectrum eigenstates $\ket{\Psi_n}$ in ergodic systems is close to $\mathcal{F}^{\mathrm{typ}}_k$ given by~\eqref{eq:leadingtyp}.

We begin our analysis with the ANNNI model, setting $\lambda=1$ in Eq.~\eqref{eq:ani}. Throughout this and the next section, we use OBC; the results for PBC are fully analogous.
The FAF of eigenstates $\ket{\Psi_n}$ of the ANNNI model as a function of their energy $E$ is shown in Fig.~\ref{fig:aniED}~(top). The FAF is suppressed at the edges of the spectrum, near the ground state and the highest excited state, and increases toward the center of the spectrum, where the density of states is maximal. As a function of the energy, $\mathcal{F}_k$ displays the characteristic shape of an inverted parabola, similar to the behavior of the entanglement entropy $S_{\mathrm{ent}}$, which is typically observed in ergodic many-body systems~\cite{Santos10, Vidmar17, Bianchi22volume} and shown in the bottom panel of Fig.~\ref{fig:aniED} for the ANNNI model. 

We observe that fluctuations of $\mathcal{F}_k$ between neighboring eigenstates decrease toward the middle of the spectrum and diminish with increasing system size. By averaging $\mathcal{F}_k$ over $\min\{1000, 2^N/20\}$ eigenstates $\ket{\Psi_n}$ with energies closest to the center of the spectrum, $E=0$, we obtain the mean FAF, $\overline {\mathcal{F}}_1$. 
This average approaches the FAF of Haar-random states $\mathcal{F}^{\mathrm{typ}}_k$ as $\mathcal{F}^{\mathrm{typ}}_1 - \overline{\mathcal{F}}_1 \propto N^{-\beta}$, cf. the inset of Fig.~\ref{fig:aniED}. This indicates that the FAF in middle-spectrum eigenstates of ergodic systems becomes nearly maximal in the large $N$ limit, scaling as
\begin{equation}
\label{eq:FAFerg}
\mathcal{F}_k = D_k N + O(N^{-\beta}),
\end{equation}
with the leading coefficient $D_k=1$.

The results for the FAF of highly excited eigenstates of the ANNNI model, particularly their agreement with typical-state behavior, can be expected to hold for any generic choice of parameters of the system, as well as for other ergodic systems that satisfy the ETH~\eqref{eq:ETH1}. However, the rate of convergence to the large $N$ limit may depend on the specific choice of parameters. 
For the ANNNI model with $\lambda = 1$, the spread of FAF among middle-spectrum eigenstates is already strongly suppressed at $N=13$. Other indicators of ergodicity, such as the features of $S_{\mathrm{ent}}$, also exhibit good convergence to the large-$N$ behavior at this system size.
In contrast, the FAF in highly excited eigenstates of the impurity model, Eq.~\eqref{eq:imp1}, with $\lambda =1$, shows significantly larger deviations from the typical-state results at the system sizes considered.

In Fig.~\ref{fig:impuED}, we observe a significant spread in $\mathcal{F}_k$ between neighboring eigenstates, even in the middle of the spectrum. The results for the entanglement entropy $S_{\mathrm{ent}}$ parallel the findings for FAF, underscoring the deviation of the impurity model from the behavior characteristic of fully ergodic systems. 
This behavior is expected, since the impurity model includes a single interaction term beyond the free-fermion (Gaussian) structure.
In contrast to the ground state, where the FAF converges to a system size independent constant, Eq.~\eqref{eq:f1}, the single impurity induces an extensive scaling of FAF, $\mathcal{F}_k\propto N$, in the highly excited eigenstates. 
This is consistent with previous studies of impurity models~\cite{Santos04,Barisic09,  Metavitsiadis10, Santos11, Torres14, Brenes18,Brenes2020impu,  Krause21, Brighi22, Sierant23slow}, which indicate that even a single impurity term can be sufficient to restore ergodicity in sufficiently large systems. 
The difference between the average FAF in mid-spectrum eigenstates, $\overline{\mathcal{F}}_k$, and the typical-state value $\mathcal{F}^{\mathrm{typ}}_k$ is suppressed with system size as a power-law $N^{-\beta}$, as shown in the inset in Fig.~\ref{fig:impuED}. This indicates that the FAF in eigenstates of sufficiently large impurity models follows the scaling by Eq.~\eqref{eq:FAFerg}.

The extensive scaling in Eq.~\eqref{eq:FAFerg}, with coefficient $D_k=1$, indicates an abundance of fermionic magic resources in highly excited eigenstates of ergodic many-body systems, comparable to that of featureless Haar-random states. In particular, even a single non-Gaussian impurity term may be sufficient to introduce nearly maximal FAF in the middle spectrum eigenstates. In the following, we examine how these findings impact the time evolution of FAF in ergodic systems.

\begin{figure*}
    \centering
    \includegraphics[width=1.\linewidth]{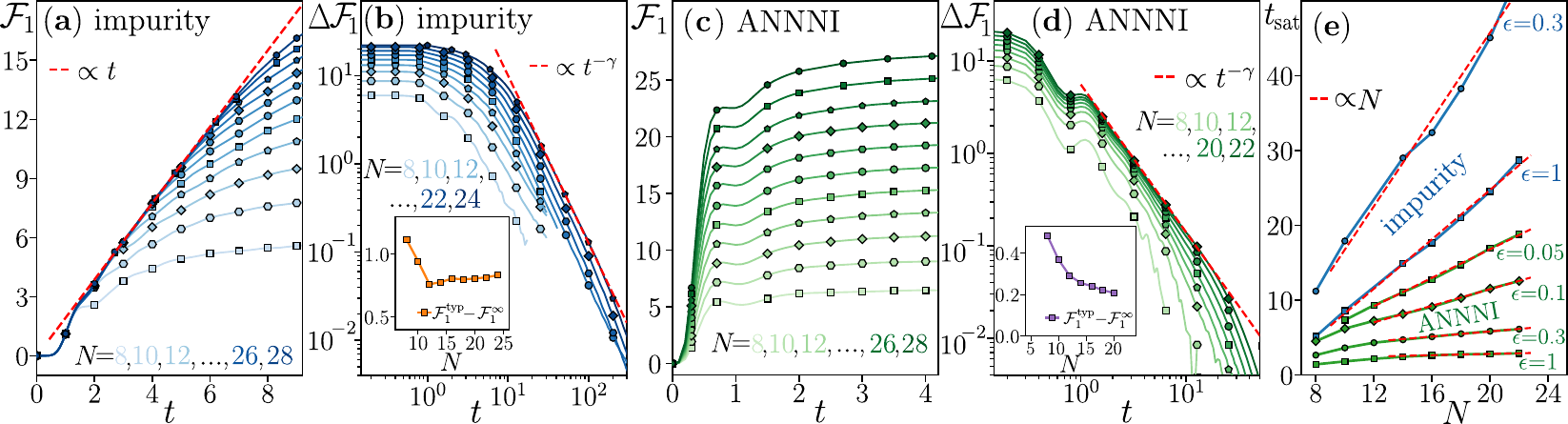}
    \caption{    
    Fermionic antiflatness $\mathcal{F}_k$ in ergodic many-body dynamics. (a) Ballistic growth of FAF, $\mathcal{F}_1\propto t$, in the impurity model, Eq.~\eqref{eq:imp1} as a function of time $t$ for fixed system size $N$ and $\lambda=1$. (b) The difference $\Delta \mathcal{F}_1=\mathcal{F}^{\infty}_1-\mathcal{F}_1(t)$ in the impurity model is well fitted by the power-law dependence $\Delta\mathcal{F}_k \propto t^{-\gamma}$ with $\gamma=1.93(10)$. Inset: the difference between $ \mathcal{F}^{\mathrm{typ}}_1$ and $\mathcal{F}^{\infty}_1$ as a function of $N$. (c) Rapid growth of FAF in the ANNNI model, cf. Eq.~\eqref{eq:ani} with $\lambda=1$. (d) Power-law approach of FAF to its saturation value in the ANNNI model, $\Delta \mathcal{F}_1 \propto t^{-\gamma} $ with $\gamma = 1.6(2)$. Inset: the difference $\mathcal{F}^{\mathrm{typ}}_1 - \mathcal{F}^{\infty}_1$ as a function of $N$. (e) Saturation time $t_{\mathrm{sat}}$, defined via $\Delta\mathcal{F}(t_{\mathrm{sat}})=\epsilon$ in both the ANNNI and impurity models, is well fitted by $t_{\mathrm{sat}} \propto N$.
    }
    \label{fig:tevol}
\end{figure*}

\subsection{Fermionic antiflatness in ergodic many-body dynamics}
Initialized in an out-of-equilibrium state, ergodic many-body systems undergo thermalization. During this process, quantum correlations between distant regions of the system are generated. In ergodic systems with local Hamiltonians, this is typically manifested by a ballistic growth of entanglement entropy~\cite{Lauchli08, Kim13ballistic}, and a fast saturation of SRE~\cite{turkeshi2024magic} and PE~\cite{turkeshi2024hilbert} to their long-time values. 
In the following, we employ the method of~\cite{Tal‐Ezer84}, which relies on expanding the time-evolution operator $e^{-i Ht}$ into Chebyshev polynomials of the Hamiltonian $H$ -- an approach that is computationally efficient due to the sparsity of $H$. We use this method to study the time evolution of FAF, $\mathcal{F}_k$, in the impurity and ANNNI models~\cite{Robertson23}, using the same parameters as in the preceding section.

As the initial state, we take a random state of the computational basis $\ket{\Psi_0} = \ket{\sigma}$ in the positive fermionic parity sector, $\mathcal{P}=1$. Such a state is a free-fermionic state, and, thus, initially $\mathcal{F}_k=0$. 
In the impurity model, we observe a linear growth of FAF, $\mathcal{F}_k \propto t$, followed by a crossover to a slower growth regime at times that increase with system size $N$, as shown in Fig.~\ref{fig:tevol}(a). This ballistic growth of FAF is consistent with the Lieb-Robinson bound~\cite{Lieb1972}, showing that the information about the presence of the impurity, encoded in beyond-Gaussian Majorana correlations, propagates within a linear light-cone. 
In contrast, the initial growth of FAF in the ANNNI model is much faster, as shown in Fig.~\ref{fig:tevol}(c), with $\mathcal{F}_k$ becoming extensive in $N$ already at times $t=O(1)$. This behavior can be intuitively understood as the combined effect of extensively many non-Gaussian impurities, each contributing a finite amount to the total FAF. 

After the initial growth, fermionic magic resources in both the impurity and ANNNI models approach their long-time saturation values, denoted by $\mathcal{F}^{\infty}_k$. These saturation values, computed numerically by averaging $\mathcal{F}_k(t)$ over the time window $t\in[1000,2000]$, are, at the accessible system sizes, close to the typical-state value given by Eq.~\eqref{eq:typfaf}, with $\mathcal{F}^{\mathrm{typ}}_k- \mathcal{F}^{\infty}_k=O(1)$, as presented in the insets in Fig.~\ref{fig:tevol}(b,d). 

To probe the saturation of FAF to its long-time value, we consider the difference $\Delta \mathcal{F}_k=\mathcal{F}^{\infty}_k- \mathcal{F}_k(t)$. As shown in Fig.~\ref{fig:tevol}(b),(d), this quantity decays as a power-law, $\Delta \mathcal{F}_1 \propto t^{-\gamma}$, with the exponent $\gamma$ depending on the model (analogous results hold for $k\geq 2$, data not shown). The combination of initial growth and power-law relaxation leads to the saturation of $\mathcal{F}_k$ up to a fixed accuracy $\epsilon$ at a saturation time $t_{\mathrm{sat}}$ that scales linearly with system size, $t_{\mathrm{sat}}\propto N$, for both the ANNNI and the impurity models, as shown in Fig.~\ref{fig:tevol}(e).

In conclusion, this section demonstrates that fermionic magic resources are rapidly generated by ergodic quantum dynamics. While the short-time behavior of $\mathcal{F}_k$ in the ANNNI model and in the random quantum circuits (cf. Fig.~\ref{fig:circuits}(a)) is similar, exhibiting extensive $\mathcal{F}_k$ already at times $t=O(1)$, the long-time behavior differs significantly between ergodic many-body systems and random circuits. This difference is particularly evident in the saturation time: for random circuits, $t_{\mathrm{sat}} \propto \ln(N)$, whereas in the ergodic many-body systems, it scales extensively, $t_{\mathrm{sat}} \propto N$. A fully analogous difference in the saturation properties of the SRE was observed between random circuits~\cite{turkeshi2024magic} and dynamics of ergodic many-body systems~\cite{Tirrito24anti}.

The discrepancy between the dynamics of Haar-random circuits and ergodic many-body systems can be traced back to the presence of globally conserved quantities, expressed as sums of local charges. Random circuits possess no conservation laws and can form approximate $k$-designs at logarithmic circuit depths $t$~\cite{Laracuente24approximate, Schuster2024random}, consistently with the logarithmic saturation time of FAF observed in Fig.~\ref{fig:circuits}(d). 
In contrast, in the presence of conserved quantities that are sums of local operators\footnote{Note that this argument does \textit{not} apply to $\mathbb{Z}_2$ parity conservation, which does not correspond to an extensive sum of local charges.} any local unitary circuit of depth less than the system size can be distinguished from a global Haar-random unitary~\cite{Haah25short}. 
The observed discrepancies in the saturation behavior of FAF indicate that this general result applies to the impurity and ANNNI models, where the dynamics conserve energy and the Hamiltonian acts as a sum of local operators. Consequently, evolution times that scale at least linearly with system size, $t\propto N$, are required for the saturation of FAF in these models.

\section{Conclusions}
\label{sec:conclusion}

\subsection{Summary}
In this work, we have presented a framework for investigating fermionic magic resources in quantum many-body systems. 
These resources quantify the complexity of many-body states relative to the class of classically simulable fermionic Gaussian states.
Quantum states that lie close to the manifold of fermionic Gaussian states exhibit limited fermionic magic resources, while interactions and beyond-Gaussian operations drive quantum states away from that manifold, increasing their fermionic non-Gaussianity. Our framework consists of two main components: a mathematical scheme for constructing efficiently computable measures of fermionic non-Gaussianity, and a set of results illustrating the behavior of fermionic magic resources in representative equilibrium and non-equilibrium many-body systems. Together, these two elements establish fermionic non-Gaussianity as a measure of quantum state complexity, complementary to entanglement and non-stabilizerness.

\textit{Fermionic commutant and measures of fermionic non-Gaussianity.} Our scheme for constructing measures of fermionic non-Gaussianity focuses on quantities that can be expressed as expectation values of operators acting on multiple copies of the investigated state. This formulation avoids costly minimization procedures and allows us to leverage the algebraic structure of the fermionic commutant. Operators from the fermionic commutant define measures of fermionic non-Gaussianity that are, by construction, invariant under fermionic Gaussian operations. Among the possible candidate measures, we identified the fermionic antiflatness (FAF) as an efficiently computable, experimentally accessible, and physically interpretable fermionic magic measure.

\textit{Fermionic antiflatness and its properties.} We demonstrated that FAF satisfies key properties of a non-Gaussianity measure, such as faithfulness—it vanishes if and only if the state is fermionic Gaussian—and (sub)additivity on product states. FAF is expressed as a sum of two-point Majorana correlation functions and can be computed efficiently using both state-vector simulations and tensor network methods. 
Moreover, due to its dependence on Majorana correlators, FAF lends itself to physical interpretation in terms of the underlying correlation structure. Finally, FAF can be estimated experimentally
using fermionic shadow tomography or related protocols. 
The number of measurements required to achieve an additive precision of FAF estimation scales polynomially with system size. 
This highlights its practical relevance for probing fermionic magic resources in many-body systems.

\textit{Fermionic antiflatness: simple examples.} 
The second component of our framework concerns the baseline behavior of fermionic magic resources, specifically the FAF, across a range of fundamental physical scenarios. 
To build intuition for the behavior of FAF in many-body states, we began by studying simple, analytically tractable examples. We showed that applying a local non-Gaussian operation to a fermionic Gaussian state generates a finite amount of FAF. 
Analyzing product states, we demonstrated that FAF generally scales extensively with system size. 
The study of Haar-random states further revealed that typical states exhibit nearly maximal FAF. Finally, examples involving random matrix product states and quantum circuits provided a general picture of how FAF behaves in many-body ground states and under ergodic quantum dynamics.

\textit{Fermionic magic resources in equilibrium settings.} 
The transverse-field Ising model, perturbed with non-Gaussian terms, served as a testbed for benchmarking the behavior of fermionic magic resources across different phases of matter and near quantum phase transitions. We found that a single non-Gaussian impurity induces a finite amount of FAF. In contrast, an extensive number of non-Gaussian terms --typically present in translationally invariant systems, leads to an extensive scaling of FAF, as demonstrated in the ANNNI model. Moreover, enhanced correlations near criticality further increase the value of FAF.

We found that critical points can manifest in the FAF either through a logarithmic divergence of subleading terms in its system-size scaling or through constant subleading contributions, depending on the boundary conditions. The logarithmic divergence mirrors the behavior of entanglement entropy at criticality, and the overall phenomenology of FAF in many-body ground states closely parallels that of participation entropy and stabilizer Rényi entropy.
We have shown that these universal features of FAF in ground states can be understood through its dependence on two-point Majorana correlations, particularly due to their generic behavior: exponential decay away from criticality and universal power-law scaling at critical points governed by conformal field theory.

Importantly, FAF not only captures information about phases and critical points, but also is able to identify special points in the phase diagram where the ground state becomes particularly simple from the perspective of fermionic Gaussian states. A prominent example is the Peschel-Emery point in the ANNNI model. 
While other quantities, such as entanglement entropy, vary smoothly in its vicinity, the FAF reveals a hidden structure in the ground state, associated with its proximity to a fermionic Gaussian state.

\textit{Fermionic magic resources in out-of-equilibrium scenarios.} 
We then investigated the behavior of FAF in out-of-equilibrium settings. Focusing on the impurity and ANNNI models, we found that FAF varies across the many-body spectrum in a manner similar to entanglement entropy, exhibiting a characteristic inverted-parabola shape as a function of energy. The value of FAF reaches its maximum in mid-spectrum eigenstates and approaches the value expected for Haar-random states as the system size increases. These results indicate that many-body eigenstates at finite energy densities are significantly more complex, relative to fermionic Gaussian states, than ground states.

We investigated the growth of FAF in ergodic many-body systems. In the model with a single non-Gaussian impurity, FAF exhibits linear growth in time, consistent with the finite-velocity spreading of beyond-Gaussian correlations throughout the system. 
In contrast, the ANNNI model, with non-Gaussian impurities distributed throughout the system, exhibits extensive FAF at times of order unity, similarly to the behavior of random quantum circuits. Regardless of the short-time behavior, in both types of systems, the FAF saturates at times scaling linearly with the system size, mirroring the behavior of 
participation entropy and SRE in ergodic many-body systems.
This stands in sharp contrast to Haar-random circuit models, where FAF saturates much more rapidly, at time scales logarithmic in system size. 
We argued that this difference is a generic feature of systems with conservation laws expressed as sums of local operators.
In the impurity and ANNNI models, energy conservation exemplifies this effect, slowing down the approach of FAF to values characteristic of featureless, typical states.

\subsection{Outlook}

The results of this work raise several questions about the structure of quantum many-body states and their implications for their classical simulability.

\textit{Theory of fermionic magic resource measures.}
The fermionic commutant offers a natural source of operators that can be used to define candidate measures of fermionic non-Gaussianity. 
While our construction relies on a specific family of such operators, a full mathematical characterization of the fermionic commutant, analogous to the well-established structure of the Clifford commutant~\cite{gross2021schurweylduality, Bittel25commutant}, remains an open problem. Developing such a framework would pave the way for defining a broader class of non-Gaussianity measures and for achieving a more complete understanding of non-Gaussian correlations in fermionic systems~\cite{Semenyakin_2025}.

Although the FAF studied in this work satisfies the basic criteria of a measure of non-Gaussianity, alternative measures may offer additional advantages, such as monotonicity under more general free-fermionic operations or a direct operational interpretation. An important open question is whether one can establish concrete connections between such directly computable measures of non-Gaussianity and existing measures of fermionic magic resources, including fermionic rank and fermionic Gaussian extent~\cite{Hebenstreit19, Cudby24gaussian, Reardon24extent, Bittel24optimal, Lyu24NGE, Coffman25magic}. Furthermore, since our analysis has focused exclusively on pure states, extending this framework to mixed states and identifying computationally meaningful non-Gaussianity measures in that setting remains an important direction for future work.
Two main directions are motivated by the close analogy between non-stabilizerness and fermionic magic. On the one hand, one can generalize the convex-roof construction introduced for the stabilizer entropy in Ref.\cite{leone2024stabilizer} to the case of fermionic antiflatness. Since this approach involves an optimization procedure and is therefore impractical for large systems, a complementary path is to construct non-Gaussianity witnesses for mixed states, following the ideas of Ref.\cite{haug2025efficientwitnessingtestingmagic}.

Fermionic Gaussian states and stabilizer states are two prominent classes of quantum states that are efficiently simulable on classical computers, yet they are fundamentally distinct~\cite{Projansky25}. Understanding the relationship between the resource theories of fermionic non-Gaussianity, non-stabilizerness, and entanglement remains an open question. For instance, are there meaningful connections between classes of states with limited non-Gaussianity and those with limited non-stabilizerness? A first step in this direction was taken in~\cite{Collura24fermionicgaussian}, which showed that random fermionic Gaussian states exhibit near-maximal values of stabilizer Rényi entropy.

\textit{Phenomenology of non-Gaussianity in many-body systems.}
Our investigation of the behavior of fermionic magic resources in many-body systems can be extended along several directions.

We have demonstrated that the features of FAF in many-body ground states can be understood through universal behavior of two-point Majorana correlation functions. A promising next step is to develop a more complete description of FAF and other measures of fermionic non-Gaussianity at quantum phase transitions, within the framework of conformal field theory, building on recent analogous progress for non-stabilizerness~\cite{Hoshino25SREcft}. Furthermore, it would be interesting to study fermionic magic resources in systems exhibiting Berezinskii–Kosterlitz–Thouless transitions, topological phenomena, or systems relevant for high energy physics, like the Sachdev-Ye-Kitaev model~\cite{Bera2025SYK, Sachdev_1993}. 

The fermionic magic resources studied here in ergodic many-body systems rapidly become abundant, causing states to quickly become complex from the perspective of fermionic Gaussian states. In contrast, in settings where ergodicity is \emph{broken}, such as in many-body localized systems~\cite{Nandkishore15, Alet18, abanin2019colloquium, sierant2024mbl}, models featuring quantum many-body scars~\cite{Serbyn21}, or lattice gauge theories~\cite{Brenes18lgt}, the fermionic magic resources may be limited and their growth hindered, similarly to entanglement entropy~\cite{DeChiara06, Znidaric08, Bardarson12}. This direction is especially compelling in light of recent arguments regarding the proximity of many-body localized models to free-fermionic systems~\cite{Vidmar21phenomenology, Krajewski22Restoring}.

Another important set of questions concerns possible transitions between dynamical regimes characterized by distinct behaviors of fermionic magic resources. In the presence of measurements, there exist phase transitions between regimes exhibiting different entanglement scaling~\cite{Yaodong19, Skinner19, Turkeshi2024coherent}, as well as transitions distinguished by stabilizer Rényi entropy scaling~\cite{Niroula24transition, Turkeshi2024coherent, Bejan24, Fux24separation, Tarabunga24transition}. Whether similar phenomena can be identified in the behavior of fermionic non-Gaussianity remains an open question for future studies.

\textit{Implications for classical simulation of many-body states.}
A better understanding of fermionic magic resources may improve our ability to perform classical simulations of quantum many-body systems. 
For example, a promising future direction is to explore how fermionic antiflatness connects to variational approaches in condensed matter physics~\cite{Bellomia25}, such as those based on Gutzwiller-projection paradigms~\cite{Becca2017}. 
While these methods are self-contained, they currently lack \textit{ab initio} validation regarding the regimes in which they perform reliably. 
Non-Gaussianity measures like FAF may provide a general framework to address these questions, which we leave for future investigation.
Furthermore, recently developed frameworks for simulating quantum states~\cite{Dias24classical, ReardonSmith24improved, Wille2025} or operators~\cite{Miller25majoranapropagation} evolving under fermionic Gaussian unitaries and sporadic non-Gaussian gates can be particularly useful for simulations of many-body systems with limited non-Gaussianity. Insight into the phenomenology of fermionic magic resources across various physical scenarios may help identify scenarios where such approaches offer significant advantages~\cite{Bravyi17impurity}.

Another interesting set of questions involves combining tensor-network methods with fermionic Gaussian operations. Such approaches aim to exploit fermionic Gaussian operators to reduce the entanglement in tensor-network states~\cite{Fishman15, Krumnow16, Krumnow21, Wu25fermDisent, fux2024disentanglingunitarydynamicsclassically}. Since fermionic Gaussian unitaries preserve fermionic non-Gaussianity, understanding the non-Gaussian features of many-body states could clarify both the opportunities and limitations of these techniques. Another challenge is to directly construct hybrid setups analogous to Clifford-augmented matrix product states~\cite{masotllima2024stabilizertensornetworks, Qian24prl, Lami25rmps, qian2024cliffordcircuitsaugmentedtimedependent, nakhl2024sta, liu2024classical}, where fermionic Gaussian unitaries would act alongside MPS, with the latter capturing the non-Gaussian resources of the state.
We leave these open questions for future investigation.

\vspace{1cm}
\begin{acknowledgments}
P. Sierant acknowledges conversations about fermionic states with M. Oszmaniec. X.T. thanks N. Dowling, L. Leone, A. Mele for discussions, and G. White, and thanks M. Collura, G. Lami, L. Lumia, A. Paviglianiti, and E. Tirrito for collaborations on related topics. 
P.S. and P.S. acknowledge insightful discussions with the members of BSC’s Quantic group.
P. Sierant acknowledges fellowship within the “Generación D” initiative, Red.es, Ministerio para la Transformación Digital y de la Función Pública, for talent attraction (C005/24-ED CV1), funded by the European Union NextGenerationEU funds, through PRTR.
P. Stornati acknowledges funding from the Spanish Ministry for Digital Transformation and the Civil Service of the Spanish Government through the QUANTUM ENIA project call - Quantum Spain, EU, through the Recovery, Transformation and Resilience Plan – NextGenerationEU, within the framework of Digital Spain 2026.
X.T. acknowledges support from DFG under Germany's Excellence Strategy – Cluster of Excellence Matter and Light for Quantum Computing (ML4Q) EXC 2004/1 – 390534769, and DFG Collaborative Research Center (CRC) 183 Project No. 277101999 - project B01.
\end{acknowledgments}

\textit{Data Availability. ---}
Numerical data are available in Ref.~\cite{sierant_2025_17641582}.

\vspace{1cm}
\appendix 
\begin{widetext}
\section{Additional information about fermionic commutant and construction of non-Gaussianity measures}
\label{app:commutant}

\subsection{Elements of fermionic commutant}
To show that the operators $|\Upsilon_{r_1,\dots,r_k}^{(k)}\rrangle$, given by Eq.~\eqref{eq:comop} belong to the $k$-th fermionic commutant, it suffices to show that the condition in Eq.~\eqref{eq:condition} holds for any generator in $\mathrm{O}(2N)$. 

We consider the set of reflections along the $m$-th axis, defined by $\mathcal{U}_{\mathrm{R}_m} |\gamma_n\rrangle=(-1)^{\delta_{m,n}}|\gamma_{n}\rrangle$, together with the rotations $\mathcal{V}_{m,m+1}(\theta)\equiv V_{m,m+1}(\theta)\otimes V^T_{m,m+1}(\theta)$ in the $(m,m+1)$ plane for $m\in [2N-1]$, cf. Eq.~\eqref{eq:rotat}.
First, consider the reflections. 
It is straightforward to see that $\mathcal{U}_{\mathrm{R}_l}^{\otimes k}\bigotimes_{m=1}^{k}|\gamma_{A_m}\gamma_{A_{m+1}}\rrangle  = \bigotimes_{m=1}^{k}|\gamma_{A_m}\gamma_{A_{m+1}}\rrangle $, since each reflection axis $l$ either appears in one of the disjoint subsets $A_1,A_2,\dots,A_k$ -- in which case the reflection introduces two minus signs that cancel -- or it appears in none of them, leaving the state $\bigotimes_{m=1}^{k}|\gamma_{A_m}\gamma_{A_{m+1}}\rrangle $ unchanged. This shows that reflections satisfy Eq.~\eqref{eq:condition}. 

It remains to verify Eq.~\eqref{eq:condition} for arbitrary rotation $V_{m,m+1}(\theta)$ for any angle $\theta$. 
If $m$ and $m+1$ are not present in any $A_1,A_2,\dots A_k$, the rotation acts trivially and the state $\bigotimes_{m=1}^{k}|\gamma_{A_m}\gamma_{A_{m+1}}\rrangle $ is left unchanged. 
Instead, for any Majorana string  $\gamma_S$ such that $S$ contains both $m$ and $m+1$, the rotation acts nontrivially only on $|\gamma_m \gamma_{m+1}\rrangle$. However, by direct inspection, we find the Majorana string is left unchanged, since 

\begin{equation}
    \mathcal{V}_{m,m+1} |\gamma_m\gamma_{m+1}\rrangle = (\cos(\theta)|\gamma_m\rrangle + \sin(\theta)|\gamma_{m+1}\rrangle) (-\sin(\theta)|\gamma_m\rrangle + \cos(\theta)|\gamma_{m+1}\rrangle) = |\gamma_m \gamma_{m+1}\rrangle\;,
\end{equation}
where we used $\gamma_n^2=I$ and noted that $\cos(\theta)\sin(\theta)|I\rrangle-\cos(\theta)\sin(\theta)|I\rrangle=0$.

We now consider two representative cases; all others follow by symmetry: (i) $m\in A_1$ and $m+1\notin A_2,A_3,\dots,A_k$ and (ii) $m\in A_1$ and $m+1\in A_2$. 

Let us first consider the case where $m\in A_1$ and $m+1$ does not appear in any other $A_j$. 
In this case, the sum in Eq.~\eqref{eq:comop} includes a term corresponding to the set $\tilde{A}^{(m)}_1=\{j \in A_1\;|\;j<m\}\cup \{m+1\} \cup \{ j\in A_1 \;|\; j>m+1\}$ where the element $m$ in $A_1$ is replaced by $m+1$. 
Then, expanding the rotations and performing straightforward algebra, we obtain
\begin{equation}
\begin{split}
        \mathcal{V}_{m,m+1}^{\otimes k}(\theta) (|\gamma_{A_1}\gamma_{A_2}\rrangle\cdots |\gamma_{A_k}\gamma_{A_1}\rrangle +|\gamma_{\tilde{A}^{(m)}_1}\gamma_{A_2}\rrangle\cdots |\gamma_{A_k}\gamma_{\tilde{A}^{(m)}_1}\rrangle)=|\gamma_{A_1}\gamma_{A_2}\rrangle\cdots |\gamma_{A_k}\gamma_{A_1}\rrangle +|\gamma_{\tilde{A}^{(m)}_1}\gamma_{A_2}\rrangle\cdots |\gamma_{A_k}\gamma_{\tilde{A}^{(m)}_1}\rrangle\;.
\end{split}
\end{equation}
Now, consider the second case where $m\in A_1$ and $m+1\in A_2$. Define $\tilde{A}^{(m)}_2=\{j \in A_2\;|\;j<m\}\cup \{m\} \cup \{ j\in A_2 \;|\; j>m+1\}$ where the element $m+1$ in $A_2$ is replaced by $m$. Performing similar algebraic manipulation, we find
\begin{equation}
    \begin{split}
        &\mathcal{V}_{m,m+1}^{\otimes k}(\theta) (|\gamma_{A_1}\gamma_{A_2}\rrangle\cdots |\gamma_{A_k}\gamma_{A_1}\rrangle +|\gamma_{\tilde{A}^{(m)}_1}\gamma_{\tilde{A}^{(m)}_2}\rrangle|\gamma_{\tilde{A}^{(m)}_2}\gamma_{A_3}\rrangle\cdots |\gamma_{A_k}\gamma_{\tilde{A}^{(m)}_1}\rrangle)=\\
        &\qquad =|\gamma_{A_1}\gamma_{A_2}\rrangle\cdots |\gamma_{A_k}\gamma_{A_1}\rrangle +|\gamma_{\tilde{A}^{(m)}_1}\gamma_{\tilde{A}^{(m)}_2}\rrangle|\gamma_{\tilde{A}^{(m)}_2}\gamma_{A_3}\rrangle\cdots |\gamma_{A_k}\gamma_{\tilde{A}^{(m)}_1}\rrangle\;.
\end{split}
\end{equation}
\end{widetext}
Summing over all possible configurations of disjoint sets $A_1,\dots,A_k\subset [2N]$ with given sizes $|A_n|=r_n $ for all $n\in [k]$, our results conclude that Eq.~\eqref{eq:condition} holds for all generators $\mathrm{O}(2N)$, as required. 

\subsection{Examples}
\label{app:COMMexampl}
In the following, we illustrate the structure of the fermionic commutant with a few simple examples focusing on \( k = 1 \) and \( k = 2 \).  

For \( k = 1 \), the only operators invariant under all Gaussian unitaries \( U_G \) with \( G \in \mathrm{SO}(2N) \) are the identity \( \gamma_\emptyset = \mathbb{1} \) on \( N \) qubits and the fermionic parity operator \( \mathcal{P} = (-1)^N \prod_{m=1}^{2N} \gamma_m \).  
However, only the identity is invariant under the reflection operator $X_N$, which provides the additional \( \mathbb{Z}_2 \) symmetry required to extend \( \mathrm{SO}(2N) \) to \( \mathrm{O}(2N) \), equivalent to the full matchgate group $\mathfrak{G}_N$.  
Hence, the fermionic commutant for \( k = 1 \) contains a single element, $\mathrm{Comm}_1(\mathfrak{G}_N)=\{\mathbb{1}\}$.

For \( k = 2 \), using Eq.~\eqref{eq:comop} and taking into account the condition $A_{k+1}=A_1$, we find
\begin{equation}\label{eq:comopK2}
    |\Upsilon_{r_1,r_2}^{(2)}\rrangle \equiv \sum_{\substack{A_1, A_2\subset [2N]\\  A_1\cap A_2=\varnothing\; \\ |A_1|=r_1, \, |A_2|=r_2}} |\gamma_{A_1}\gamma_{A_{2}}\rrangle \otimes  |\gamma_{A_2}\gamma_{A_{1}}\rrangle. 
\end{equation}
While all choices of $r_1,r_2\in\{0,\ldots,2N\}$ are allowed, not all operators $|\Upsilon_{r_1,r_2}^{(2)}\rrangle$ are independent.
Since the sets $A_1$ and $A_2$ are disjoint, we have
\[
\gamma_{A_1}\gamma_{A_2}=(-1)^{I(A_1,A_2)}\,\gamma_{A_1\cup A_2},
\]
where $I(A_1,A_2)=\bigl|\{(a_1,a_2)\in A_1\times A_2:\, a_1>a_2\}\bigr|$ is the number of inversions needed to restore ascending order of the Majorana string $\gamma_{A_1\cup A_2}$, and
$I(A_1,A_2)+I(A_2,A_1)=|A_1|\,|A_2|$.
These identities imply that
\begin{equation}\label{eq:comopK2_1}
    \bigl|\Upsilon_{m-r,r}^{(2)}\bigr\rrangle
    =(-1)^{|A_1|\,|A_2|} 
    \sum_{\substack{A_1,A_2\subset[2N]\\ A_1\cap A_2=\varnothing\\ |A_1|=m-r,\;|A_2|=r}}
    \bigl|\gamma_{A_1\cup A_2}\bigr\rrangle \otimes \bigl|\gamma_{A_1\cup A_2}\bigr\rrangle,
\end{equation}
showing that $\bigl|\Upsilon_{r,m-r}^{(2)}\bigr\rrangle \propto \bigl|\Upsilon_{m,0}^{(2)}\bigr\rrangle$ for all integers $0\le r\le m\le 2N$.
Therefore, there are exactly $2N+1$ independent elements of the $k=2$ fermionic commutant $\mathrm{Comm}_2(\mathfrak{G}_N)$, which can be chosen as the operators $\bigl|\Upsilon_{m,0}^{(2)}\bigr\rrangle$ for $0\le m\le 2N$.
In particular, the operator \(\bigl|\Upsilon_{1,1}^{(2)}\bigr\rrangle \propto \bigl|\Upsilon_{2,0}^{(2)}\bigr\rrangle\), which enters the fermionic overlap in Eq.~\eqref{eq:fermOV1} used to define FAF \(\mathcal{F}_1\), can be written explicitly as
\begin{equation}\label{eq:comopK2_11}
    \bigl|\Upsilon_{1,1}^{(2)}\bigr\rrangle
    = \sum_{\substack{0\le i<j\le 2N}}
      \bigl|\gamma_i \gamma_j\bigr\rrangle \otimes \bigl|\gamma_i \gamma_j\bigr\rrangle.
\end{equation}
In Appendix~\ref{app:faithfu} we discuss another element of the \(k=2\) fermionic commutant, the operator \(\bigl|\Upsilon^{(2)}_{2,2}\bigr\rrangle\).

To further illustrate the elements of the $k=2$ fermionic commutant, we specialize to $N=2$ qubits and obtain explicitly
$\mathrm{Comm}_2(\mathfrak{G}_2)=\{Q_0,Q_1,Q_2,Q_3,Q_4\}$, where
\begin{equation}
        Q_0 = (\mathbb{1} \otimes \mathbb{1})\otimes (\mathbb{1}\otimes \mathbb{1}),
      \nonumber  \end{equation}
\begin{eqnarray}
Q_1\!\!\!\!\!\!\! \!\! &= (X\otimes \mathbb{1}) \otimes  (X\otimes \mathbb{1}) + (Y\otimes \mathbb{1}) \otimes  (Y\otimes \mathbb{1})  \nonumber\\ & \qquad + (Z\otimes X) \otimes  (Z\otimes X)+(Z\otimes Y) \otimes  (Z\otimes Y),
\nonumber \end{eqnarray}
\begin{eqnarray}
 Q_2 \!\!\!\!\!\!\! \!\!&= (Z\otimes \mathbb{1})\otimes (Z\otimes \mathbb{1}) +  (\mathbb{1} \otimes Z)\otimes (\mathbb{1} \otimes Z )
        \nonumber \\ & \qquad+ (X\otimes X)\otimes( X\otimes X) + (X\otimes Y)\otimes (X\otimes Y) \nonumber \\& \qquad+ (Y\otimes Y) \otimes (Y \otimes Y) + (Y\otimes X)\otimes (Y \otimes X),\nonumber \end{eqnarray}
\begin{eqnarray}
Q_3 \!\!\!\!\!\!\! \!\! &= (\mathbb{1}\otimes X)\otimes (\mathbb{1}\otimes X) + (\mathbb{1}\otimes Y)\otimes (\mathbb{1}\otimes Y) \nonumber \\& \qquad+(X\otimes Z)\otimes (X \otimes Z) + (Y \otimes Z)\otimes (Y \otimes Z),\nonumber \end{eqnarray}
\begin{equation}
       Q_4= (Z\otimes Z)\otimes (Z\otimes Z)\;.
\nonumber  \end{equation}
In the expressions above, the Pauli operators acting on the two qubits are written using tensor product, e.g., $(Z\otimes X) \equiv Z_1 X_2$ and the round brackets group operators acting on the same replica.

As \(k\) increases beyond \(2\), the set of operators \(\bigl|\Upsilon_{r_1,\ldots,r_k}^{(k)}\bigr\rrangle\) in Eq.~\eqref{eq:comop} proliferates. 
A particularly transparent subclass in \(\mathrm{Comm}_{2k}(\mathfrak{G}_N)\) is provided by the operators employed by us to define the FAF \(\mathcal{F}_k\), namely
\begin{equation}\label{eq:2kCOMPO}
    |\Upsilon_{1,\dots,1}^{(2k)}\rrangle \equiv \sum_{ \substack{0\leq i_1, i_2 , \ldots , i_{2k}  \leq 2N \\ i_m \neq i_n \,\,\forall m,n} } \,\,\bigotimes_{m=1}^{k}|\gamma_{i_m}\gamma_{i_{m+1}}\rrangle, 
\end{equation}
with $i_{2k+1} \equiv i_1$.

\subsection{A remark on faithfulness}
\label{app:faithfu}
Each measure of non-Gaussianity \( \varphi_{r_1,r_2,\dots,r_k}^{(k)} \), defined in Eq.~\eqref{eq:gfa} using operators from the fermionic Gaussian commutant, is invariant under fermionic Gaussian unitaries by construction. However, the faithfulness and additivity of \( \varphi_{r_1,r_2,\dots,r_k}^{(k)} \) are not guaranteed.

We define a Choi state on \( k = 2 \) replicas as  
\begin{equation}
    |\Upsilon^{(2)}_{2,2}\rrangle = \sum_{\substack{A_1, A_2 \subset [2N] \\ A_1 \cap A_2 = \varnothing,\; |A_1| = |A_2| = 2}} |\gamma_{A_1} \gamma_{A_2} \rrangle \otimes |\gamma_{A_2} \gamma_{A_1} \rrangle \;.
\end{equation}
The measure \( \varphi^{(2)}_{2,2} \) induced by $|\Upsilon^{(2)}_{2,2}\rrangle$  is not faithful. To illustrate this, consider a system of \( N = 6 \) qubits in the state
\begin{equation}
\ket{\psi} = \ket{\Psi_\theta} \otimes \ket{00},
\end{equation}
where $\ket{\Psi_\theta}$ is the four-qubit fermionic non-Gaussian state defined in Eq.~\eqref{eq:psi_theta}. Direct calculation shows that $\varphi^{(2)}_{2,2}(\ket{\psi}) = 0$, even though the state $\ket{\psi}$, being a tensor product of $\ket{\Psi_\theta}$ with $\ket{00}$ is clearly \emph{not} fermionic Gaussian. Indeed, the FAF for this state is non-zero:
\begin{equation}
\mathcal{F}_k(\ket{\psi}) = 4\left(1 - \cos^{2k}\frac{\theta}{2}\right).
\end{equation}
Thus, the measure $\varphi^{(2)}_{2,2}$ vanishes for a state which is not fermionic Gaussian, demonstrating explicitly that it does not satisfy the requirement of faithfulness.

A numerical investigation of \( \varphi^{(2)}_{2,2} \) suggests that it vanishes for all states of \( N=6 \) qubits with fixed parity. Moreover, this behavior appears to be unique to the \( N=6 \) case; for \( N \neq 6 \), we did not find any example of a non-fermionic Gaussian state for which \( \varphi^{(2)}_{2,2} \) would be vanishing. This observation indicates that there might exist an additional criterion ensuring that all measures derived from the fermionic commutant are faithful. Identifying such a criterion remains an open question for future studies.

\section{Relation of fermionic antiflatness with other measures of non-Gaussianity}
\label{app:relation}

In this work, we focused on FAF as a measure of fermionic magic resources. In the following, we examine the relationship between FAF and the other measures of fermionic non-Gaussianity. We begin by discussing links between the FAF and non-Gaussian Entropy (NGE), which was recently introduced in~\cite{Lyu24NGE, Coffman25magic} and studied for systems with \( U(1) \) symmetry in~\cite{Lumia24}. Subsequently, we compare FAF with  fermionic rank, Gaussian fidelity and extent on several illustrative examples.

\subsection{Non-Gaussian Entropy}
The NGE is defined using a fermionic beam splitter, which is an operator acting on two copies of the state (i.e., on \( 2N \) qubits) and is implemented by a fermionic Gaussian unitary
\begin{equation}
W= \exp \left( \frac{\pi}{8} \sum_{j=1}^{2N} \gamma_j \gamma_{2N+j}\right),
\label{eq:W}
\end{equation}
where the Majorana operators for the first copy, $\gamma_{1},..., \gamma_{2N}$,  are defined as in Eq.~\eqref{eq:JW}, and the Majorana operators for the second copy, $\gamma_{2N+1},..., \gamma_{4N}$,  are defined analogously\footnote{We note that \cite{Lyu24NGE} considers states with even fermionic parity, hence, the Jordan-Wigner string in operators $\gamma_{2N+k}$ may start at the first qubit of the first or of the second copy of the system.}. Fermionic self-convolution transforms the density matrix $\rho$ of $N$ qubit system into
\begin{equation}
      \boxtimes^1 \rho = \mathrm{tr}_{2} \left[W(\rho \otimes \rho) W^{\dagger}\right],
      \label{eq:ngeBox1}
\end{equation}
where $\mathrm{tr}_{2} \left[.\right]$ denotes partial trace over the second copy of the system containing $N$ qubits. Iteratively, one defines $\boxtimes^q \rho$ for any integer $q \geq 0$, 
via $\boxtimes^q \rho = \boxtimes^1 (\boxtimes^{q-1} \rho)$ and $ \boxtimes^0 \rho = \rho$.

Even if \( \rho \) corresponds to a pure state, \( \boxtimes^1 \rho \) may be a mixed state. Analysis of the properties of fermionic self-convolution indicates that the purity of the resulting state serves as a measure of fermionic non-Gaussianity. For this reason, Refs.~\cite{Lyu24NGE,Coffman25magic} define the NGE as a measure of non-Gaussianity of state $\ket{\Psi}$ as 
\begin{equation}
    NGE_q( \ket{\Psi} )=-\log_2\left( \mathrm{tr}\left[ \left( \boxtimes^q \rho\right)^2 \right] \right), 
    \label{eq:nge1}
\end{equation}
where \( \rho = \ket{\Psi}\bra{\Psi} \)\footnote{Here, we focus on the Rényi-2 variant of the NGE. The behavior of the NGE for different R\'{e}nyi indices and in the von Neumann limit is analogous.} and \( q \geq 1 \) is an integer. As shown in~\cite{Lyu24NGE}, the NGE satisfies the properties of (i) Gaussian invariance, (ii) faithfulness, and (iii) additivity on product states with fixed fermionic parity, for any index \( q \). Being a R\'{e}nyi entropy, the NGE varies monotonically with \( q \).

Evaluating Eq.~\eqref{eq:nge1} is numerically challenging, as it requires applying the long-range operator $W$~\eqref{eq:W} to two copies of the state $\ket{\Psi}\otimes \ket{\Psi}$. Benchmarking on the ground state of the ANNNI model~\eqref{eq:ani}, we find that the state $W\ket{\Psi}\otimes\ket{\Psi}$ appearing in Eq.~\eqref{eq:ngeBox1} exhibits volume-law entanglement.
This limits the computability of $NGE_q(\ket{\Psi})$ to systems where a full state-vector representation of the $2N$-qubit state $W\ket{\Psi}\otimes\ket{\Psi}$ is feasible, i.e., to $N\lesssim 12$.
For this reason, we focus in the following on the limit \( q = \infty \), which was also considered in~\cite{Lumia24} for systems with fermion number conservation. In this limit, the calculation of the NGE simplifies, as Eq.~\eqref{eq:nge1} reduces to computing the R\'{e}nyi-2 entropy of a fermionic Gaussian state with a covariance matrix \( M = (M_{ij}) \) identical to that of the investigated state \( \ket{\Psi} \), which is given by
\begin{equation}
\label{eq:NGE}
    NGE_{\infty}(\ket{\Psi})= -\sum_i \log_2 \left( \left(\frac{1+\lambda_i}{2}\right)^2 +\left(\frac{1-\lambda_i}{2}\right)^2 \right),
\end{equation}
where $\lambda_i$ are the Williamson's eigenvalues of the correlation matrix $M$.

The expression for FAF in terms of Williamson's eigenvalues, Eq.~\eqref{eq:fafWilliam}, closely resembles that of \( \mathrm{NGE}_\infty \), Eq.~\eqref{eq:NGE}. This similarity immediately indicates that the NGE exhibits behavior analogous to FAF in several examples: the fermionic magic product state (Sec.~\ref{subsec:prod1}), the action of local gates on fermionic Gaussian states (Sec.~\ref{subsec:localGate}), and products of single-qubit states (Sec.~\ref{subsec:producsingle}).

The parallels extend to other settings considered in this work, as most examples involve states that are either close to fermionic Gaussian states (e.g., ground states with limited fermionic magic resources) or approximate typical Haar-random states. The latter include states obtained for large values of \( \chi \) in RMPS, highly excited eigenstates of ergodic many-body systems, and states in regimes where FAF approaches its saturation value characteristic of Haar-random states.

For states close to fermionic Gaussian states, the relationship between FAF and NGE can be established by noting that for such states \( \lambda_i = 1 - \epsilon_i \) with \( \epsilon_i \approx 0 \). In this regime, one can perform a Taylor expansion of Eq.~\eqref{eq:NGE}:
\begin{equation}
\label{eq:NGEexp}
    NGE_{\infty}(\ket{\Psi})=
    \sum_i \epsilon_i+ O(\epsilon_i^2) = \frac{1}{2}\mathcal{F}_1+O(\epsilon_i^2),
\end{equation}
showing that FAF and NGE differ only by a constant factor when \( \lambda_i \approx 1 \), i.e., when the state is close to being fermionic Gaussian.

When the fermionic magic resources are abundant, the Williamson's eigenvalues are close to zero, \( \lambda_i \approx 0 \), cf.~\eqref{eq:density}. This results in the following expansion:
\begin{equation}
\label{eq:NGEexp2}
    NGE_{\infty}(\ket{\Psi})= N - \sum_i \lambda_i^2 + O(\lambda_i^4)= \mathcal{F}_1 + O(\lambda_i^4).
    \end{equation}
Therefore, up to negligible \( O(\lambda_i^4) \) corrections, FAF and NGE coincide when the state is close to a typical Haar-random state. This shows that the saturation behavior of FAF and NGE toward their Haar-state values is fully analogous.

Finally, we note that experimental measurement of $NGE_q$ is possible when multiple copies of the state are available~\cite{Lyu24NGE}. In contrast, $NGE_{\infty}(\ket{\Psi})$ can be accessed via the eigenvalues of the covariance matrix in Eq.~\eqref{eq:NGE}, at a cost comparable to FAF.
Despite the similarities in the phenomenology of FAF and NGE, FAF offers several advantages over NGE. The FAF can be expressed as a simple polynomial in terms of two-point Majorana fermion correlation functions, which facilitates its physical interpretation, a feature not shared by NGE. Furthermore, FAF can be written as the expectation value of an operator acting on \( k \) copies of the state, whereas no such formulation is available for NGE. This property not only enables analytical calculations of FAF for typical states, but also allows for replacing two-qubit Haar-random gates with Clifford gates when computing averages of FAF for \( k = 1 \), significantly simplifying the computations.

\subsection{Fermionic non-Gaussianity measures requiring minimization}

Here we briefly recall several measures of non-Gaussianity that require optimization. 
\textit{Gaussian fidelity}~\cite{Dias24classical} is the maximum overlap of the state \(\ket{\Psi}\) with a fermionic Gaussian state:
\begin{equation}
    \label{eq:GausFid}
    f_G(\ket{\Psi})=\max_{U_G\in\mathfrak{G}_N}\bigl|\braket{\Psi|U_G|\mathbf{0}}\bigr|^2,
\end{equation}
where the maximization is over all fermionic Gaussian states \(U_G\ket{\mathbf{0}}\).
The \textit{fermionic rank} \(\chi(\ket{\Psi})\)~\cite{Dias24classical} is the minimal number of terms in a decomposition of \(\ket{\Psi}\) as a linear combination of fermionic Gaussian states \(\{\ket{\phi_j}\}\):
\begin{equation}
    \label{eq:fermRANK}
    \chi(\ket{\Psi})=\min\Bigl\{\chi\in\mathbb{N}:\ \ket{\Psi}=\sum_{j=1}^{\chi} a_j \ket{\phi_j}\Bigr\}.
\end{equation}
Finally, the \textit{fermionic Gaussian extent} \(\xi(\ket{\Psi})\)~\cite{ReardonSmith24improved,Dias24classical} is
\begin{equation}
    \label{eq:fermEXT}
    \xi(\ket{\Psi})=\min\Bigl\{ \|a\|_1:\ \ket{\Psi}=\sum_{j=1}^{\chi} a_j \ket{\phi_j} \Bigr\},
\end{equation}
where \(\|a\|_1=\sum_{j=1}^{\chi}|a_j|\). Unlike FAF and NGE, the evaluation of \(f_G(\ket{\Psi})\), \(\chi(\ket{\Psi})\) and \(\xi(\ket{\Psi})\) requires optimization, which distinguishes these quantities from FAF and NGE. The complexity of such optimization procedures as the system size \(N\) increases remains an open question.

The Gaussian fidelity, fermionic rank, and Gaussian extent are invariant under Gaussian unitaries \(U_G\), i.e., \(f_G(U_G\ket{\Psi})=f_G(\ket{\Psi})\), \(\chi(U_G\ket{\Psi})=\chi(\ket{\Psi})\), and \(\xi(U_G\ket{\Psi})=\xi(\ket{\Psi})\) for \(U_G\in\mathfrak{G}_N\), which follows directly from their definitions. Moreover, they are faithful measures of fermionic non-Gaussianity: \(f_G(\ket{\Psi})=1\) if and only if \(\ket{\Psi}\) is a fermionic Gaussian state; the same holds when \(\chi(\ket{\Psi})=1\) or \(\xi(\ket{\Psi})=1\).
In contrast with FAF and NGE, the (sub)multiplicativity properties of \(f_G(\ket{\Psi})\), \(\xi(\ket{\Psi})\), and \(\chi(\ket{\Psi})\) (equivalently, the (sub)additivity of their logarithms) remain unclear, with two exceptions. First, the Gaussian extent is multiplicative for four-qubit parity eigenstates~\cite{Cudby24gaussian,Reardon24extent}. Second, the multiplicativity of the Gaussian fidelity implies the multiplicativity of the Gaussian extent~\cite{Dias24classical}.

We close this section by discussing the relation between FAF and optimization-based measures of fermionic non-Gaussianity on two simple examples. As a first example, the fermionic magic state \(\ket{\Psi_\theta}\) from Eq.~\eqref{eq:psi_theta} admits the decomposition~\cite{ReardonSmith24improved}
\begin{align}
  \ket{\Psi_\theta} = &\cos\left(\frac{\theta}{4}\right)\ket{A_\theta} + i\sin\left(\frac{\theta}{4}\right)\ket{B_\theta},\label{eq:magic-state-orthogonal-decomposition}
\end{align}
where 
\begin{align}
    \ket{A_\theta} &=  \frac{e^{-i\frac{\theta}{4}}}{2}\ket{0000} +  \frac{e^{i\frac{\theta}{4}}}{2}\ket{0011} +  \frac{e^{i\frac{\theta}{4}}}{2}\ket{1100} +  \frac{e^{i\frac{3\theta}{4}}}{2}\ket{1111}, \nonumber \\
    \ket{B_\theta} &= \frac{e^{-i\frac{\theta}{4}}}{2}\ket{0000} -  \frac{e^{i\frac{\theta}{4}}}{2}\ket{0011} - \frac{e^{i\frac{\theta}{4}}}{2}\ket{1100} +  \frac{e^{i\frac{3\theta}{4}}}{2}\ket{1111},\nonumber 
\end{align}
with \(\mathcal{F}_k(\ket{A_\theta})=0=\mathcal{F}_k(\ket{B_\theta})\).
Equation~\eqref{eq:magic-state-orthogonal-decomposition} immediately yields the lower bound on the Gaussian fidelity
\(f_G(\ket{\Psi_\theta}) \ge \max\{\cos^2(\theta/4),\,\sin^2(\theta/4)\}\),
and shows that the fermionic rank satisfies \(\chi(\ket{\Psi_\theta})\le 2\).
Together with the fact that \(\mathcal{F}_k(\ket{\Psi_\theta})>0\) for any \(\theta \neq 2\pi m\) (with \(m\in\mathbb{Z}\)), this implies
\(\chi(\ket{\Psi_\theta})=2\) for any \(\theta\) that is not an integer multiple of \(2\pi\). An explicit optimization in Ref.~\cite{ReardonSmith24improved} shows that the fermionic Gaussian extent of \(\ket{\Psi_\theta}\) is
\(\xi(\ket{\Psi_\theta}) = 1 + |\sin(\theta/2)|\).
Combined with the multiplicativity of the Gaussian extent for products of four-qubit parity eigenstates~\cite{Reardon24extent}, this yields for the product of \(N/4\) fermionic magic states (see Eq.~\eqref{eq:psitheta_N})
\begin{align}
  \xi\bigl(\ket{\Psi_{\theta,N}}\bigr)
  = \bigl(1 + |\sin(\theta/2)|\bigr)^{N/4}.
\end{align}
Hence, the \emph{logarithm of the Gaussian extent} satisfies
\(\ln \xi\bigl(\ket{\Psi_{\theta,N}}\bigr) \propto N\),
exhibiting the same extensive scaling with system size \(N\) as FAF; cf. Eq.~\eqref{eq:FkProdN}.

In Sec.~\ref{subsec:localGate}, we analyzed the action of a local, parity-preserving, non-Gaussian gate \(O_a\) acting on at most \(a\) qubits applied to a fermionic Gaussian state \(\ket{\Psi_G}\). We showed that
\(\mathcal{F}_k(O_a\ket{\Psi_G}) \le 2a\),
i.e., a local non-Gaussian gate increases FAF only by a constant independent of the system size \(N\).
Since a local non-Gaussian gate can be expressed as a superposition of a finite number of Gaussian unitaries, the fermionic rank \(\chi(O_a\ket{\Psi_G})\) and the Gaussian extent \(\xi(O_a\ket{\Psi_G})\) are bounded above by a constant that depends on the chosen gate. 
For instance, the non-Gaussian unitary \(O_2=(1+i Z_1 Z_2)/\sqrt{2}\) acting on \(\ket{\Psi_G}\) yields
\(\chi(O_2\ket{\Psi_G}) \le 2\) and \(\xi(O_2\ket{\Psi_G}) \le \sqrt{2}\), both upper bounded by a system size independent constant, similarly to FAF.

Summarizing, the Gaussian fidelity, fermionic rank, and fermionic Gaussian extent are measures based on optomization procedures with intuitive interpretations in terms of decomposing a given state into superpositions of fermionic Gaussian states. Their practical computability at scale remains an open question; while they satisfy the basic requirements of faithfulness and invariance under Gaussian unitaries, their multiplicativity (and the additivity of their logarithms) is not established. This contrasts with FAF and NGE, whose additivity/subadditivity properties are well characterized. In basic examples such as the product states and states generated by local non-Gaussian gates, these measures share several qualitative features with FAF. Clarifying the relations among FAF, NGE, and other non-Gaussianity measures, including establishing inequalities analogous to those known in the non-stabilizerness framework~\cite{leone2024stabilizer}, remains an important open direction.

\section{Majorana correlation functions at the Ising critical point}
\label{app:IsingCrit}
In Sec.~\ref{subsec:critical}, we discussed how the universal conformal-field-theory form of the two-point Majorana correlation function, Eq.~\eqref{eq:pow}, fixes the behavior of FAF at the critical point, leading to the results shown in Fig.~\ref{fig:aniPBC}(b). Here, we discuss how interactions in the ANNNI model impact the two-point Majorana correlation functions at the Ising critical point.

In the ground state of the Ising model ($\lambda=0$) with PBC, the two-point Majorana correlation function for $r>0$ is~\cite{Mbeng24,Surace22}
\begin{equation}
\bigl|\langle \gamma_i\,\gamma_{i+2r} \rangle\bigr|=
\begin{cases}
0, & r \in \mathbb{Z},\\[6pt]
\displaystyle \bigl[N\,\sin\!\bigl(\pi r/N\bigr)\bigr]^{-1}, 
& r \in \mathbb{Z}+\tfrac{1}{2},
\end{cases}
\label{eq:isingPBCcorr}
\end{equation}
which, for half-integer $r \ll N$, reduces to $\bigl|\langle \gamma_i\,\gamma_{i+2r} \rangle\bigr| = c_0/r$ with $c_0=1/\pi$.

\begin{figure}[h]
    \centering
    \includegraphics[width=1\linewidth]{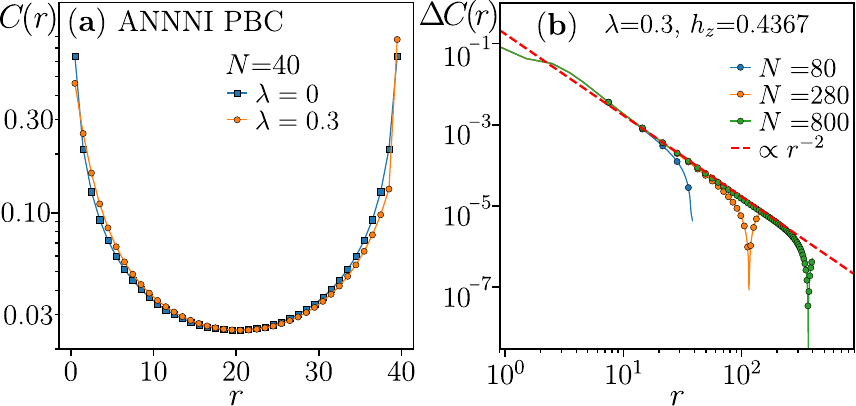}
    \caption{(a) Absolute value of two-point Majorana correlation function $C(r)\equiv \bigl|\langle \gamma_i\,\gamma_{i+2r} \rangle\bigr|$ at the critical point of the Ising model ($\lambda=0$, $h_z=1$) and the ANNNI model ($\lambda=0.3$, $h_z=0.4367$) with PBC.  (b) The difference $\Delta C(r)\equiv |C_{\mathrm{Ising}}(r) - C_{\mathrm{ANNNI}}(r)|$ between the two-point Majorana correlation functions in the Ising model $C_{\mathrm{Ising}}(r)$ ($\lambda=0$, $h_z=1$) and ANNNI model $C_{\mathrm{ANNNI}}(r)$ ($\lambda=0.3$, $h_z=0.4367$), decays according to a power-law $\Delta C(r) \approx r^{-2}$, consistently with $\Delta_{\mathrm{irr}}=2$.
    }
    \label{fig:IsiCOR}
\end{figure}

The introduction of interactions ($\lambda>0$) perturbs the two-point Majorana correlation function at the critical point, as shown in Fig.~\ref{fig:IsiCOR}(a).
In the ANNNI model the Ising critical point exists for any $0<\lambda<1/2$, with its position shifting from $h_z=1$ at $\lambda=0$ toward $h_z\!\to\!0$ as $\lambda\!\to\!1/2$.
For all $\lambda$ in this interval, interactions produce an \emph{additive} correction to the two-point Majorana correlation function that decays with distance as $r^{-\Delta_{\mathrm{irr}}}$ with $\Delta_{\mathrm{irr}}=2$; see Fig.~\ref{fig:IsiCOR}(b), consistent with Eq.~\eqref{eq:pow}.


%

\end{document}